\documentclass[superscriptaddress,prd,showpacs,twocolumn,nofootinbib]{revtex4-2}
\usepackage{graphicx}
\usepackage{xcolor}
\usepackage{epsf}
\usepackage{bm}
\usepackage{amsmath}
\usepackage{amsfonts}
\usepackage{amssymb}
\usepackage{epstopdf}
\usepackage{makecell}
\usepackage{booktabs}
\newcommand{\lsim}   {\mathrel{\mathop{\kern 0pt \rlap
  {\raise.2ex\hbox{$<$}}}
  \lower.9ex\hbox{\kern-.190em $\sim$}}}
\newcommand{\gsim}   {\mathrel{\mathop{\kern 0pt \rlap
  {\raise.2ex\hbox{$>$}}}
  \lower.9ex\hbox{\kern-.190em $\sim$}}}
\newcommand{\bw}{\begin{widetext}\begin{equation}}
\newcommand{\ew}{\end{equation}\end{widetext}}
\newcommand{\be}{\begin{equation}}
\newcommand{\ee}{\end{equation}}
\newcommand{\bea}{\begin{eqnarray}}
\newcommand{\eea}{\end{eqnarray}}

\begin{document}

\title{Bose-Einstein Condensate dark matter with logarithmic nonlinearity}

\author{Zahra Haghani}
\email{z.haghani@du.ac.ir}
\affiliation{School of Physics, Damghan University, Damghan, 36716-45667, Iran}
\author{Tiberiu Harko}
\email{tiberiu.harko@aira.astro.ro}
\affiliation{Faculty of Physics, Babe\c s-Bolyai University, 1 Kog\u alniceanu Street,
400084 Cluj-Napoca, Romania}
\affiliation{Astronomical Observatory, 19 Cire\c silor Street, 400487
Cluj-Napoca, Romania}

\begin{abstract}
If dark matter is composed of massive bosons, a Bose-Einstein Condensation
process must have occurred during the cosmological evolution. Therefore,
galactic dark matter may be in a form of a self-gravitating condensate, in the presence of 
 self-interactions. We consider the possibility that the
self-interacting potential of the condensate dark matter is of the
logarithmic form. In order to describe the condensate dark matter we use the
Gross-Pitaevskii equation with a logarithmic nonlinearity, and the
Thomas-Fermi approximation. With the use of the hydrodynamic representation
of the Gross-Pitaevskii equation we obtain the equation of state of the
condensate, which has the form of the ideal gas equation of state, with the
pressure proportional to the dark matter density. The basic equation
describing the density distribution of the static condensate is derived, and
its solution is obtained in the form of a series solution, constructed with the help of the Adomian Decomposition Method.  To test the model we consider the properties of the galactic rotation curves in the logarithmic Bose-Einstein Condensate dark matter scenario, by using a sample from the Spitzer Photometry and Accurate Rotation Curves (SPARC) data. The fit of the theoretical predictions of the rotation curves  with the observational data indicate that the logarithmic Bose-Einstein Condensate dark matter model gives an acceptable description of the SPARC data, and thus it may be considered as a possible candidate for the in depth understanding of the dark matter properties.
\end{abstract}

\date{\today }
\maketitle
\tableofcontents

\section{Introduction}

The possibility of the existence of dark matter (DM) is one of the
basic assumptions of present day astrophysics and cosmology \cite{d1a,d2a,d3a,d4a,d5a,d6a,d7a,d8a,d9a}. The first evidence for its presence in the Universe was provided by the study of the galactic rotation curves by Zwicky \cite{Zwicky1,Zwicky2}. More exactly, dark matter is necessary to explain the rotation curves of spiral galaxies, whose rotation curves decay much more slowly than predicted  by 
considering the effects of baryonic matter (stars and galactic gas) only. The observations show the rotational velocity of hydrogen clouds in stable circular orbits gravitating around the galactic center, the rotational velocities first increase close to the galactic center, and thus they can be described by the Newtonian theory. However, at larger distances from the galaxy they remain approximately constant, with the velocity reaching an asymptotic value of the order of $v_{tg\infty} \sim 200 - 300$ km/s. Thus the rotation curves lead to a mass profile of the form $M(r) =rv^2_ {tg\infty}/G$, where $G$ is the Newtonian gravitational constant. This result suggest that within a distance $r$ from the galactic center, the mass profile increases linearly with $r$, and this result is also valid at large distances where very little luminous (baryonic) matter can be found.

This observational result is considered as the basic evidence for the existence of a new form of matter, probably consisting of particle(s) that do not belong to the present day standard model of particle physics. Hence, the observations of the rotation curves provides the most convincing and powerful support for the existence of DM \cite{Persic,Read,Salucci,Haghi}. 

Other important cosmological and astrophysical observations have also supported the hypothesis of the existence of dark matter. The recent determinations of the cosmological
parameters from the measurements by the Planck satellite  of the temperature fluctuations of the Cosmic Microwave Background
Radiation \cite{Planck} have shown that dark matter must exist as an independent form of matter, together with baryonic matter. Determinations of the cosmological parameters by using
the Planck data have convincingly shown that the Universe is composed of 4\% baryons, 22\%  dark matter (which is nonbaryonic), and
74\% dark energy \cite{Planck}.

These findings did confirm the basic assumptions of the standard $%
\Lambda $ Cold Dark Matter ($\Lambda$CDM) cosmological model, which require the presence of a cosmological constant and of dark matter to fully explain the cosmological observations. An alternative to the $\Lambda$CDM model is represented by the dark energy and modified gravity theories \cite{PeRa03,Pa03,DEreviews, acc, Od,Od1,Od2}. For a consistent interpretation of the data gravitational lensing also requires the existence of dark matter  \cite{17, Wegg, Munoz, Chuda, Liu, Liu1}. 

A very powerful evidence for the presence of dark matter at cosmological scales  is  presented by the observations of the Bullet Cluster (1E 0657-56), a cluster in which the dark matter and the baryonic matter components have become separated as the result of a past collision of its two components  \cite{massey2007dark, Bul1, Bul2}. 

From a theoretical point of view dark matter models can generally be classified into three types, cold, warm and hot dark
matter models, respectively \cite{Overduin}. This classification  is based on the values of the energies of the particles composing dark matter. There are many candidates for dark matter, like, for example the WIMPs (Weakly Interacting Massive Particles) and the axions \cite{Overduin}. WIMPs are massive, yet undetected particles, which interact via the weak force \cite{Cui,Matsumoto}.
Axions are bosons that were first proposed in the framework of the physics of the elementary particles to solve the strong CP problem
\cite{Mielke, Schwabe}. The attempts for their experimental detection did not prove successful yet. If dark matter is composed of axions, then at low temperatures the axion gas could experience a phase transition to form a Bose-Einstein Condensate (BEC).

Theories that try to explain the galactic observations without postulating the existence of dark matter have also been investigated. These theories assume that the
law of gravity is modified at the galactic scales. One of the first theories explaining dark matter without dark matter is the MOND
theory (Modified Newtonian Dynamics) \cite{Milgrom}. Many other modified gravity
theories have been investigated as an alternative to dark matter
\cite{alt1,alt2,alt3,alt4,alt5,alt6,alt7,alt8,alt9,alt10,alt11,alt12,alt13,alt14,alt15,alt16,alt17,alt18,alt19,alt20}.

When analyzing the properties of dark matter one must take into account several aspects. The mass and nature of the dark matter particle is certainly a fundamental issue, but equally important is the thermodynamic state in which dark matter exists. An intriguing  possibility, very attractive both theoretically and observationally, is that dark matter is in the form of a Bose-Einstein  Condensate. This possibility was initially considered in \cite{Mem1,Mem2}, and later on in \cite{sin}. It is a standard result in statistical physics that if the temperature of a bosonic system becomes lower than the critical temperature given by
\begin{equation}\label{Tcr}
	T_C=\frac{2\pi \hbar}{mk_B} \left[\frac{n}{\zeta (3/2)} \right] ^{2/3},
\end{equation}
 a phase transition occurs in the system, with most of the bosons occupying the lowest quantum state \cite{Dalfovo, Pita, Pethick,ZNG}. In Eq.~(\ref{Tcr}) $n$ is the bosonic particle number density, $\hbar$ is the Planck constant, $k_B$ is Boltzmann's constant, while $\zeta(3/2)$ denotes the Riemann zeta function.
Thus, if dark matter is made of bosonic particles, such as fundamental scalars, then the possibility of a self-gravitating BEC on an astrophysical scale cannot be excluded a priori. In \cite{Sik} it was pointed out that if dark matter is formed of axions, they can form Bose-Einstein Condensate not only because of their self interactions,  but as a result of the gravitational interactions. Even the phase space density is enough large, axions will form Bose-Einstein Condensate only if they reach thermal equilibrium. Since the axions are very weakly coupled, it is the gravitational interaction that rethermalizes the axion Bose-Einstein Condensate time scales shorter than the age of the Universe.

In the Bose-Einstein condensate state bosons occupy the lowest possible quantum state simultaneously. This is in contrast to the behavior of the fermions, whose states are restricted by the Pauli exclusion principle. The Bose-Einstein Condensation process is well modeled by the nonlinear Schr\"{o}dinger equation (SE) for the macroscopic wave function of the system. The nonlinear Schr\"{o}dinger for the bosonic condensate phase  is also known as the Gross-Pitaevskii equation (GP). The Gross-Pitaevskii equation is obtained in the long-wavelength approximation, and it is successfully used to describe theoretically dilute Bose-Einstein Condensates.

The general form of the Gross-Pitaevskii equation is given by \cite{Dalfovo, Pita, Pethick,ZNG}
\begin{equation}
	i \hbar \frac{\partial \Psi}{\partial t}=\left[-\frac{\hbar^2}{2m}\nabla^2 + V +\int{|\Psi \left(x',t \right)|^2 U \left(|x-x'| \right)}dx'\right]\Psi,
\end{equation}
where by $m$ we denote the mass of the particle in the BEC,  $U \left(|x-x'| \right)$ is the self-interaction potential between bosons, and $V$ is the external potential. In the case of a weakly interacting system, the self-interaction potential $U$ can be approximated by a contact potential $U \left(|x-x'| \right) = U_0 \delta \left(|x-x'| \right)$, where $U_0$ is a constant, and thus we obtain the Gross-Pitaevskii equation with quadratic nonlinearity
\begin{equation}\label{eq1}
	i \hbar \frac{\partial \Psi}{\partial t}=\left[-\frac{\hbar^2}{2m}\nabla^2+V\right]\Psi + U_0 |\Psi|^2 \Psi.
\end{equation}

The density of the condensed bosonic dark matter is given by $\rho=m\left|\Psi\right|^2$.  

Assuming the bosonic system is confined by the gravitational potential, $V$ must satisfy the Poisson equation, given by \cite{BoHa07},
\begin{equation}\label{eq2}
\Delta V=4\pi G \rho=4\pi G m\left|\Psi\right|^2.
\end{equation}
 
 The system of coupled nonlinear differential equations (\ref{eq1}) and (\ref{eq2}) allow us to obtain a full description of the properties of the dark matter halos, including the dynamical evolution of the galactic rotation curves \cite{BoHa07,Cra, HaM}. The density distribution of the static spherically symmetric galactic dark matter condensate can be obtained as \cite{BoHa07}
\begin{equation}
\rho (r)=\rho_c\frac{\sin (kr)}{kr},
\end{equation}
where $\rho_c$ is the central density, and $k=\sqrt{Gm^3/\hbar ^2l_a}$, where $l_a$ is the scattering length of the particles in the condensate,  is a constant. A powerful method for discriminating between the  condensate dark matter model and other approaches to the dark matter problem is represented by the study of the light deflection by galaxies and the gravitational lensing. Some of the theoretical tools necessary to compare the predictions of the condensate dark matter model and the observational data on the galactic rotation curves were presented in \cite{BoHa07}.

 The linear Schr\"{o}dinger-Poisson system of equations was also used to investigate the evolution of cosmic voids, and for finding wave mechanical solutions for the dynamics in a standard cosmological background with appropriate boundary conditions \cite{Brook}. Assuming that dark matter does indeed exist in the form of a Bose-Einstein Condensate, its description should be done with the use of the coupled nonlinear Schr\"{o}dinger-Poisson system.

Another relevant example of a Bose-Einstein Condensate,  driven by the time-dependent Gross-Pitaevskii system of equations, is represented by  a condensate of scalar field particles at zero temperature in the mean field approximation. In cylindrical coordinates the dynamics of the system is described by the equations \cite{gonzalez}
\begin{equation}
	i \frac{\partial \psi}{\partial t} = -\frac{1}{2}\left(\frac{\partial^2 \psi}{\partial r^2} + \frac{1}{r}\frac{\partial \psi}{\partial r}+\frac{\partial^2 \psi}{\partial z^2}\right) + U \psi + \Lambda |\psi|^2 \psi ,
\end{equation}
\begin{equation}
	\frac{\partial^2 U}{\partial r^2} + \frac{1}{r}\frac{\partial U}{\partial r}+\frac{\partial^2 U}{\partial z^2} = \psi^{*} \psi,
\end{equation}
where $\psi$ and $U$ are the wave function and the gravitational potential, respectively.

The properties of the Bose-Einstein Condensate dark matter with quadratic nonlinearity have been intensively investigated in the literature, and their applications in astrophysics and cosmology have been considered in detail.  In particular, in \cite{harko2011b} it was pointed out that Bose-Einstein condensation of dark matter can successfully address and solve the core-cusp problem of the galactic structure.

The cosmological evolution of the finite temperature condensed dark matter was investigated in \cite{harkoandgabi}, by adopting for the space-time geometry the flat Friedmann-Robertson-Walker metric, and by including in the model a nonlinear term related to the density.  The cosmological perturbations of the Bose-Einstein Condensate dark matter were considered in \cite{harko2011}. 

The lightest possible class of dark matter candidates is called Ultralight dark matter (ULDM). This form of dark matter 
is described by a massive boson (of mass $m$), produced non-thermally in the early Universe. The bosonic particle can describe dark matter if it has a  mass in the range  $10^{-28}\;{\rm eV} < m < eV$ \cite{N1}. In the literature ULDM is also known under other various names as wave dark matter, ultra-light axions, scalar field dark matter etc. Wave dark matter is usually considered  as a bosonic dark matter candidate, with mass smaller than 30 eV. The de Broglie wavelength of these particle is greater than the average interparticle separation in a galaxy like the Milky Way. Therefore, this type of dark matter can be described by a set of classical waves \cite{N2}. ULDM models can be further classified into (at least) three classes, according to the different non-linear evolution and structures they form in galaxies \cite{N3}. These classes are represented by the fuzzy dark matter (FDM), the self-interacting fuzzy dark matter (SIFDM), and the DM superfluid. Each of these classes comprise many models. For the fuzzy dark matter models the relevant mass scale of the dark matter particle is of the order of $m\approx 10^{-22}$ eV.

A scalar field description of the galactic dark matter, proposed in \cite{mielke}, can be considered as representing a gravitationally confined Bose-Einstein Condensate, having galactic dimensions.  Cold Boson stars can also form, and they can be as well considered as  viable astrophysical objects. The properties of the boson stars were considered in both the non-relativistic and relativistic limits in \cite{pierre-harko}, by describing the star as a Bose-Einstein Condensate with quadratic non-linearity. For studies of the various properties and implications of the Bose-Einstein Condensate dark matter see \cite{harko2014, harko2019, Brook1, Del, Coh, Pozo}.   

Bose-Einstein Condensates are quantum objects having macroscopic sizes. Hence they are very good candidates for testing fundamental physics phenomena at both theoretical and experimental level, including the Schr\"{o}dinger equation and gravitational theories, as well as the possible relations between these two theories. Quantum reflection of up to 60\% has been observed in Bose-Einstein Condensate experiments \cite{pasquini}. Freely evolving degenerate quantum gases have been observed in microgravity experiments for large time intervals, and thus these experiments opened the possibility of the testing of the nonlinear form of the Schr\"{o}dinger equation \cite{hermann}.

The analogy between dark matter's superfluidity and the physics of cold atomic systems was considered in \cite{DMsuperfluidity}, in a study that emphasized the importance of the cosmological tests of the theory.
 
 The experimental and theoretical study of the effects of the interaction between coherent states of matter, such as the Bose-Einstein Condensates, and the gravitational field, may lead to an answer to the important question of whether it is possible for dark matter to exist in a coherent state, with the gravitational force acting as the only form of interaction with ordinary baryonic matter.

Despite the remarkable success of the Schr\"{o}dinger equation  with quadratic nonlinearity, it turns out that in many interesting physical situations, like, for example,  high matter densities,   the approximation of the dilute bosonic systems cannot give a good physical description of the system. Therefore,  models that would include multi-body interactions, and long-range particle correlations are necessary to describe the properties of bosonic systems. Thus, to understand the behavior of the bosonic condensates in realistic physical situations, it is necessary to replace the quadratic self-interaction model, and to consider more general forms of the particles self-interaction potential, as well as of the Gross-Pitaevskii equation. One of the possibilities that have been considered in some detail, and which could open some new perspectives for the understanding of dark matter, and of its astrophysical properties,  is the Gross-Pitaevskii equation with the logarithmic nonlinearity.

The first formulation of a nonlinear quantum evolution equation, which considers the separability of non-interacting states is the logarithmic Schr\"{o}dinger equation, proposed in  \cite{creation}. The logarithmic nonlinear Schr\"{o}dinger equation has found many applications in condensed matter physics. Multi-body interactions are an important characteristic for the logarithmic Bose liquids, implying the existence of new degrees of freedom \cite{rmp}. The analysis of the behavior of two-body and three-body interactions shows that the stability in the logarithmic BECs exceeds the stability of the classical ones, with quadratic self-interaction \cite{Bouharia}.

There are many quantum systems in which the logarithmic nonlinearity plays a major role. The deformed vacuum wave dispersion and observations of the cosmic ray properties have been explained by the Schr\"{o}dinger equation with a generalized nonlinearity \cite{rmp}.  The logarithmic Schr\"{o}dinger equation has found important applications in fields such as nuclear physics \cite{kartavenki}, quantum optics \cite{buljan}, stochastic quantum mechanics \cite{lemos},  quantum liquids and superfluidity \cite{avdeenkov,Bouharia,rmp,natur}, as well as in the theories of the  gravitational field \cite{grav,acta}, respectively.

One of the important properties of logarithmic nonlinearity is that the energy additivity, and the separability of non-interacting subsystems in non-relativistic quantum mechanics are not modified as compared to standard quantum mechanics \cite{rmp}. 

The logarithmic Schr\"{o}dinger equation has in the case of the central potential and the non-zero angular momentum cases the same symmetries found in the linear case \cite{centralpotential}. Logarithmic nonlinear systems posses as an universal property the hydrodynamic description. Important  properties of the logarithmic Schr\"{o}dinger equation  are the conservation of the norm of the wave function, dimensional homogeneity, and the Galilean invariance. The nonlinearity of the system can not produce quantum entanglement, and thus the logarithmic nonlinearity guarantees the separation of the product states from the time evolution.

Bose-Einstein Condensates with logarithmic nonlinearity have been also been investigated for the possible description of the state of dark matter. By using the theory of scale relativity a generalized Gross-Pitaevskii equation, including a logarithmic term, was proposed in \cite{Chav1} and \cite{Chav2}, respectively.  The presence of the logarithmic nonlinearity in the Bose-Einstein Condensate dark matter may suggest the existence of regions in which the presence of the self-interaction leads to solitonic behavior of the galactic halo \cite{2002}.

The Jeans instability of the dark matter halos, described by a generalized Schr\"{o}dinger equation, in the presence of logarithmic non-linearity, associated with an effective temperature and a source of dissipation, was considered in \cite{Our1}. In \cite{Our2} the Benjamin-Feir type modulational instability of the nonlinear Schr\"{o}dinger equation with a logarithmic nonlinearity was analyzed.  

 The stability and free expansion of a one-dimensional logarithmic Bose-Einstein condensate was considered in \cite{rod}. A comparative study of the propagation of sound pulses in elongated Bose liquids and Bose-Einstein condensates in Gross-Pitaevskii and logarithmic models was performed in \cite{Zlo22}. In \cite{Our3}  a discussion of the properties of the logarithmic Bose-Einstein Condensates, and for their implications for dark matter, and modified gravity investigations was presented. 
The gravitational collapse of dark matter halos in the presence of a logarithmic nonlinearity was considered in \cite{Stef}.

It is the main goal of the present work to consider an important possibility for testing the validity of the logarithmic Bose-Einstein Condensate dark matter model, namely, the behavior of the galactic rotation curves, and  of their physical characteristics. As a starting point of our analysis we assume that dark matter can be described by a  logarithmic nonlinear Schr\"{o}dinger type equation,  which includes both the terms of the linear Schr\"{o}dinger equation, as well as a nonlinear term of the form $b\ln \left[\left|\Psi \left(\vec{r},t\right)\right|/\left|\Psi_0\right|^2\right]$, where $b$ is the strength of the logarithmic nonlinearity in units of energy, and which describes the self-interaction properties of dark matter. Hence, the present dark matter model can be considered as belonging to the SIFDM class of dark matter theories.  

  To obtain a clear physical picture of the galactic dynamics and dark matter distribution, we introduce for the logarithmic Gross-Pitaevskii equation the Madelung representation of the wave function \cite{mad}, which allows to represent dynamical properties  of the condensate in terms of a continuity, and Euler type fluid mechanical equation, respectively, in the presence of the gravitational potential satisfying the Poisson equation. Moreover, we also introduce the Thomas-Fermi approximation, which allows us to neglect the effects of the quantum potential in the Euler equation. Furthermore, we assume that the dark matter distribution is spherically symmetric. An important property of the logarithmic Bose-Einstein Condensate is that it satisfies the ideal gas equation of state, with its pressure proportional to the condensate density. We also take into account the possible rotation of the condensate. 
  
  In order to obtain the properties of the logarithmic dark matter halo the coupled system of the nonlinear logarithmic Schr\"{o}dinger and Poisson equations must be solved. In static spherical symmetry the system can be reduced to a single second order nonlinear differential equation. We obtain an approximate series solution of this equation by using the Adomian Decomposition Method \cite{Ad,Ad1}, or more exactly, a version of the method called the Laplace-Adomian Decomposition Method. This approach allows to obtain a series solution of the basic evolution equation that reproduces at an acceptable level the exact numerical solution. It also allows to obtain an analytical representation of the tangential velocity of the massive test particles moving in circular stable trajectories around the galactic center.
  
  To test the viability of the logarithmic Bose-Einstein Condensate dark matter model we compare the theoretical predictions on the behavior of the galactic rotation curves with a set of observational data of the SPARC database \cite{Sparc}.  The SPARC database has been extensively used to test modified gravity or alternative dark matter models \cite{Sparc1,Sparc2,Sparc3,Sparc4,Sparc5,Sparc6,Sparc7,Sparc8,Sparc9}. The rotation curves of high-resolution low surface brightness  and SPARC galaxies have been investigated in \cite{Sparc10}, by considering two scalar field dark matter profiles, the soliton and Navarro-Frenk-White profile in the fuzzy dark matter model, arising from cosmological simulations of real, non-interacting scalar fields at zero temperature, and  the multistate scalar field dark matter profile, representing an exact solution to the Einstein–Klein–Gordon equations for a real, self-interacting scalar field, with finite temperature into the field potential, introducing several quantum states as a realistic model for a galactic halo. The analytical scalar field dark matter models fits the observations as well as or better than the empirical soliton plus NFW profiles, and they reproduce naturally the wiggles present in some galaxies. Isolated fuzzy-dark-matter lumps, made of ultralight axion particles with masses arising due to distinct SU(2) Yang–Mills scales and the Planck mass $M_P$ were investigated in \cite{Sparc11}, by fitting the  model  to the observed rotation curves of the low-surface-brightness galaxies of the SPARC catalogue.
  
  Our results show that the logarithmic BEC model can give a satisfactory description of the observational data, and hence it can be considered as a viable dark matter model.  The comparison with the observational data also allows us to obtain tight constraints on the parameters of the logarithmic BEC model, including the rotation velocity of the dark matter halo.   
   
The present paper is organized as follows. We present the basic theoretical model of the Bose-Einstein Condensate with logarithmic nonlinearity, as well as its hydrodynamical representation, in Section~\ref{sect1}. The equilibrium equation and the generalized Lane-Emden equation are also obtained. The density, mass and tangential velocity profiles of the logarithmic BEC condensate are obtained, via the Adomian Decomposition Method, in Section~\ref{sect2}. The comparison of the theoretical predictions of the logarithmic Bose-Einstein Condensate dark matter model and a sample of observational galactic rotation curves is presented in Section~\ref{sect3}. We discuss and conclude our results in Section~\ref{sect4}. The details of the derivation of the hydrodynamic representation of the Gross-Pitaevskii equation are given in Appendix~\ref{app1}.

\section{Bose-Einstein Condensate Dark matter with logarithmic nonlinearity}\label{sect1}

The general nonlinear Schrodinger equation (also known as Gross-Pitaevskii
equation) used to describe a BEC is given by \cite{Bar}
\begin{equation}
i\hbar \dfrac{\partial \Psi \left( \vec{r},t\right) }{\partial t}=\left[ -%
\dfrac{\hbar ^{2}}{2m}\nabla ^{2}+mV_{ext}+g^{\prime }\left( |\Psi \left(
\vec{r},t\right) |^{2}\right) \right] \Psi \left( \vec{r},t\right) ,
\label{eq:nonlinearSE}
\end{equation}%
where $m$ is the mass of the condensate particle, and $V_{ext}$ is the external
potential. Moreover, the term $g^{\prime }\left( |\Psi |^{2}\right) $ corresponds
the self-interaction of the particles in the bosonic system. In the following we will denote by a prime
the derivative with respect to the independent variable.

For the logarithmic BEC, the self-interaction term is given by \cite{creation, Zlo22}
\begin{equation}\
g^{\prime }\left( |\Psi \left( \vec{r},t\right) |^{2}\right) =b\ln \frac{%
|\Psi \left( \vec{r},t\right) |^{2}}{|\Psi _{0}|^{2}},
\label{eq:Log interaction}
\end{equation}%
where $b$ and $\Psi _{0}$ are constants. The constant term $|\Psi _{0}|^{2}$, having the physical interpretation as a particle number density, is introduced to make the logarithmic term  dimensionless. The normalization condition of the wave function 
imposes the constraint $\int |\Psi \left( \vec{r},t\right) |^{2}d^{3}\vec{r}=N$,
where $N$ is the total number of particles in the condensate.

\subsection{The equation of state of the logarithmic BEC}

Hence, by adopting for $g^{\prime }\left( |\Psi \left( \vec{r},t\right) |^{2}\right)$ the expression given by Eq.~(\ref{eq:Log interaction}),  the Gross-Pitaevskii Eq.~(\ref{eq:nonlinearSE}) in the presence of a
logarithmic self-interaction term takes the form
\begin{equation}
i\hbar \dfrac{\partial \Psi \left( \vec{r},t\right) }{\partial t}=\left[ -%
\dfrac{\hbar ^{2}}{2m}\nabla ^{2}+mV_{ext}+b\ln \frac{|\Psi \left( \vec{r}%
,t\right) |^{2}}{|\Psi _{0}|^{2}}\right] \Psi \left( \vec{r},t\right) .
\label{eq:log SE}
\end{equation}

We consider now the hydrodynamic representation  of Eq.~(\ref{eq:log SE}), which allows us to find the expression of the quantum pressure of the logarithmic BEC
dark matter. Firstly, we introduce for the wave function of the BEC dark matter with logarithmic nonlinearity the Madelung representation \cite{mad}
\begin{equation}
\Psi \left( \vec{r},t\right) =\sqrt{n\left( \vec{r},t\right) }e^{i\Phi(\vec{r}%
,t)/\hbar },  \label{eq:wavefunction}
\end{equation}%
where $n(\vec{r},t)=|\Psi \left( \vec{r},t\right) |^{2}$ is the particle
number density of  the bosonic system, and $\Phi(\vec{r},t)$ is a phase factor. 

This specific representation  of the wave function immediately leads to the hydrodynamic or Madelung representation of
quantum mechanics, and of the nonlinear Schr\"{o}dinger equation. We also introduce the matter density of the condensate,
defined as 
\be
\rho_{m}\left( \vec{r},t\right) =mn(\vec{r},t)=m|\Psi \left(
\vec{r},t\right) |^{2}. 
\ee

With the wave function $\Psi \left( \vec{r}%
,t\right) $ as given in Eq.~(\ref{eq:log SE}) replaced by Eq.~(\ref{eq:wavefunction}), the logarithmic Bose-Einstein Condensate
Eq.~(\ref{eq:log SE}) becomes
\begin{eqnarray}
\hspace{-0.5cm} &&i\hbar \dfrac{\partial }{\partial t}\left[ \sqrt{n\left(
\vec{r},t\right) }e^{i\frac{\Phi \left( \vec{r},t\right) }{\hbar }}\right] =-%
\dfrac{\hbar ^{2}}{2m}\nabla ^{2}\Bigg[\sqrt{n\left( \vec{r},t\right) }e^{i%
\frac{\Phi\left( \vec{r},t\right) }{\hbar }}\Bigg]+  \notag
\label{eq:LogSEwithwavefunction} \\
\hspace{-0.5cm} &&\left(m  V_{ext}+b\ln \frac{|\Psi |^{2}}{|\Psi _{0}|^{2}}%
\right) \sqrt{n\left( \vec{r},t\right) }e^{i\frac{\Phi\left( \vec{r},t\right) }{%
\hbar }}=-\dfrac{\hbar ^{2}}{2m}\times  \notag \\
\hspace{-0.5cm} &&\nabla ^{2}\Bigg[\sqrt{n\left( \vec{r},t\right) }e^{i\frac{%
\Phi\left( \vec{r},t\right) }{\hbar }}\Bigg]+V\left( \vec{r},t\right) \left(
\sqrt{n\left( \vec{r},t\right) }e^{i\frac{\Phi\left( \vec{r},t\right) }{\hbar }%
}\right) ,
\end{eqnarray}%
where we have denoted
\begin{equation}
V\left( \vec{r},t\right) =mV_{ext}\left( \vec{r},t\right) +b\ln \frac{|\Psi
\left( \vec{r},t\right) |^{2}}{|\Psi _{0}|^{2}}.
\end{equation}%

After separating the real and imaginary parts of Eq.~(\ref%
{eq:LogSEwithwavefunction}), we obtain  the hydrodynamic representation of the logarithmic Bose-Einstein Condensate as given by  (for the
details of the calculations see Appendix~\ref{app1})
\begin{eqnarray}
&&\dfrac{\partial \rho _{m}}{\partial t}+\nabla \cdot (\rho _{m}\vec{v})=0,
\label{eq:continum equation} \\
&&\dfrac{\partial \vec{v}}{\partial t}=-\frac{1}{m}\nabla (Q+K+V),
\label{eq:madelung eq2}
\end{eqnarray}%
where by $\vec{v}=\nabla \Phi/m$ we have denoted the velocity of the quantum fluid, $K=\dfrac{m%
}{2}v^{2}$ is the kinetic energy of the system, and
\begin{equation}
Q=-\dfrac{\hbar ^{2}}{2m}\dfrac{\nabla ^{2}\sqrt{\rho _{m}}}{\sqrt{\rho _{m}}%
},
\end{equation}%
is the quantum potential. From a physical point of view, Eq.~(\ref%
{eq:continum equation}) can be immediately interpreted as the continuity equation of the
quantum fluid moving with a velocity $\vec{v}$. On the other hand, Eq.~(\ref{eq:madelung eq2}) has the same form as
the Euler equation in standard Newtonian fluid mechanics. 

In the following
we will adopt the Thomas-Fermi approximation, according to which if the number of particles in the system is
large enough, the quantum
potential $Q$ can be neglected, \cite{Dalfovo, Pita, Pethick, ZNG}.

Hence with the use of the Thomas-Fermi approximation Eq.~(\ref{eq:madelung eq2}) takes the form
\begin{eqnarray}\label{eq:madelung eq2 without Q}
\hspace{-0.5cm}\dfrac{\partial \vec{v}}{\partial t}+\left( \vec{v}\cdot \nabla \right) \vec{%
v} =-\frac{1}{m}\nabla V 
=-\nabla V_{ext}-\frac{b}{m}\frac{1}{\rho _{m}}\nabla \rho _{m},
\end{eqnarray}
where the constant term $|\Psi _{0}|^{2}$ does not appear anymore in the definition of the physical quantities.

Assuming that the logarithmic BEC is an inviscid fluid, we can compare Eq.~(%
\ref{eq:madelung eq2 without Q}) with the standard Euler equation in fluid dynamics, given by
\begin{equation}
\frac{\partial\vec{v}}{\partial t}+(\vec{v}\cdot \nabla )\vec{v}=\vec{F}-\dfrac{1}{%
\rho _{m}}\nabla p,
\end{equation}%
where  $\vec{F}=-\nabla V_{ext}$ is the
force term,  and $p$ is the thermodynamic
pressure. 

The comparison of the two Euler type equations makes is easy to infer
the functional form of the quantum pressure (the equation of state) of the logarithmic BEC, which is 
given by the linear relation \cite{Chav2,Our3}
\begin{equation}\label{eq:quantum pressure}
p=\frac{b}{m}\rho _m .  
\end{equation}

Therefore we obtain the important result that the BEC with logarithmic nonlinearity satisfies the ideal gas
equation of state, with its thermodynamic pressure being proportional to the
fluid density.

\subsection{Equilibrium of the rotating logarithmic condensate}

We assume now that the BEC dark matter in rotating like a rigid body. In
order to simplify our analysis in the following we neglect the quantum
potential corrections. To describe the rotational properties of the condensate  we
switch to the rotating frame by setting
\begin{equation}
\vec{v}\rightarrow \vec{v}+\vec{\Omega}\times \vec{r},
\end{equation}%
where we assume that the angular velocity vector $\vec{\Omega}$ is constant.

Thus the Euler type equation corresponding to the hydrodynamic representation of the Gross-Pitaevskii equation  is
given in the rotating frame as
\begin{eqnarray}
&&\frac{\partial \vec{v}}{\partial t}-\vec{v}\times (\nabla \times \vec{v})=-%
\frac{1}{\rho _{m}}\nabla p-\nabla \left( \frac{\vec{v}^{2}}{2}%
+V_{ext}\right)  \notag  \label{eq:original NSE} \\
&&- \vec{\Omega}\times \vec{\Omega}\times \vec{r}-2\vec{\Omega}\times \vec{v}.
\end{eqnarray}%
where $\rho _m$ is the matter density of the condensate, related to the quantum probability density. For a static condensate, we can
neglect the stream velocity $\vec{v}$, and Eq.~(\ref{eq:original NSE}) becomes
\begin{equation}
\nabla V_{ext}=-\frac{1}{\rho _{m}}\nabla p-\vec{\Omega}\times \vec{\Omega}%
\times \vec{r}.  \label{eq:final NSE}
\end{equation}

As for the external potential $V_{ext}$, we consider it to be  the Newtonian gravitational
potential, which satisfies the Poisson equation
\begin{equation}
\nabla ^{2}V_{ext}=4\pi G\left( \rho _{m}+\rho _{b}\right) ,
\end{equation}%
where $\rho _{b}$ is the density of the baryonic matter. 

By taking into
account that for the BEC with logarithmic nonlinearity we have for the
quantum pressure the equation of state $p=b\rho _{m}/m$, by taking the
divergence of Eq.~(\ref{eq:final NSE}) we obtain
\begin{eqnarray}\label{eq:reduced final NSE} 
4\pi G\left( \rho _{m}+\rho _{b}\right) &=&-\nabla \cdot \left( \frac{b}{m}%
\frac{1}{\rho _{m}}\nabla \rho _{m}\right) -\nabla \cdot \left( \vec{\Omega}%
\times \vec{\Omega}\times \vec{r}\right) . \nonumber\\
\end{eqnarray}

The second term on the right side of Eq.~(\ref{eq:reduced final NSE}) can be
transformed as,
\begin{equation*}
\nabla \cdot (\vec{\Omega}\times \vec{\Omega}\times \vec{r})=\vec{\Omega}%
^{2}-3\vec{\Omega}^{2}=-2\vec{\Omega}^{2},
\end{equation*}%
and finally Eq.~(\ref{eq:reduced final NSE}), describing the equilibrium
properties of a rotating static BEC with logarithmic nonlinearity becomes
\begin{equation}\label{basic}
4\pi G\left( \rho _{m}+\rho _{b}\right) =-\nabla \cdot \left( \frac{b}{m}%
\frac{1}{\rho _{m}}\nabla \rho _{m}\right) +2\vec{\Omega}^{2}. 
\end{equation}

This equation is the basic equation describing the physical and
astrophysical properties of BEC dark matter halos with logarithmic
nonlinearity, with the effects of the baryonic matter and rotation included.

\subsubsection{The generalized Lane-Emden equation}

In the following we assume that the dark matter halos have spherical
symmetry. In spherical coordinates $(r,\theta ,\phi )$, the radial part of
Eq.~(\ref{basic}) becomes
\begin{equation*}
4\pi G\left( \rho_m +\rho _{b}\right) =-\frac{b}{mr^{2}}\frac{d}{dr}\left(
\frac{r^{2}}{\rho _{m}}\frac{d}{dr}\rho _{m}\right) +2\vec{\Omega}^{2},
\end{equation*}%
or, equivalently,
\begin{equation}
\hspace{-0.3cm}4\pi G\left( \rho _{m}+\rho _{b}\right) =-\frac{2b}{m}\frac{1%
}{r}\frac{d}{dr}\ln \rho _{m}-\frac{b}{m}\frac{d^{2}}{dr^{2}}\ln \rho _{m}+2%
\vec{\Omega}^{2}.  \label{basic1}
\end{equation}

We introduce now a new dependent variable $W$, defined as $\ln \left( \rho
_{m}/\rho _{m0}\right) =-W(r)$, where $\rho _{m0}$ is the central mass
density of the dark matter halo, $\rho _{m0}=\rho _{m}(0)$. Hence we have $%
\rho _{m}(r)=\rho _{m0}e^{-W(r)}$, and Eq.~(\ref{basic1}) becomes 
\begin{equation}
\frac{d^{2}}{dr^{2}}W+\frac{2}{r}\frac{dW}{dr}+\frac{2m}{b}\vec{\Omega}^{2}=%
\frac{4\pi Gm\rho _{m0}}{b}\left( e^{-W}+\frac{\rho _{b}}{\rho _{m0}}\right)
.  \label{basic2a}
\end{equation}

Eq.~(\ref{basic2a}), describing the equilibrium properties of Bose-Einstein
condensate dark matter halos with logarithmic nonlinearity, must be
integrated with the initial conditions $W(0)=0$, and $\left( dW/dr\right)
(0)=0$, respectively.

To simplify the mathematical formalism we rescale the radial coordinate $r$
according to $r=\alpha \theta $, where $\theta $ is the dimensionless radial
distance, and
\begin{eqnarray}
\alpha &=&\sqrt{\frac{b}{4\pi Gm\rho _{m0}}}=1.0926\times 10^{15}\times
\notag \\
&&\left( \frac{\rho _{m0}}{10^{-24}\;\mathrm{g/cm^{3}}}\right) ^{-1/2}\sqrt{%
\frac{b}{m}}\;\mathrm{cm}\text{ }= 3. 5475\times  \notag \\
&&10^{-7}\times \left( \frac{\rho _{m0}}{10^{-24}\;\mathrm{g/cm^{3}}}\right)
^{-1/2}\sqrt{\frac{b}{m}}\;\mathrm{kpc},
\end{eqnarray}%
is a characteristic length. Moreover, we denote
\begin{eqnarray}
\omega ^{2} &=&\frac{\vec{\Omega}^{2}}{2\pi G\rho _{m0}}=2.3873\times
10^{-2}\times  \notag \\
&&\left( \frac{\vec{\Omega}}{10^{-16}\;\mathrm{s}^{-1}}\right) ^{2}\left(
\frac{\rho _{m0}}{10^{-24}\;\mathrm{g/cm^{3}}}\right) ^{-1}.
\end{eqnarray}

Hence Eq.~(\ref{basic2a}) takes the dimensionless form
\begin{equation}  \label{26a}
\frac{d^{2}W(\theta)}{d\theta ^{2}}+\frac{2}{\theta }\frac{dW(\theta)}{%
d\theta }-e^{-W(\theta)}+\omega ^{2}-\frac{\rho _{b}(\theta )}{\rho _{m0}}=0.
\end{equation}%

The initial conditions for Eq.~(\ref{26a}) are $W(0)=0$, and $W^{\prime
}(0)=0 $, respectively. Equivalently, Eq.~(\ref{26a}) can be written as
\begin{equation}
\theta \frac{d^{2}W(\theta)}{d\theta ^{2}}+2\frac{dW\theta)}{d\theta }%
-\theta N\left\{ W(\theta )\right\} =\left[ \frac{\rho _{b}(\theta )}{\rho
_{m0}}-\omega ^{2}\right] \theta ,  \label{27}
\end{equation}%
where $N\left\{ W(\theta )\right\} =e^{-W}$denotes the nonlinear term in the
equation.

\section{The density, mass and tangential velocity profiles of the logarithmic dark matter BEC's}\label{sect2}

In order to solve Eq. (\ref{27}) we will the Adomian
Decomposition Method, and the Adomian polynomials \cite{Ad,Ad1}. For this we assume first
that $W\left( \theta \right) $ can be decomposed in an infinite series of
components given by
\begin{equation}
W(\theta )=\sum_{n=0}^{\infty }W_{n}\left( \theta \right) .
\end{equation}

We also assume that the nonlinear term $N$ can be represented by an infinite
series of the Adomian polynomials $A_{n}$ in the form \cite{Ad,Ad1}
\begin{equation}
N\left\{ W(\theta )\right\} =\sum_{n=0}^{\infty }A_{n}\left( \theta \right) ,
\end{equation}
where 
\begin{equation}
A_{n}\left( \theta \right) =\left.\frac{1}{n!}\left[ \frac{d^{n}}{%
d\varepsilon ^{n}}N\left( \sum_{n=0}^{\infty }\varepsilon ^{n}W_{n}\right) %
\right] _{{}}\right|_{\varepsilon =0},n=0,1,2,....
\end{equation}

The first few Adomian polynomials can be obtained as \cite{Ad,Ad1}
\begin{equation}
A_{0}=N\left[ W_{0}\right] ,
\end{equation}%
\begin{equation}
A_{1}=W_{1}N^{\prime }\left[ W_{0}\right] ,
\end{equation}%
\begin{equation}
A_{2}=W_{2}N^{\prime }\left[ W_{0}\right] +\frac{1}{2!}W_{1}^{2}N^{\prime
\prime }\left[ W_{0}\right] ,
\end{equation}%
\begin{equation}
A_{3}=W_{3}N^{\prime }\left[ W_{0}\right] +W_{1}W_{2}N^{\prime \prime }\left[
W_{0}\right] +\frac{1}{3!}W_{1}^{3}N^{\prime \prime \prime }\left[ W_{0}%
\right] ,
\end{equation}%
\begin{eqnarray}
A_{4}&=&W_{4}N^{\prime }\left[ W_{0}\right] +\left[ \frac{1}{2!}%
W_{2}^{2}+W_{1}W_{3}\right] N^{\prime \prime }\left[ W_{0}\right] +  \notag
\\
&&\frac{1}{2!}W_{1}^{2}W_{2}N^{\prime \prime \prime }\left[ W_{0}\right] +%
\frac{1}{4!}W_{1}^{4}N^{(\mathrm{iv})}\left[ W_{0}\right] ,  \label{Ad4}
\end{eqnarray}

\begin{equation*}
...
\end{equation*}

After substituting the above results into Eq. (\ref{27}) we obtain

\begin{equation}
\sum_{n=0}^{\infty }\theta \frac{d^{2}W_{n}}{d\theta ^{2}}%
+2\sum_{n=0}^{\infty }\frac{dW_{n}}{d\theta }-\sum_{n=0}^{\infty }\theta
A_{n}\left( \theta \right) =\left[ \frac{\rho _{b}(\theta )}{\rho _{m0}}%
-\omega ^{2}\right] \theta .  \label{L0}
\end{equation}

\subsection{The Laplace-Adomian Decomposition Method}

In the following we will obtain the series solution of Eq.~(\ref{L0}) by using the Laplace-Adomian Decomposition method \cite{Ad1}, a very efficient and powerful version of the general Adomian Decomposition Method.

We define the Laplace transform operator $\mathcal{L}_{x}$
of an arbitrary function $f(x)$, as $\mathcal{L}_{x}[f(x)](s)=\int_{0}^{%
\infty }{f(x)e^{-sx}dx}$. The Laplace transform has the properties
\begin{equation}
\mathcal{L}_{\theta }[\frac{d}{d\theta }W_{n}\left( \theta \right)
](s)=sF_{n}\left( s\right) -W_{n}(0),  \label{t1}
\end{equation}
\begin{equation}
\mathcal{L}_{\theta }[\frac{d^{2}}{d\theta ^{2}}W_{n}\left( \theta \right)
](s)=s^{2}F_{n}(s)-sW_{n}(0)-W_{n}^{\prime }(0),  \label{t2}
\end{equation}
\begin{eqnarray}  \label{t3}
&&\mathcal{L}_{\theta }\left[ \theta \frac{d^{2}W_{n}(\theta )}{d\theta ^{2}}%
\right] =\int_{0}^{\infty }\theta \frac{d^{2}W_{n}(\theta )}{d\theta ^{2}}%
e^{-s\theta }d\theta =  \notag \\
&&-\frac{d}{ds}\int_{0}^{\infty }\frac{d^{2}W_{n}(\theta )}{d\theta ^{2}}%
e^{-s\theta }d\theta =-\frac{d}{ds}\mathcal{L}_{\theta }\left[ \frac{%
d^{2}W_{n}(\theta )}{d\theta ^{2}}\right] (s)=  \notag \\
&&-s^{2}F_{n}^{\prime }(s)-2sF_{n}(s)+W_{n}(0),
\end{eqnarray}
where we have denoted $F_{n}(s)=\mathcal{L}_{\theta }\left[ W_{n}\left(
\theta \right) \right] (s).$

We take now the Laplace transform of Eq. (\ref{L0}). With the use of the
relations (\ref{t1})-(\ref{t3}), and by taking into account the initial
conditions and the linearity of the Laplace transform, we obtain
\begin{eqnarray}
&&-s^{2}\sum_{n=0}^{\infty }F_{n}^{\prime }(s)-\sum_{n=0}^{\infty }\mathcal{L%
}_{\theta }\left[ \theta A_{n}\left( \theta \right) \right] (s)=-\frac{%
\omega ^{2}}{s^{2}}+  \notag \\
&&\mathcal{L}_{\theta }\left[ \theta \frac{\rho _{b}(\theta )}{\rho _{m0}}%
\right] (s).  \label{L1}
\end{eqnarray}%
From Eq. (\ref{L1}) we obtain the following recursion relations
\begin{equation}
F_{0}^{\prime }(s)=\frac{\omega ^{2}}{s^{4}}-\frac{1}{s^{2}}\mathcal{L}%
_{\theta }\left[ \theta \frac{\rho _{b}(\theta )}{\rho _{m0}}\right] (s),
\end{equation}
\begin{equation}
F_{n+1}^{\prime }=-\frac{1}{s^{2}}\mathcal{L}_{\theta }\left[ \theta
A_{n}\left( \theta \right) \right] (s),\quad n\geq 0.
\end{equation}

By integration we obtain
\begin{equation}
F_{0}\left( s\right) =-\frac{\omega ^{2}}{3s^{3}}-G(s),
\end{equation}
\begin{equation}
F_{n+1}(s)=-\int \frac{1}{s^{2}}\mathcal{L}_{\theta }\left[ \theta
A_{n}\left( \theta \right) \right] (s)ds,
\end{equation}%
where we have denoted
\begin{equation}
G(s)=\int \frac{1}{s^{2}}\mathcal{L}_{\theta }\left[ \theta \frac{\rho
_{b}(\theta )}{\rho _{m0}}\right] (s)ds.
\end{equation}

With the use of the inverse Laplace transform we find
\begin{equation}
W_{0}(\theta )=-\frac{1}{6}\omega ^{2}\theta ^{2}-H(\theta ),
\end{equation}
\begin{equation}
W_{n+1}(\theta )=-\mathcal{L}_{\theta }^{-1}\left\{ \int \frac{1}{s^{2}}%
\mathcal{L}_{\theta }\left[ \theta A_{n}\left( \theta \right) \right]
(s)ds\right\}(\theta) ,
\end{equation}%
where
\begin{equation}
H\left( \theta \right) =\mathcal{L}_{\theta }^{-1}\left\{ \int \frac{1}{s^{2}%
}\mathcal{L}_{\theta }\left[ \theta \frac{\rho _{b}(\theta )}{\rho _{m0}}%
\right] (s)ds\right\} (\theta).
\end{equation}

In the following we will neglect the effects of the baryonic matter on the
condensate dark matter distribution, by assuming that $\rho _{m0}\gg\rho
_{b}(\theta ),\forall \theta \geq 0$, that is, the central density of the
dark matter is much bigger than the baryonic density at all radial distances
from the galactic center. Hence, the basic equation assumed to describe the properties of the rotating logarithmic condensate dark matter halos is given by
\begin{equation}\label{basic1}
\theta \frac{d^{2}W(\theta)}{d\theta ^{2}}+2\frac{dW\theta)}{d\theta }+\omega ^2\theta =
\theta e^{-W(\theta)}.  
\end{equation}%

The above assumptions also imply $H(\theta )\approx 0$.

\subsection{Series solution of the generalized Lane-Emden equation}

Hence, for the initial approximation of $W$ we obtain the expression $%
W_{0}(\theta )=-\omega ^{2}\theta ^{2}/6$. The first Adomian polynomial is
given by $A_{0}=e^{-W_{0}(\theta )}\approx 1+\omega ^{2}\theta ^{2}/6$.
Therefore we find for the next term the expression
\begin{eqnarray}
W_{1}(\theta )&=&-\mathcal{L}_{\theta }^{-1}\left\{ \int \frac{1}{s^{2}}%
\mathcal{L}_{\theta }\left[ \theta A_{0}\left( \theta \right) \right]
(s)ds\right\} (\theta )  \notag \\
&&=\frac{1}{120}\theta ^{2}\left( \theta ^{2}\omega ^{2}+20\right) .
\end{eqnarray}

The second Adomian polynomial is given by $A_{1}(\theta )=W_{1}(\theta
)e^{-W_{0}(\theta )}\approx -\left( 1+\omega ^{2}\theta ^{2}/6\right)
W_{1}\left( \theta \right) $. Thus we immediately find
\begin{equation}
W_{2}\left( \theta \right) =-\frac{\theta ^4 \left(7 \theta ^4 \omega ^4+312
\theta ^2 \omega ^2+3024\right)}{362880}.
\end{equation}
Then we successively obtain
\begin{equation}
W_{3}\left( \theta \right)=\frac{7 \theta ^{12} \omega ^6}{121305600}+\frac{%
389 \theta ^{10} \omega ^4}{99792000}+\frac{\theta ^8 \omega ^2}{12096}+%
\frac{\theta ^6}{1890},
\end{equation}
\begin{eqnarray}
\hspace{-0.3cm}W_{4}\left( \theta \right)&=&-\frac{191 \theta ^{16} \omega ^8%
}{989853696000}-\frac{56963 \theta ^{14} \omega ^6}{3269185920000}  \notag
\\
\hspace{-0.3cm}&& -\frac{13261 \theta ^{12} \omega ^4}{23351328000}-\frac{%
2339 \theta ^{10} \omega ^2}{299376000}-\frac{61 \theta ^8}{1632960},
\end{eqnarray}
\begin{eqnarray}
W_{5}\left( \theta \right)&=& \frac{803 \theta ^{20} \omega ^{10}}{%
1169264678400000}+\frac{6636457 \theta ^{18} \omega ^8}{85531711224960000}
\notag \\
&& +\frac{909413 \theta ^{16} \omega ^6}{266765571072000}+\frac{225709 \theta
^{14} \omega ^4}{3120586560000} \notag \\
&& +\frac{154213 \theta ^{12} \omega ^2}{210161952000}+\frac{629 \theta ^{10}%
}{224532000},
\end{eqnarray}
\begin{eqnarray}
W_6(\theta)&=&-\frac{428299 \theta ^{24} \omega ^{12}}{168374113689600000000}%
 \notag \\
&&-\frac{35836904401 \theta ^{22} \omega ^{10}}{103869710111591424000000}
\notag \\
&&- \frac{1118301257 \theta ^{20} \omega ^8}{58650316268544000000}  \notag \\
&& -\frac{15731649991 \theta ^{18} \omega ^6}{28738654971586560000}  \notag
\\
&& -\frac{719709593 \theta ^{16} \omega ^4}{84031154887680000}-\frac{884671
\theta ^{14} \omega ^2}{12872419560000}-  \notag \\
&& \frac{2869 \theta ^{12}}{13135122000}.
\end{eqnarray}

The next terms of the Adomian series can be calculated similarly in a
straightforward manner, giving $W(\theta)=\sum _{k=0}^n{W_k(\theta)}$.
However, in the following we approximate the series representation of $%
W(\theta)$ by its Pad\'{e} approximants $W[m/n](\theta)$ \cite{Pade}. Such an approach
is necessary to avoid the possible divergence of the Adomian series, which
may occur for large values of $\theta$.

If we have a power series of the form $f(z)=\sum_{z=0}^{\infty}f_kz^k$ the
Pad\'{e} approximant of the order $(m,n)$ in the vicinity of the point $z=0$
is given by the rational function $\Pi_{m.n}\in R_{m,n}$, with the important
property that it has the closest numerical values to the given series near
the point $z=0$ \cite{Pade}. $R_{m,n}$ denotes the set of rational functions having the
form $P/Q$, with$P$ and $Q$ polynomials in $z$ having the degree $p\leq m$
and $q\leq n$, respectively \cite{Pade}. 

\begin{figure*}[htbp]
\centering
\includegraphics[width=7.8cm]{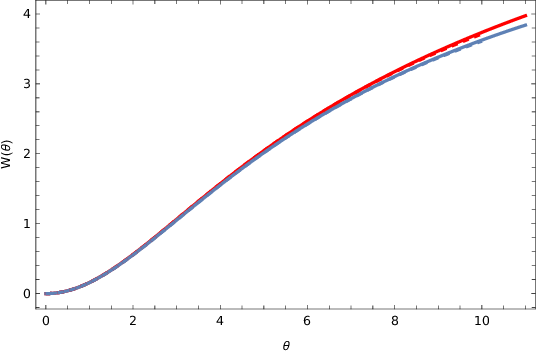}\hspace{.4cm} %
\includegraphics[width=7.8cm]{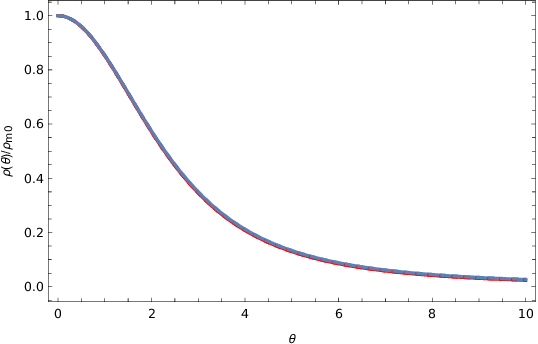}
\caption{Comparison between the full numerical solution of the generalized
Lane-Emden equation for $W(\theta)$ (solid curve), with the $W[8/8](%
\theta)$ Pad\'{e} approximant of the Adomian series solution (dotted
curve) (left panel), and between the full numerical density distribution $%
\rho (\theta)/\rho _{m0}=\exp [-W(\theta)]$
with the Pad\'{e} approximant expression $\exp \left[-W[8/8](\theta%
)\right)$ (dotted curve) (right panel), for $\omega =0$ (red curves) and $%
\omega =0.10$ (blue curves), respectively. }
\label{fig1}
\end{figure*}

The comparisons of the $%
W[8/8](\theta)$ Pad\'{e} approximant of the Adomian series solution, and of
the density $\exp \left[-W[8/8](\theta)\right]$ and the full numerical
solutions are represented in Fig.~\ref{fig1}. 

Even that small differences do appear between the numerical and the Adomian
series solution for $W(\theta )$, there is a very good concordance for the
case of the density. Also in the present set of dimensionless variables the
differences in the density distribution corresponding to different values of
$\omega $ is very small. 

\paragraph{The $W[2/2](\theta )$ approximation.} However, in the following we will investigate the
properties of the logarithmic Bose-Einstein Condensate dark matter halos by
using only the lowest order Pad\'{e} approximation of the Adomian series,
given by
\begin{equation}
W(\theta )=W[2/2](\theta )\approx -\frac{10\,\theta ^{2}\left( \omega
^{2}-1\right) }{3\theta^{2}+60 }.  \label{Wa}
\end{equation}

\begin{figure*}[htbp]
\centering
\includegraphics[width=7.8cm]{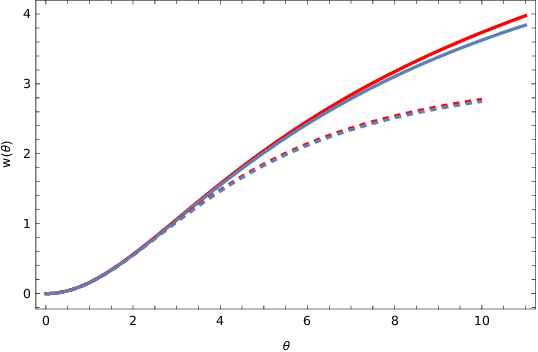}\hspace{.4cm} %
\includegraphics[width=7.8cm]{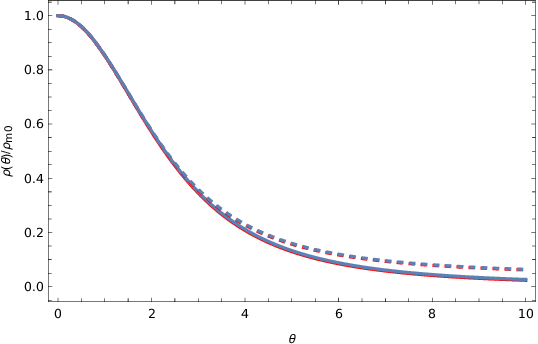}
\caption{Comparison between the full numerical solution of the generalized
Lane-Emden equation for $W(\theta)$ (solid curve), with the $W[2/2](%
\theta)$ Pad\'{e} approximant of the Adomian series solution (dotted
curve) (left panel), and between the full numerical density distribution $%
\rho (\theta)/\rho _{m0}=\exp [-W(\theta)]$
with the Pad\'{e} approximant expression $\exp \left[-W[2/2](\theta%
)\right)$ (dotted curve) (right panel), for $\omega =0$ (red curves) and $%
\omega =0.10$ (blue curves), respectively. }
\label{fig2}
\end{figure*}

The comparison between the full numerical solution and the lowest order Pad%
\'{e} approximant is presented in Fig.~\ref{fig2}. As one can see from the
Figure, even that some differences do appear for $W(\theta)$ between this
approximation and the full numerical solution, the density distribution of
the logarithmic Bose-Einstein Condensate is relatively well described by the
equation
\begin{equation}\label{rho1}
\rho _m(\theta)\approx \rho _{m0}\,e^{-\frac{10 \theta ^2 \left(1-\omega ^2\right)}{%
3 \theta ^2+60}},
\end{equation}
with the difference slightly increasing when approaching the vacuum boundary
of the galaxy. This expression allows for a simple mathematical description of the basic properties of the galactic dark matter halos, assumed to be in the form of a logarithmic Bose-Einstein Condensate. For the nonrotating galactic halo its density distribution can be approximated as
\be\label{rho2}
\rho _m(\theta)\approx \rho _{m0}\,e^{-\frac{10 \theta ^2 }{%
3 \theta ^2+60}}.
\ee 

\subsection{Comparison with dark matter profiles from numerical simulations}

The density distribution of ULDM  has been extensively investigated by considering both dark matter described by the simple Schr\"{o}dinger-Poisson system, or by more general mathematical models that include self-interaction \cite{N4,N5,N6,N7,N8,N9,N10,N11,N12}. Usually numerical methods and detailed simulations are used to construct galactic models in which dark matter is in the form of a condensate. The simulation results also allow to reconstruct the density profile of dark matter through fitting with the observational data. The large scale simulations lead to galactic structures that are basically identical with the one predicted by the standard Cold Dark Matter models. However, inside galaxies, a very different structure is formed, consisting  of a gravitationally self-bounded solitonic core inside every galaxy, which is surrounded by extended haloes of fluctuating density granules \cite{N4}. The soliton density profile can be approximated by \cite{N4,N5}
\bea\label{Schive}
\rho_s (r)&\approx & \frac{1.9 \left(m_B/10^{-23}\;{\rm eV}\right)^{-2}\left(r_c/{\rm kpc}\right)^{-4}}{\left[1+1.91\times 10^{-2}\left(r/r_c\right)^2\right]^8}\nonumber\\
&=&\frac{\rho _c}{\left[1+0.091\left(\frac{r}{r_c}\right)^2\right]^8}, \quad r<r_1,
\eea  
where $m_B$ is the dark matter particle mass, $r_c$ is the core radius, and $\rho_c$ is the central density. Outside the solitonic core for the dark matter a Navarro-Frenk-White (NFW) profile of the form
\be
\rho (r)=\frac{\rho_s}{\left(r/r_s\right)\left[1+\left(r/r_s\right)\right]^2},\quad r>r_1,
\ee 
is assumed. Massive halos were constructed from the wave distribution function in \cite{N6}, by adopting as a key ingredient the distribution function of the wave power. Several halos produced by structure formation simulations as templates were used to determine the wave distribution function. The fermionic King model presented the best fits,  and it was used to construct the wave-halo model.

An approximate analytical form for the soliton density profile was proposed in \cite{N11,N12}, by considering the effect of repulsive self-interactions of moderate strength on fuzzy dark matter halos. The oscillation frequency of the central soliton core, the granule size, the spatial dependence of the field’s coherence, and the turbulent vortex angle and  were also considered. The soliton density profile was approximated by a super-Gaussian function \cite{N11,N12},
\be
\rho(r)=\rho_c\left(\Gamma _g\right)\exp \left[-\ln 2\left(\frac{r}{r_c\left(\Gamma _g\right)}\right)^{n \left(\Gamma _g\right)}\right],
\ee 
where $\Gamma _g$ is the relative strength interaction parameter, and $n\left(\Gamma _g\right)$ is a parameter that can be obtained from a semi-analytical relation, given by
\begin{equation}
    n(\Gamma_g) = \frac{\vartheta_0+\vartheta_{TF}}{2}-\frac{\vartheta_0 - \vartheta_{TF}}{2} \tanh\left( \frac{\log_{10}\left(\Gamma_g\right)-0.6}{1.5}  \right), 
\end{equation}
with $\vartheta_0 = 1.62 $ and $\vartheta_{TF} = 2.3 $ the exponent values at the non-interacting and Thomas-Fermi limits, respectively.  The super-Gaussian profile as considered in \cite{N11,N12} is closely related to the Einasto profile $\rho(r)=d^{3n}/\left(4\pi n\Gamma (3n)\right)\left(M/r_h^3\right)\exp \left[-d\left(r/r_h\right)^{1/n}\right]$ \cite{Ei1,Ei2}, which depends on three free parameters: the Einasto index $n$, the total mass $M$, and the half mass radius $r_h$, respectively. The parameter $d$ is not a free parameter, but a dimensionless constant depending on $n$.

A simple Gaussian solitonic density profile for the galactic core of the form
\be\label{Chav}
\rho (r)=M\left(\frac{1}{\pi R^2}\right)e^{-r^2/R^2},
\ee  
where $M$ is the mass and $R$ is the condensate radius, was proposed in \cite{N13}, and was used to model astrophysical objects such as boson stars and dark matter galactic halos. Later, a similar profile for self-interacting superfluid dark matter droplets was considered in \cite{N14}. This model predicts cored galactic haloes with rotation curves that obey a single universal equation in the inner region $r\leq 1$ kpc.
  kpc).  
  
   A solitonic density profile was proposed in \cite{Che}, by assuming that the self-interaction of the Bose-Einstein Condensate is non-negligible. The variation of the density os given by
  \be\label{Che}
  \rho(r)=\rho_0\left[1+\left(2^{1/\beta}-1\right)\left(\frac{r}{r_{core}}\right)^\alpha\right]^{-\beta},
  \ee
where $\alpha$ and $\beta$ are constants. When the self-interaction is negligible, $\alpha \rightarrow 2$ and $\beta \rightarrow 8$, and we recover  the profile given by Eq.~(\ref{Schive}).

As compared to the numerical simulation results, the density profiles given by Eqs.~(\ref{rho1}) and (\ref{rho2}) suggest a much simpler structure. The density profile has a solitonic character, with a nonsingular center, but which extends to infinity. The central soliton basically contains the entire dark matter distribution. Hence, in the present model, there is no need to consider, and match, two distinct density profile, having rather distinct characteristics. Moreover, no complex, fluctuating  structures, or filaments can be obtained via the present approach. One explanation for the lack of these effects is that our model is one-dimensional, while the previously summarized results are obtained from full 3D simulations of the Schr\"{o}dinger-Poisson or of the Gross-Pitaevskii-Poisson system.  Even though the initial condensation of dark matter took place in the very early Universe, it is not physically realistic to assume that the cosmic environment did consist only of dark matter, and that other particles that did not condense were not present during the condensation  period. These particles (for example, hydrogen atoms) can be considered  as impurities in the condensate at the moment of condensation, as well as at later times. The impurities locally interact with the condensate dark matter, and generate random fluctuations, and a complex dissipative structure, in the condensate distribution. These effects can be described by adding a supplementary term to the potential  $V\left( \vec{r},t\right)$, which thus takes the form
\be
V\left( \vec{r},t\right) =mV_{ext}\left( \vec{r},t\right) +V_b\left( \vec{r},t\right)+b\ln \frac{|\Psi
\left( \vec{r},t\right) |^{2}}{|\Psi _{0}|^{2}}+V_{d}\left( \vec{r},t\right), 
\ee 
where $V_b\left( \vec{r},t\right)$ is the gravitational potential of the baryonic matter, and $V_{d}\left( \vec{r},t\right)$ is a random potential satisfying the conditions $\left<V_{d}\left( \vec{r},t\right)\right> =0$ and $\left<V_{d}\left( \vec{r},t\right)V_{d}\left( \vec{r}\;',t'\right)\right>=\kappa ^2\delta \left(t-t'\right)\delta \left(\vec{r}-\vec{r}\;'\right)$, where $\kappa ^2$ is a constant. The inclusion of relevant physical aspects into the theoretical models may contribute to a better understanding of the properties of galactic dark matter halos in the presence of a logarithmic condensate, and could give a better description of the realistic structures and features observed in galaxies.

In order to compare the density profile of the logarithmic Bose-Einstein Condensate with the previously mentioned solitonic density profiles we rescale in all cases the radial coordinate with respect to the core or halo radius, by introducing a new coordinate $\theta$, so that $\theta =r/r_c$, $\theta =r/R$, or $\theta =r/r_{core}$. Moreover, we normalize all density profiles with respect to the central density, and introduce the dimensionless density variable $\Pi (\theta)=\rho (\theta)/\rho_{center}$.  For the super-Gausssian profile we adopt the expression $\Pi (\theta)=e^{-2a\theta ^n}$, where $a$ and $n$ are constants. The simple Gaussian profile is given by $\Pi (\theta)=e^{-a\theta ^2}$. For the  power law type density profile (\ref{Schive}) we adopt the form $\Pi (\theta)=\left(1+0.091\theta ^2\right)^{-8}$. For the solitonic core profile given by Eq.~(\ref{Che}) we consider the form $\Pi (\theta)=\left[1+\left(2^{1/\beta}-1\right)\theta ^\alpha\right]^{-\beta}$. Finally, the logarithmic Bose-Einstein Condensate dark matter profile is given by $\Pi (\theta)=\exp \left]-10\theta ^2/\left(3\theta ^2+60\right)\right]$.

\begin{figure}[htbp]
\centering
\includegraphics[width=8cm]{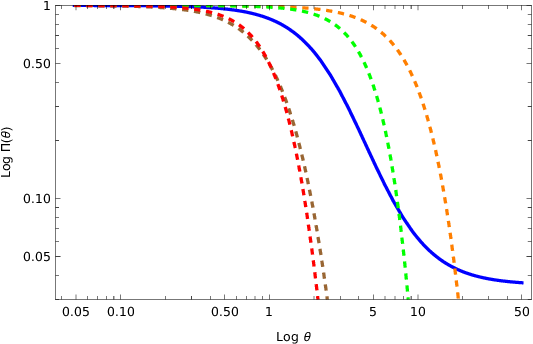}
\caption{Comparison between the logarithmic Bose-Einstein Condensate density profile (\ref{rho2}), $\Pi (\theta)=\exp \left]-10\theta ^2/\left(3\theta ^2+60\right)\right]$ (solid blue line), the Gaussian  profile (\ref{Chav}),  $\Pi (\theta)=e^{-0.01\theta ^2}$ (orange dashed curve), the super-Gaussian profile $\Pi (\theta)=e^{-0.02\theta ^{2.4}}$ (green dashed line),  the power law potential (\ref{Schive}), $\Pi (\theta)=\left(1+0.091\theta ^2\right)^{-8}$ (brown dashed curve), and the solitonic core potential (\ref{Che}) in the presence of dark matter self interaction, $\Pi (\theta)=\left[1+\left(2^{1/\beta}-1\right)\theta ^\alpha\right]^{-\beta}$ with $\beta=7.8$ and $\alpha=2.4$ (red dashed curve), respectively.}
\label{fignew}
\end{figure}

The variation of these potentials with respect to the dimensionless distance $\theta$ is represented as a log-log plot in Fig.~\ref{fignew}.  By assuming that the mass $m$ of the dark matter particle is the same in all these models, the normalization with respect to the central density $\rho_c=mn_c$ implies a common central particle number density. However, the particle number varies inside the condensate dark matter halo according to different laws.  On the other hand, we have assumed that the range of values and variation of the variable $\theta$ is the same for all considered models.

All the presented density profiles indicate the presence of constant density core. However, the core extension is model dependent,  with the most massive core corresponding to the Gaussian profile. As compared to the other density profiles, the logarithmic dark matter Bose-Einstein Condensate density profile has a smooth decrease, while the other profiles present a sharp decrease in density. Hence, while the solitonic dark matter profiles require a match with an external dark matter distribution, the logarithmic Bose-Einstein Condensate dark matter distribution provides a full description of the galactic halo from it center to the vacuum boundary.

\subsection{Mass distribution, and tangential velocity of the logarithmic BEC's}

The mass distribution inside radius $r$ of the logarithmic condensate dark
matter halo is given by
\begin{equation}
M(r)=4\pi \int_{0}^{r}{r^{2}\rho _{m}(r)dr}=4\pi \alpha ^{3}\rho
_{m0}\int_{0}^{\theta }{\theta ^{2}e^{-W(\theta )}d\theta }.
\end{equation}

To estimate the integral we proceed as follows. By neglecting the baryonic
matter distribution, Eq.~(\ref{26a}) can be reformulated as
\begin{equation}
\theta ^2e^{-W(\theta)}=\frac{d}{d\theta}\left(\theta ^2\frac{dW(\theta)}{%
d\theta}\right)+\omega ^2\theta ^2.
\end{equation}

By integrating both sides of the above equation we find
\begin{equation}
\int_0^{\theta}{\theta ^2 e^{-W(\theta)}d\theta}=\theta ^2\frac{dW(\theta)}{%
d\theta}+\frac{\omega ^2\theta ^3}{3}.
\end{equation}

Hence for the mass distribution of the logarithmic condensate BEC we obtain
the expression
\begin{equation}
M(\theta )=4\pi \alpha ^{3}\rho _{m0}\left[ \theta ^{2}\frac{dW(\theta )}{%
d\theta }+\frac{\omega ^{2}\theta ^{3}}{3}\right] .
\end{equation}

By taking into account the approximation (\ref{Wa}) of the function $%
W(\theta )$, we obtain for the mass distribution the relation
\begin{eqnarray}
M(\theta )=M_0
\frac{\theta ^{3}\left( \omega^2 \theta^4+ 40 \,\omega^2 \theta^2+400\right) }{3\left(\theta ^{2}+20\right) ^{2}},  
\end{eqnarray}
where we have denoted $M_0=4\pi \alpha ^{3}\rho _{m0}$. The mass distribution for the non-rotating condensate with $\omega =0$ is given by
\be
M(\theta)=\frac{400}{3}M_0\frac{\theta ^{3} }{\left(\theta ^{2}+20\right) ^{2}}.
\ee

The tangential velocity of the massive test particles moving in the dark
matter halo is given by
\begin{equation}
V_{DM}^{2}(r)=\frac{GM(r)}{r},
\end{equation}%
or,
\bea\label{VDM}
V^2_{DM}(\theta)&=&4\pi G\alpha ^2\rho _{m0}\frac{\theta ^{2}\left( \omega^2 \theta^4+ 40 \,\omega^2 \theta^2+400\right) }{3\left(\theta ^{2}+20\right) ^{2}}\nonumber\\
&=&\frac{b}{m}V^2(\theta).
\eea

For the nonrotating condensate we obtain for the tangential velocity the expression
\be
V^2_{DM}(\theta)=\frac{1600}{3}\pi G\alpha ^2\rho _{m0}\frac{\theta ^{2} }{\left(\theta ^{2}+20\right) ^{2}}
\ee

\begin{figure*}[htbp]
\centering
\includegraphics[width=7.8cm]{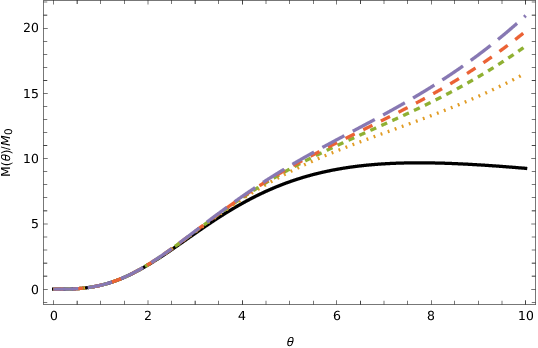}\hspace{.4cm} %
\includegraphics[width=7.8cm]{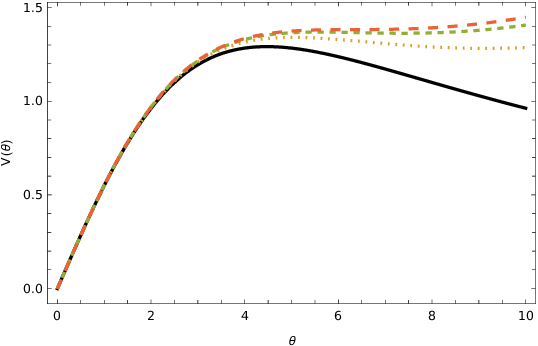}
\caption{Variations of the dimensionless mass $M(\theta)/M_0$ of the
logarithmic BEC halo (left panel), and of the tangential velocity $V(%
\theta)$ of a massive particle in rotational motion around the galactic
center (right panel), for different values of $\omega$: $%
\omega =0$ (solid curve), $\omega =0.15$ (dotted curve), $%
\omega =0.17$ (short dashed curve), $\omega =0.18$ (dashed curve),
and $\omega =0.19$ (long dashed curve), respectively. }
\label{fig3}
\end{figure*}

The variations of the dark matter halo mass, and of the dimensionless
tangential velocity $V(\theta)$ are represented, for different values of $%
\omega$, in Fig.~\ref{fig3}. 

\section{Comparison with observational data}\label{sect3}

In order to compare the velocity of the massive test particles moving on circular
orbits in the logarithmic BEC dark matter halos, as given by Eq.~ (%
\ref{VDM}), we need first to restore the dimensional dependence, by substituting $\theta \rightarrow r/\alpha$, and $\omega ^2=\vec{\Omega} ^2/2\pi G \rho _{m0}$, respectively. Thus we obtain 
\bea\label{VDMr}
V_{DM}^{2}(r)&=&2\pi G\rho _{m0}r^{2}\frac{\left(10 B^2+\pi G \rho_{m0}r^2\right)\vec{\Omega}^2r^2+50\,B^4}{3\left(5 B^2+\pi G \rho_{m0}r^2\right)^2},\nonumber\\ 
\eea
where we have denoted $B=\sqrt{b/m}$. Hence the properties of the
logarithmic Bose-Einstein Condensate dark matter halos are fully determined
by three parameters only, the central density of the halo $\rho _{m0}$, the
rotational angular velocity of the halo $\vec{\Omega}$, and the parameter $B$, describing the particle mass, and the self-interaction properties of the condensate. 

It is
interesting to note that the mass of the dark matter particle appears in the
tangential velocity of massive particles only in combination with $b$, and
thus its independent determination is impossible in the present model.

\paragraph{The fitting procedure.} We fit Eq.~(\ref{VDMr})  with the total observational velocity of the hydrogen
clouds that can be obtained from the SPARC database, and which is given by
\begin{eqnarray}  \label{v_obs}
v_{obs}= \sqrt{v_{gas}^2\!+\!\Upsilon_d\!\times\!
v_{disk}^2\!+\!\Upsilon_b\!\times\!
v_{bulge}^2\!+\!V_{DM}^{2}},  
\end{eqnarray}
where $v_{obs}$ is the total velocity, including the contributions of both
baryonic and dark matter, and $v_{gas}$, $v_{disk}$, $v_{bulge}$ and $V_{DM}$
denote the contributions from the velocity of the gas, of the disk, of the
bulge, and of the logarithmic BEC dark matter halo, respectively. Moreover, $\!\Upsilon_d\!$
and $\!\Upsilon_b\!$ denote the stellar mass-to-light ratios for the disk
and the stellar bulge, respectively, which must also be obtained from the fitting of the rotation curves.

SPARC \cite{Sparc} is a sample of 175 rotationally supported galaxies, containing the radial profiles of the tangential velocity of gas clouds moving around the galactic center, and determined in the near-infrared region. The SPARC dataset has been extensively used for the comparison of the theoretical models with the observational data. The galactic sample includes not only the tangential velocity, but also presents the mass models of the individual baryonic components, that is, of the disk, bulge, and of the galactic hydrogen gas. SPARC contains data for galaxies having various morphological types, including of spiral, late-time, early, and starburst galaxies.

The free parameters of the rotating logarithmic Bose-Einstein
Condensate dark halo model are $\left(\!\Upsilon_d\,, \!\Upsilon_b\,, B, |\vec{\Omega}|,\rho
_{m0}\right)$.

 In the following, we will use the SPARC data for different galaxies to obtain the best fit values of these parameters by performing the Likelihood analysis. The Likelihood function is defined as
 \begin{align}
 	L=L_0e^{-\chi^2/2},
 \end{align}
 where $L_0$ is the normalization constant. The loss functions $\chi^2$ for the observed velocity can be written as
 \begin{align}
 \chi^2=\sum_i\left(\frac{{v}_{\text{obs},i}-{v}_{\text{th},i}}{\sigma_i}\right)^2,
\end{align}

 Here $i$ counts data points, $``obs"$ are the values of observed velocity, $``th"$ are the theoretical values obtained from the model \eqref{v_obs}, and $\sigma_i$ are the errors associated with the $i$th data obtained from observations. 
 By maximizing the likelihood function, the best fit values of the parameters  $\left(\!\Upsilon_d\,, \!\Upsilon_b\,, B, |\vec{\Omega}|,\rho
 _{m0}\right)$, together with their confidence intervals, can be obtained for each galaxy. 
 
 To constrain the model parameters, we employed a Markov Chain Monte Carlo (MCMC) fitting procedure.

\subsection{Fitting results}

We present now the results of the fitting of the logarithmic Bose-Einstein Condensate model with the observational data. We consider independently the cases of the galaxies with bulge and without bulge data. 

\subsubsection{Bulgeless galaxies}

Since the number of model parameters depends on the presence of bulge velocity data, we consider two separate cases when fitting the model to the SPARC rotation curve data, galaxies with bulge components and galaxies without it.

\begin{figure*}[htbp]
	\centering
	\includegraphics[width=0.23\textwidth]{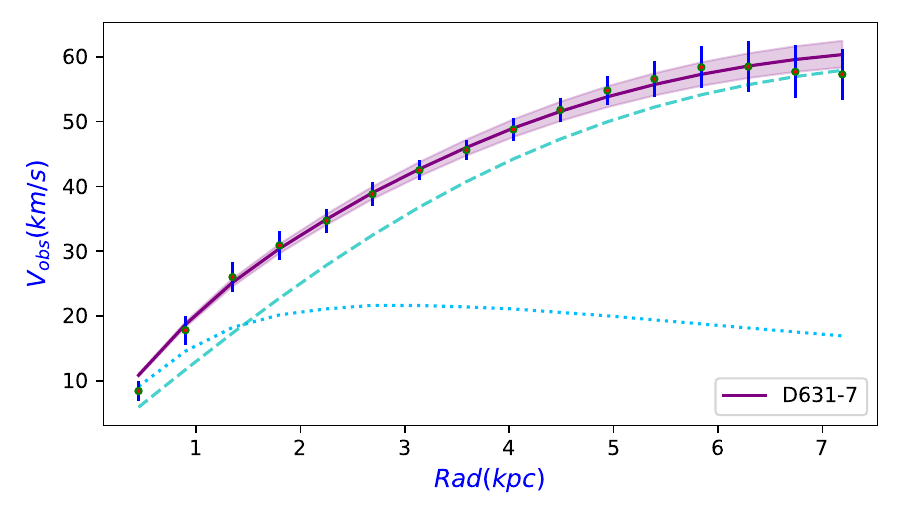} \hspace{0.01\textwidth}
	\includegraphics[width=0.23\textwidth]{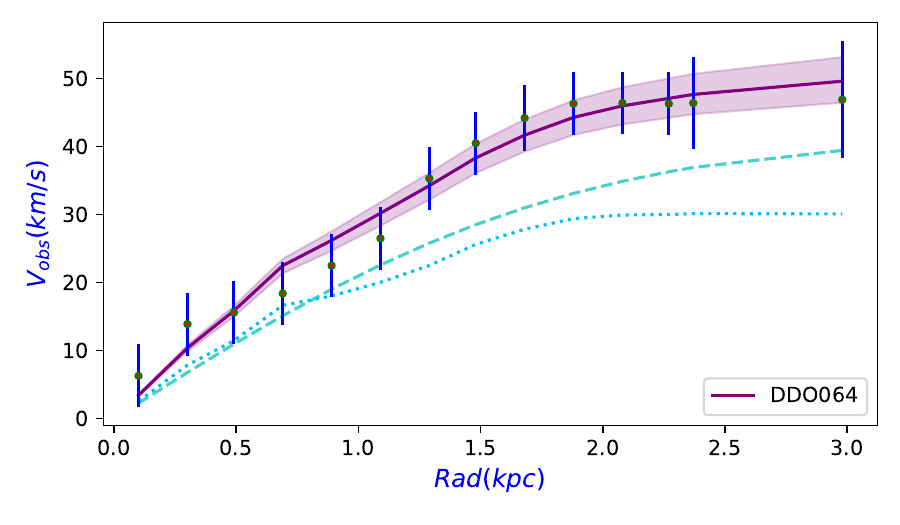} \hspace{0.01\textwidth}
	\includegraphics[width=0.23\textwidth]{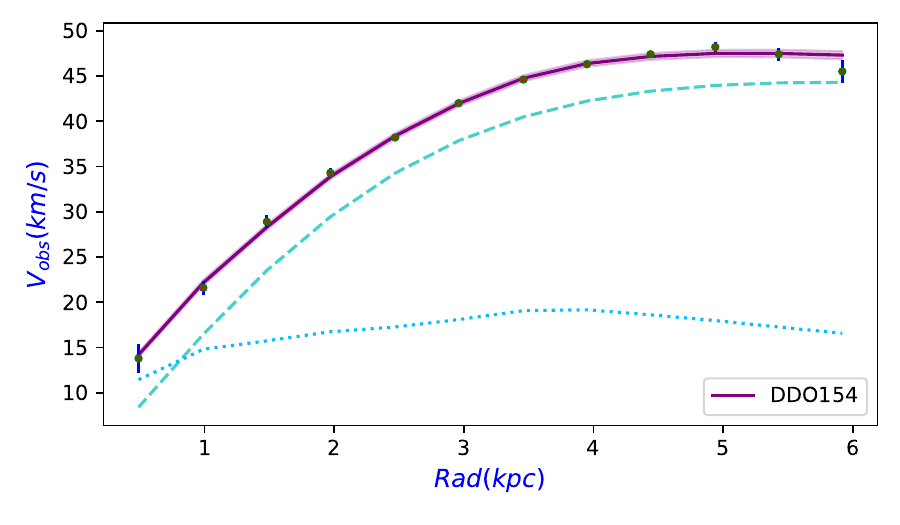} \hspace{0.01\textwidth}
	\includegraphics[width=0.23\textwidth]{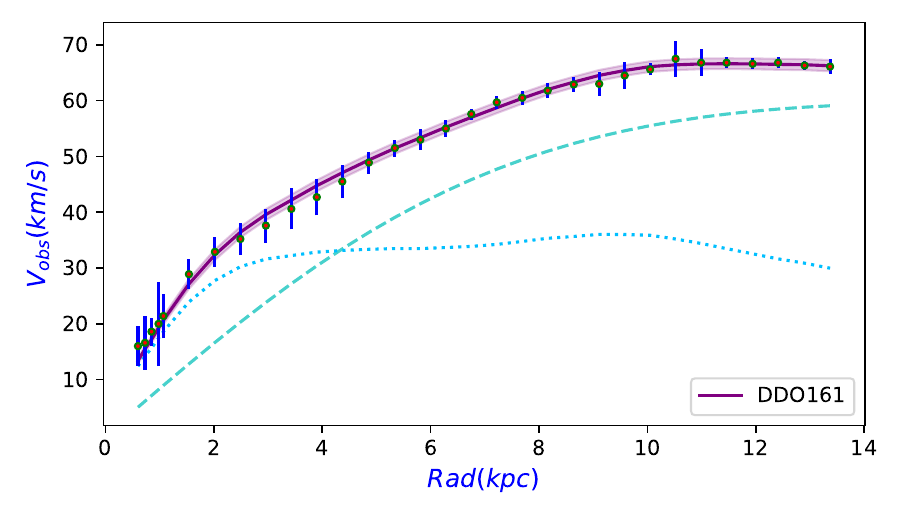} \\[0.35cm]
	\includegraphics[width=0.23\textwidth]{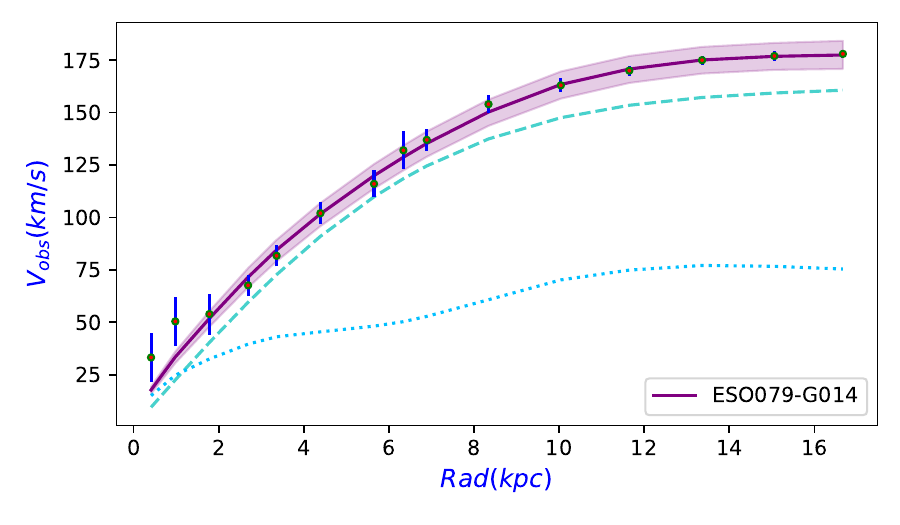} \hspace{0.01\textwidth}
	\includegraphics[width=0.23\textwidth]{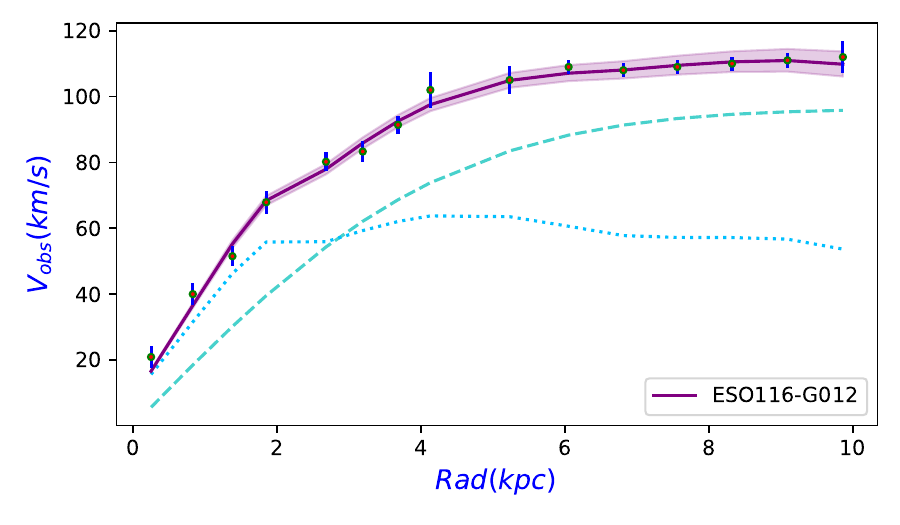} \hspace{0.01\textwidth}
	\includegraphics[width=0.23\textwidth]{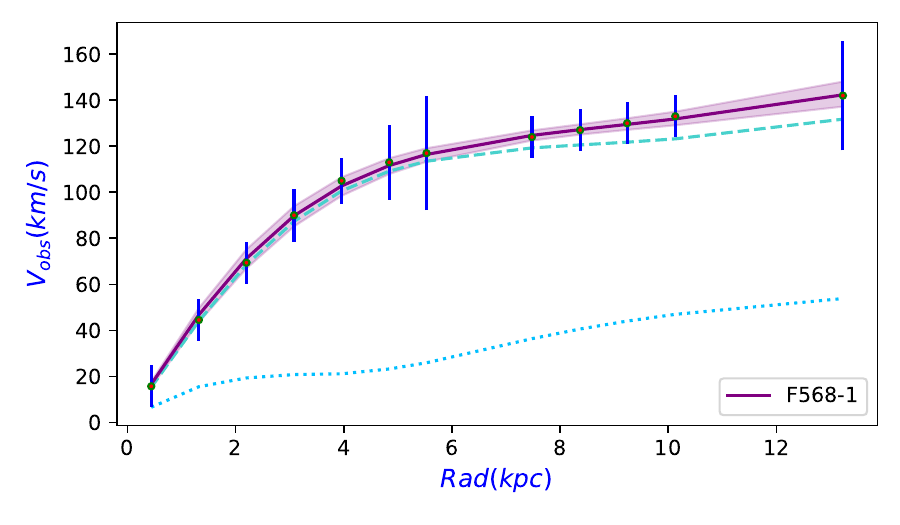} \hspace{0.01\textwidth}
	\includegraphics[width=0.23\textwidth]{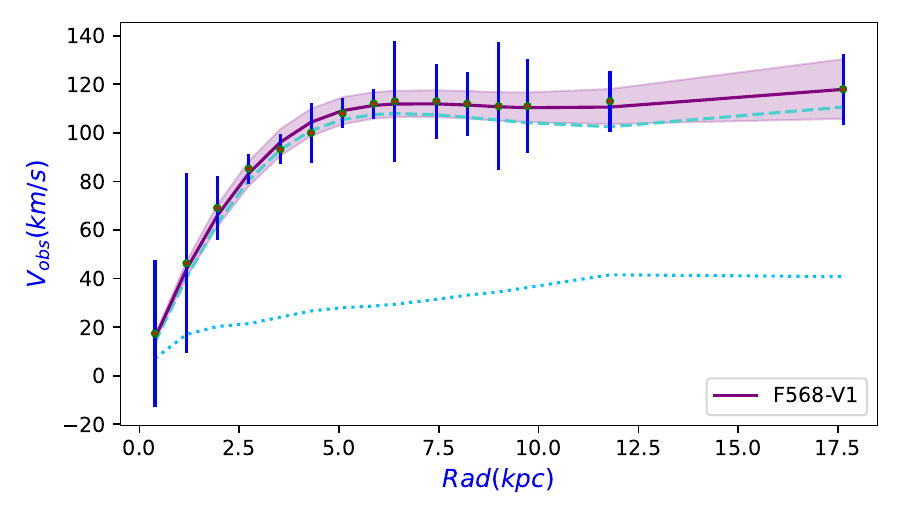}\\[0.35cm] 
	\includegraphics[width=0.23\textwidth]{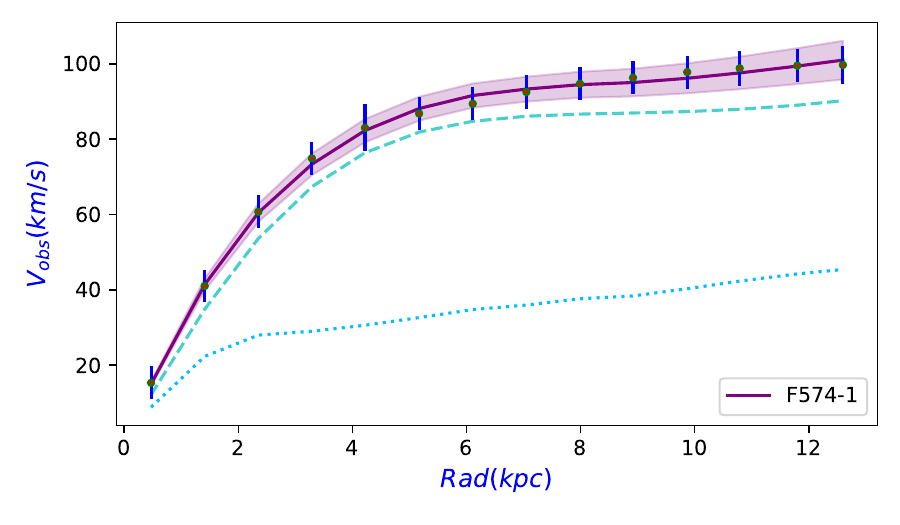} \hspace{0.01\textwidth}
	\includegraphics[width=0.23\textwidth]{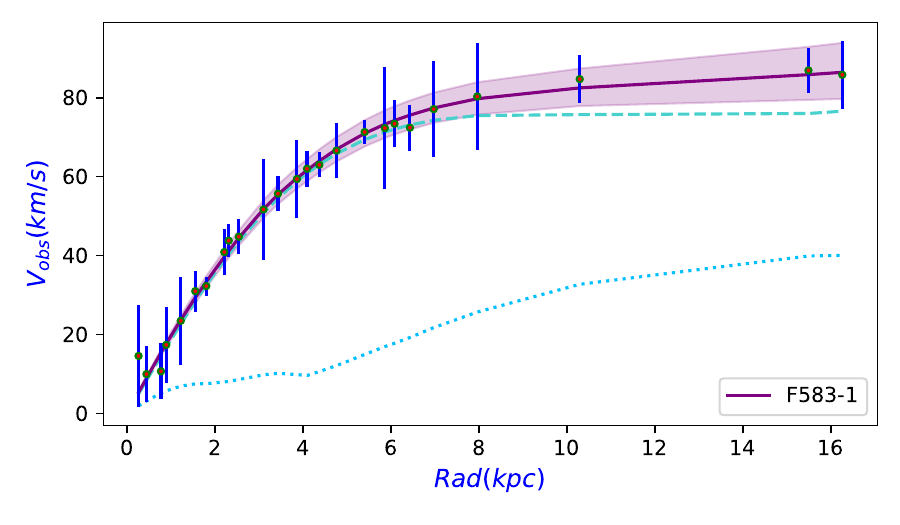} \hspace{0.01\textwidth}
	\includegraphics[width=0.23\textwidth]{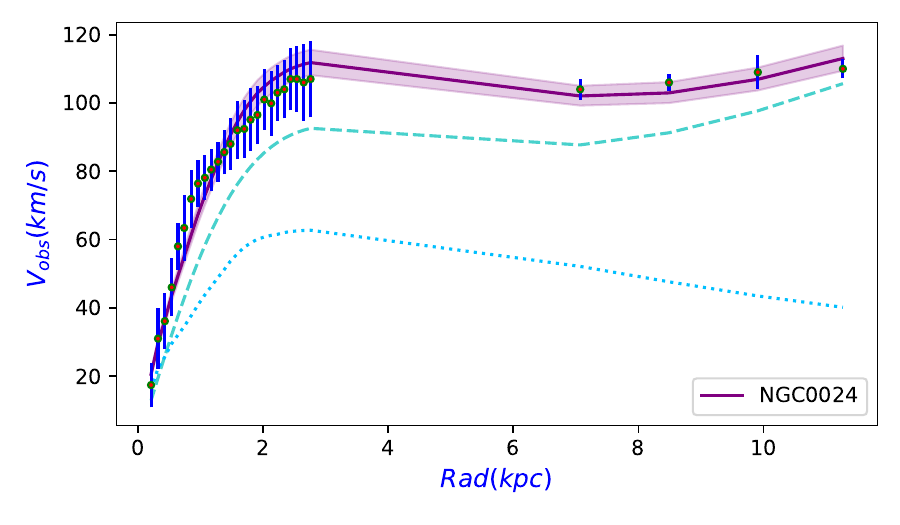} \hspace{0.01\textwidth}
	\includegraphics[width=0.23\textwidth]{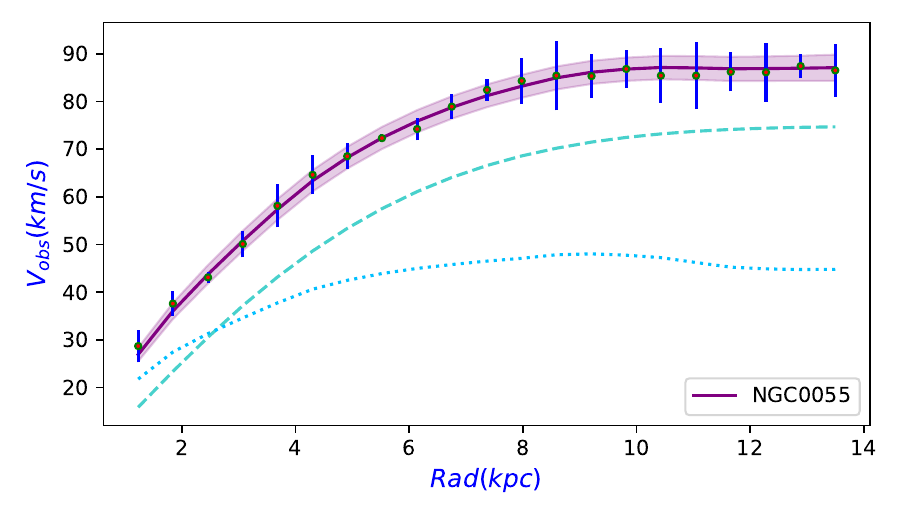}  \\[0.35cm]
	\includegraphics[width=0.23\textwidth]{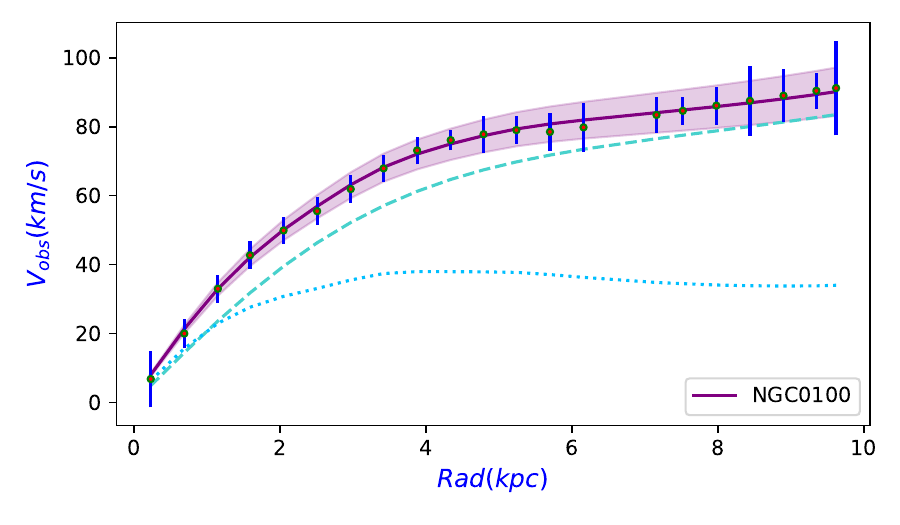} \hspace{0.01\textwidth}
	\includegraphics[width=0.23\textwidth]{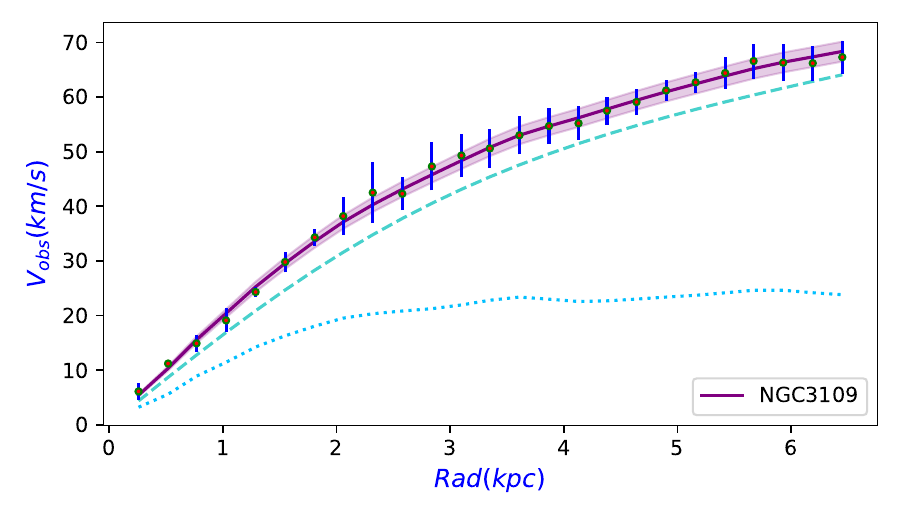} \hspace{0.01\textwidth}
	\includegraphics[width=0.23\textwidth]{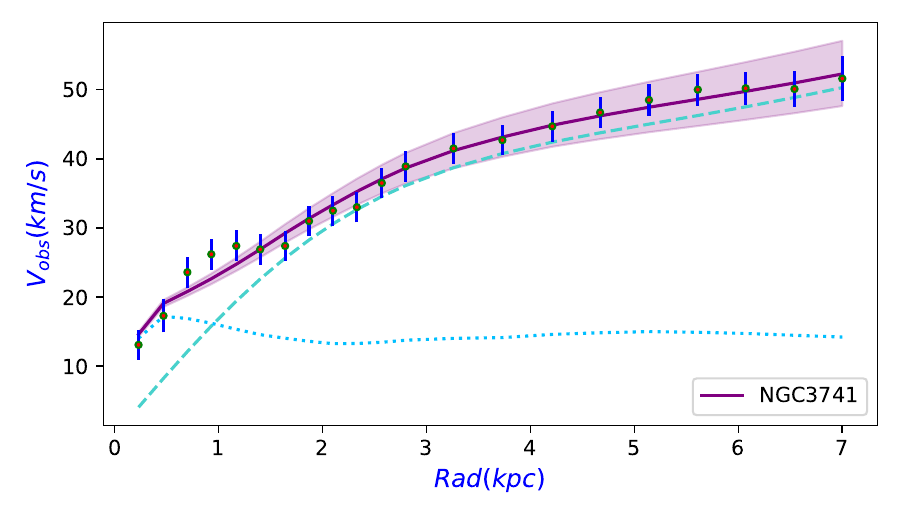} \hspace{0.01\textwidth} 
	\includegraphics[width=0.23\textwidth]{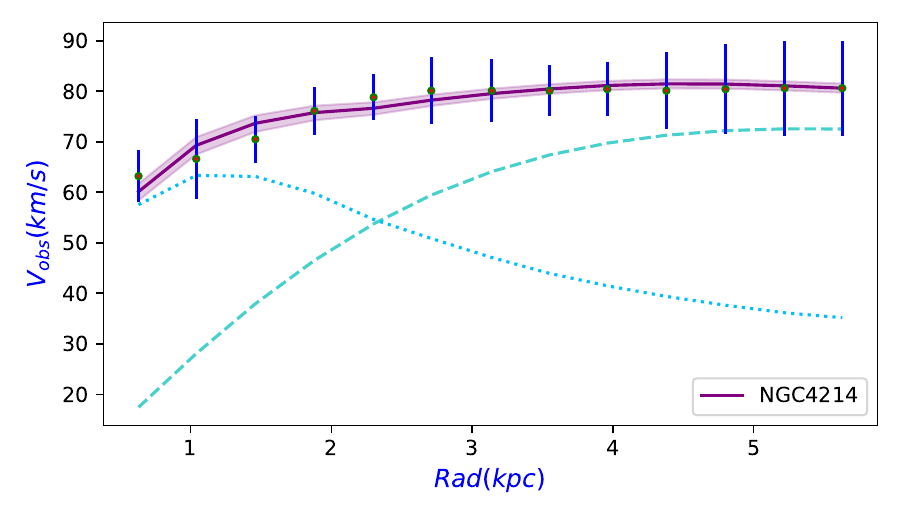}\\[0.35cm]
	\includegraphics[width=0.23\textwidth]{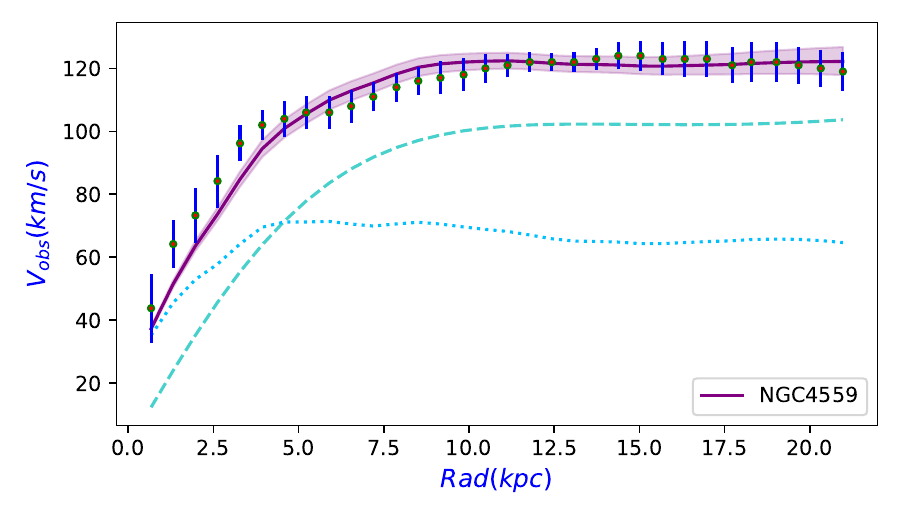} \hspace{0.01\textwidth}
	\includegraphics[width=0.23\textwidth]{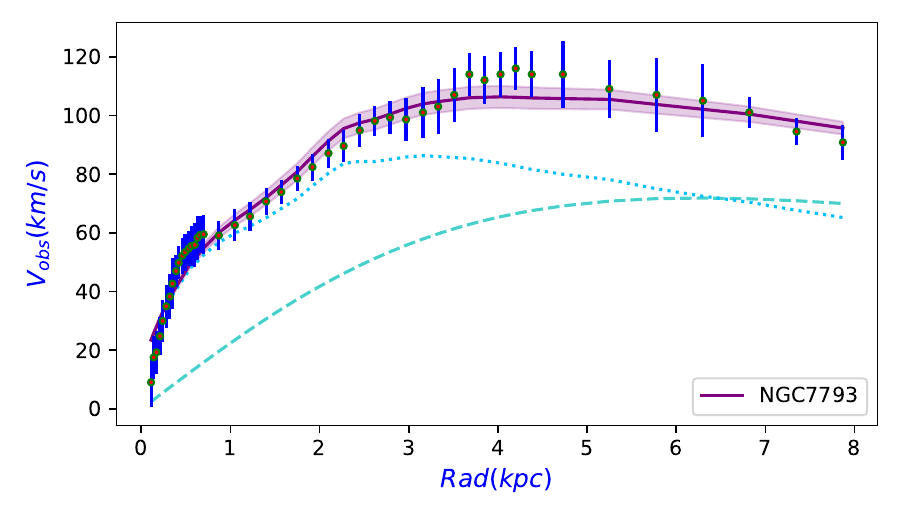} \hspace{0.01\textwidth}
	\includegraphics[width=0.23\textwidth]{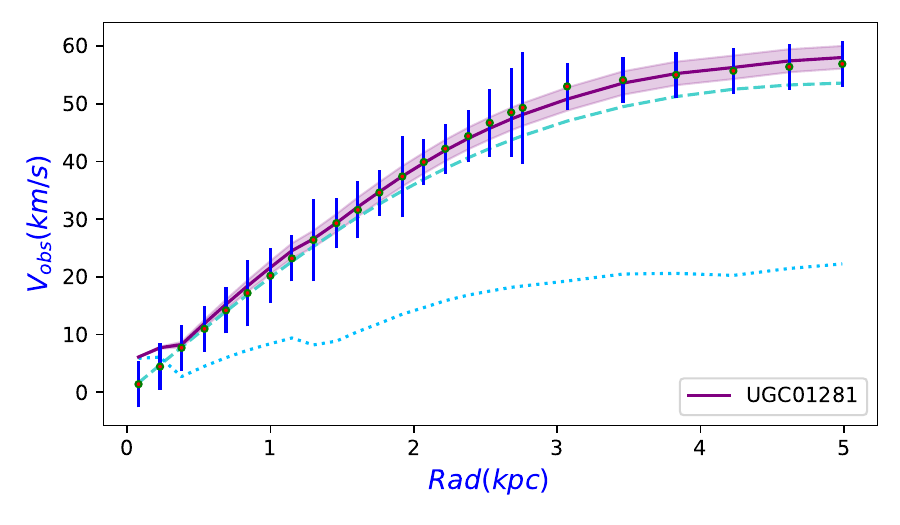} \hspace{0.01\textwidth}
	\includegraphics[width=0.23\textwidth]{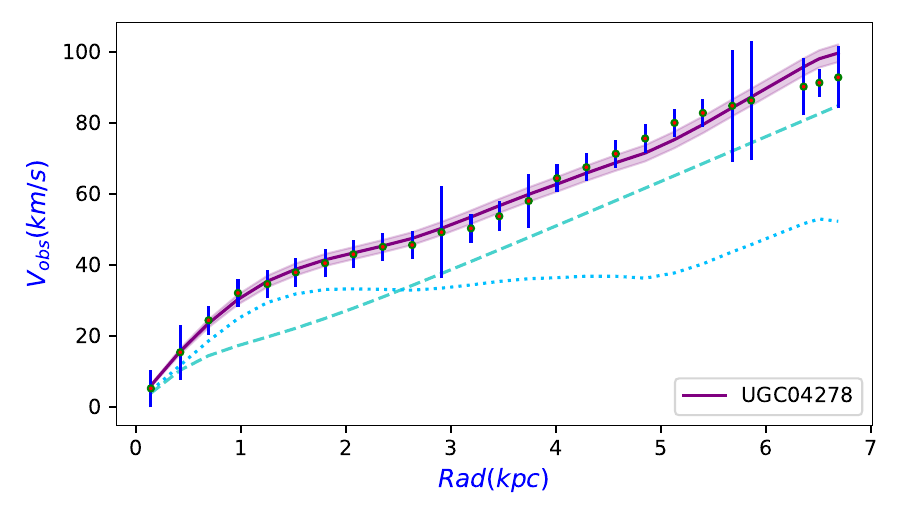}  \\[0.35cm]
	\includegraphics[width=0.23\textwidth]{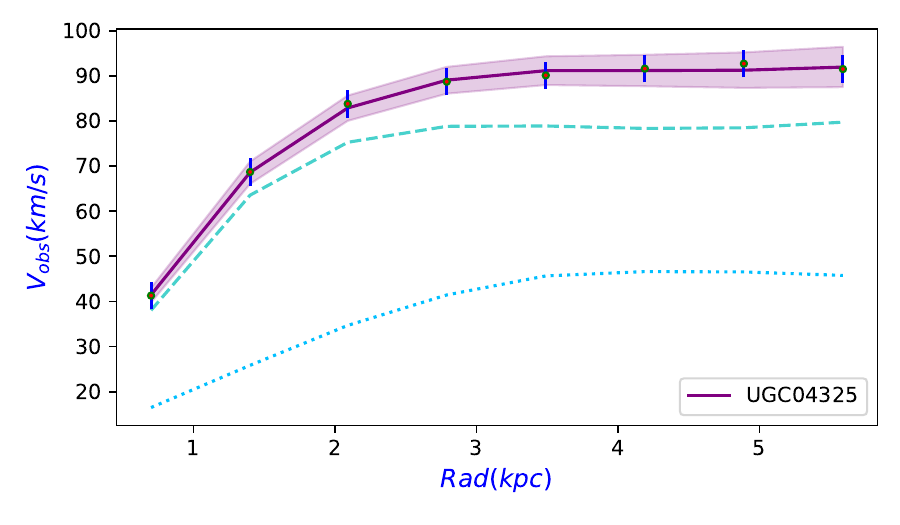} \hspace{0.01\textwidth}
	\includegraphics[width=0.23\textwidth]{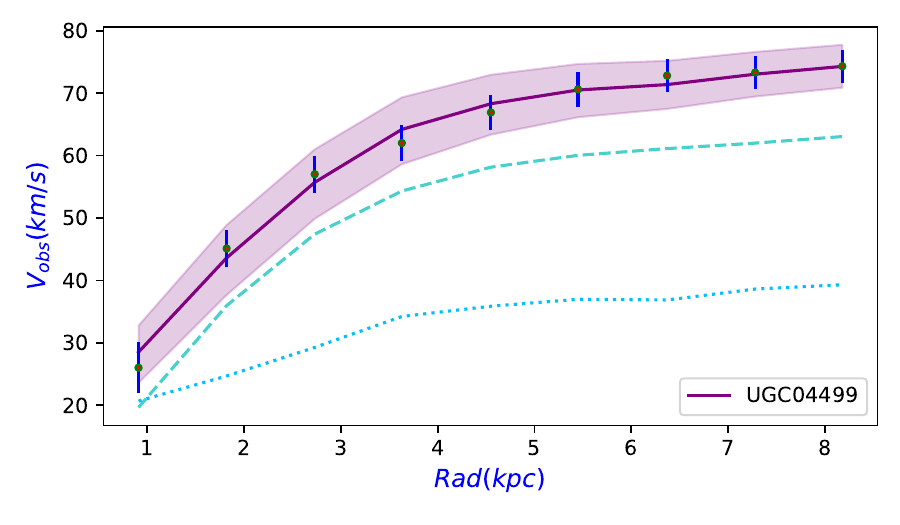} \hspace{0.01\textwidth}
	\includegraphics[width=0.23\textwidth]{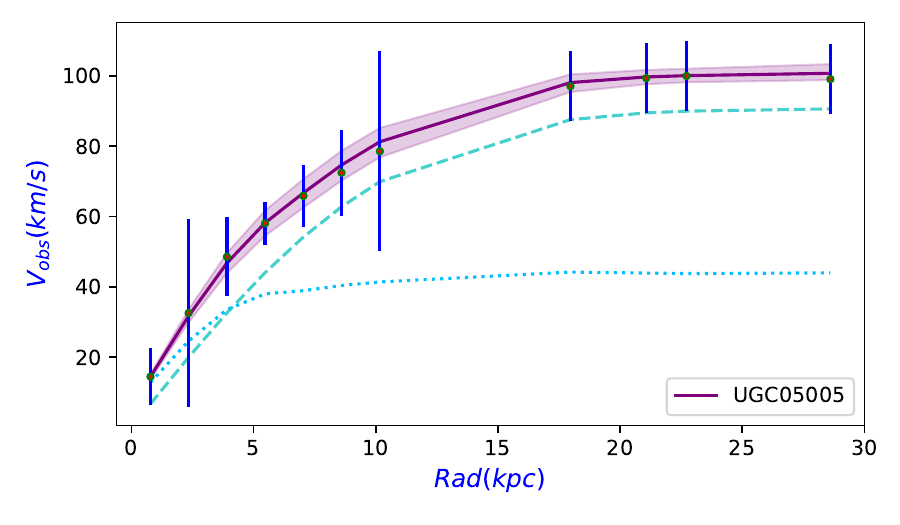} \hspace{0.01\textwidth}
	\includegraphics[width=0.23\textwidth]{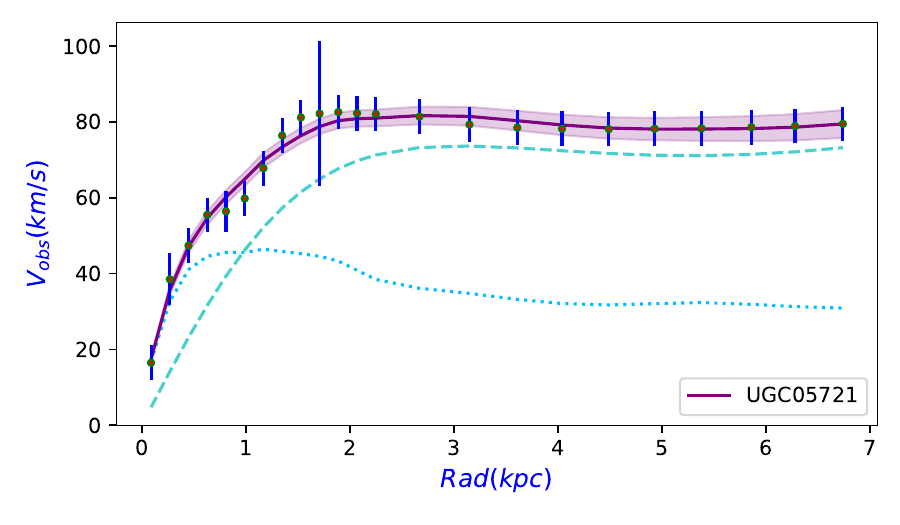} \\[0.35cm]
	\includegraphics[width=0.23\textwidth]{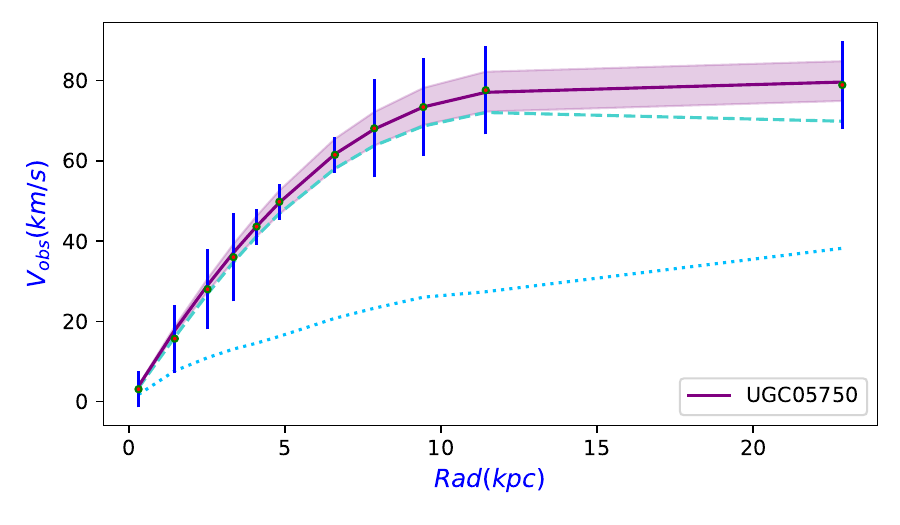}\hspace{0.01\textwidth}
	\includegraphics[width=0.23\textwidth]{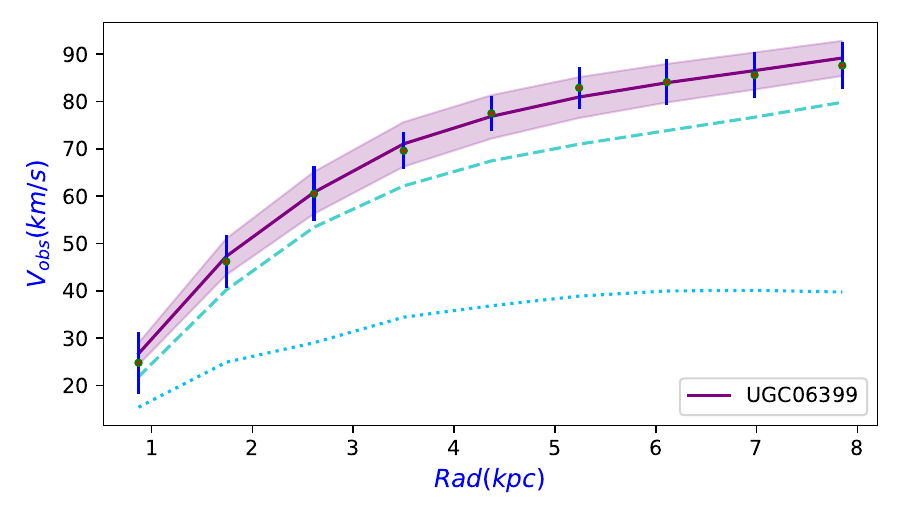} \hspace{0.01\textwidth}
	\includegraphics[width=0.23\textwidth]{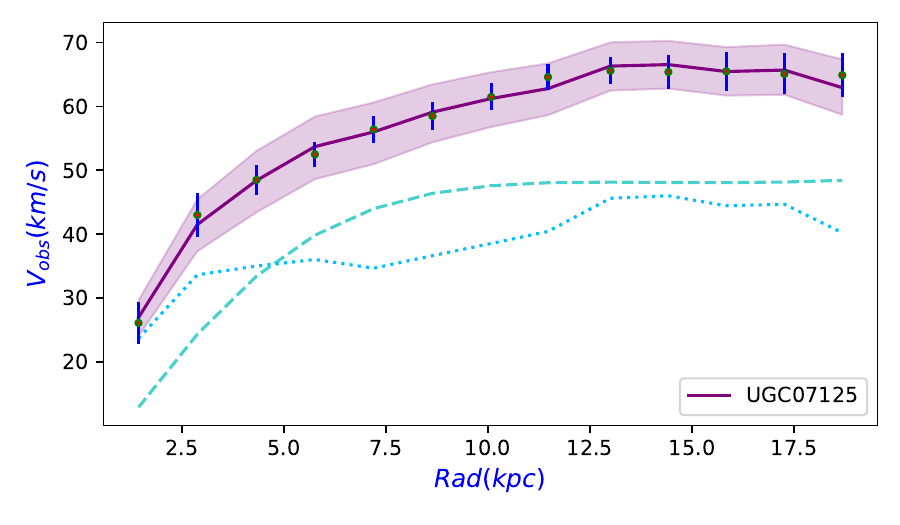} \hspace{0.01\textwidth}
	\includegraphics[width=0.23\textwidth]{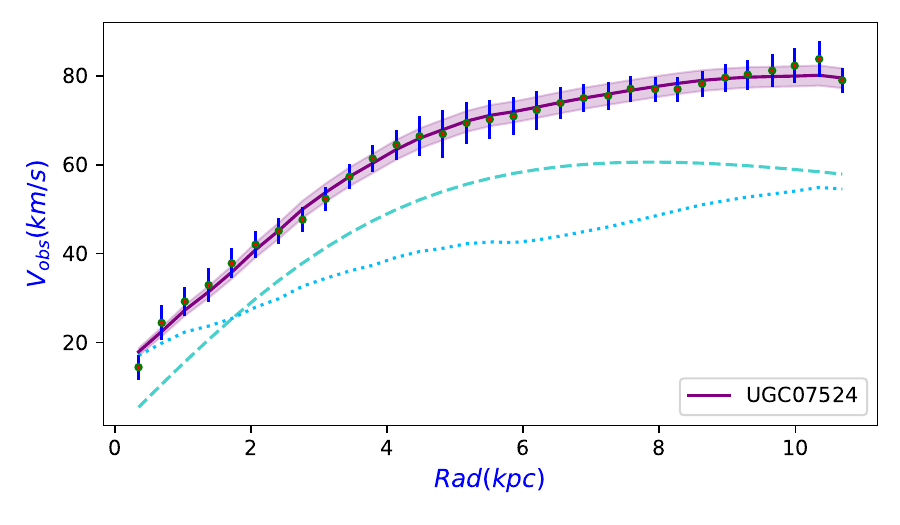} \\[0.35cm]
	\includegraphics[width=0.23\textwidth]{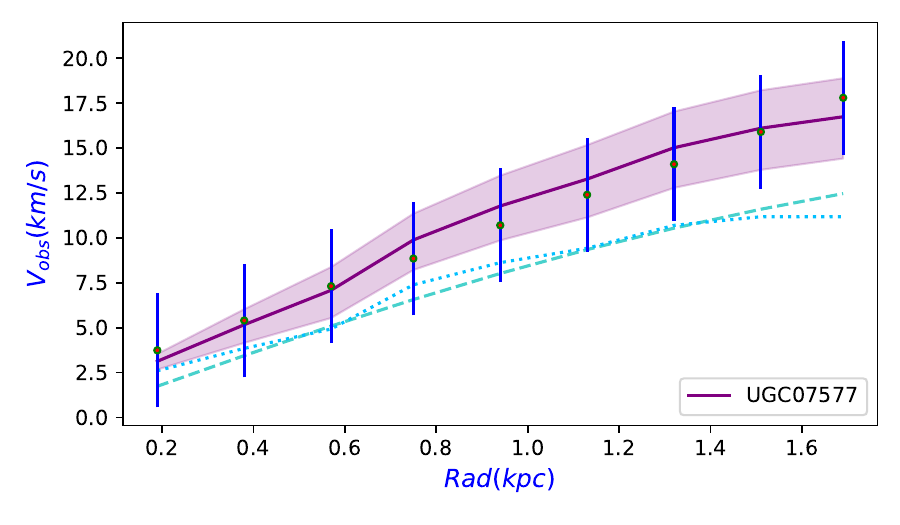} \hspace{0.01\textwidth}
	\includegraphics[width=0.23\textwidth]{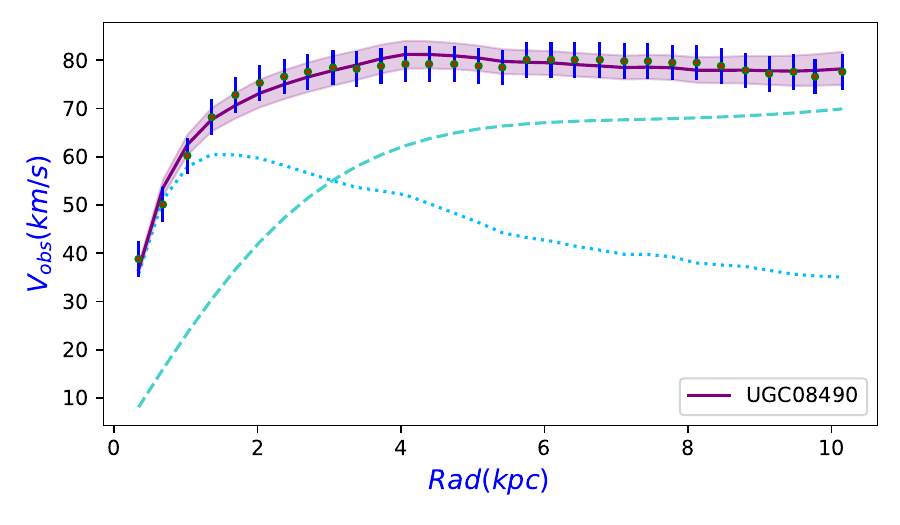} \hspace{0.01\textwidth}
    \includegraphics[width=0.23\textwidth]{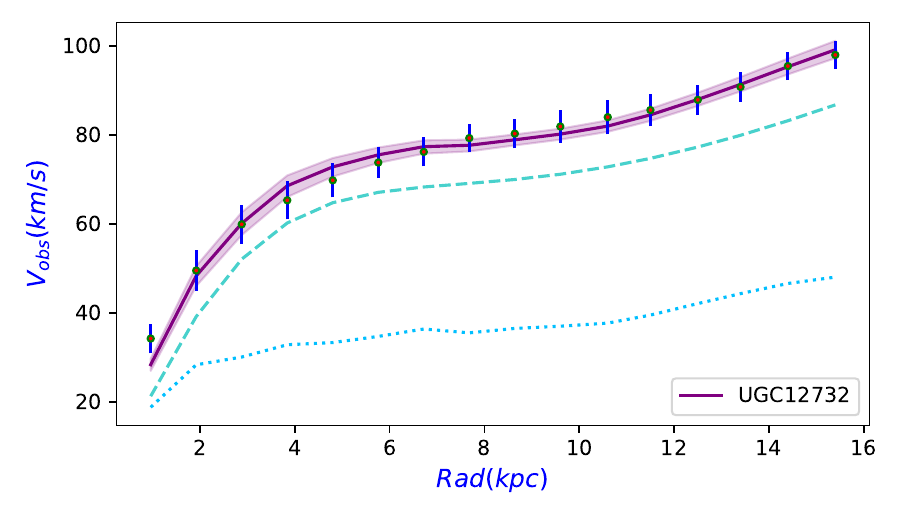} \hspace{0.01\textwidth}
    \includegraphics[width=0.23\textwidth]{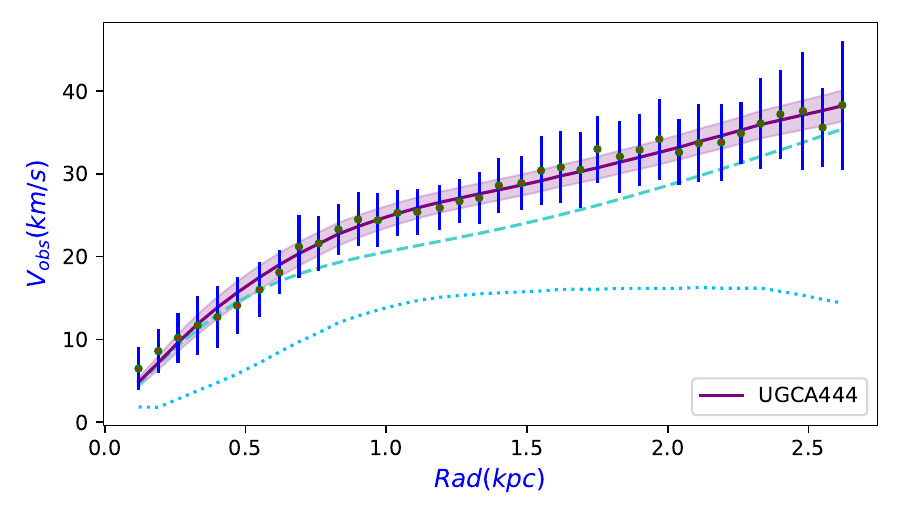}  
	\caption{Rotation curves for 32 SPARC galaxies without bulge velocity component. The observed data points are shown with their respective uncertainties, while solid lines represent the best-fit model rotation curves based on SPARC data, including the $1\sigma$ confidence interval. The dashed curves represent the contribution from logarithmic BEC dark matter, whereas the dotted curves indicate the contribution from baryonic matter. }
	\label{velo}
\end{figure*}

\paragraph{Fitting results for a small selected sample of galaxies.} In Fig.~\ref{velo}, we present the fitting results for a sample of 32 galaxies without bulge components. For each galaxy, the observed rotation curve data with error bars are shown alongside the best-fit prediction of the logarithmic BEC dark matter model, represented by a solid line. The shaded region indicates the $1\sigma$ confidence interval of the fit.

	
The best-fit values of the four parameters $\left(\Upsilon_d, B, |\vec{\Omega}|, \rho_{m0}\right)$, along with their $1\sigma$ confidence intervals, are presented in Table~\ref{tab1}. The Table also includes the reduced chi-squared values ($\chi^2_{\text{red}}$) for each galaxy, indicating how well the model fits the data. It is evident that for all cases, $\chi^2_{\text{red}} < 1$, which signifies a good fit of the logarithmic BEC dark matter model to the observational data. 

\begin{figure*}[htbp]
\centering
	\includegraphics[width=0.50\textwidth]{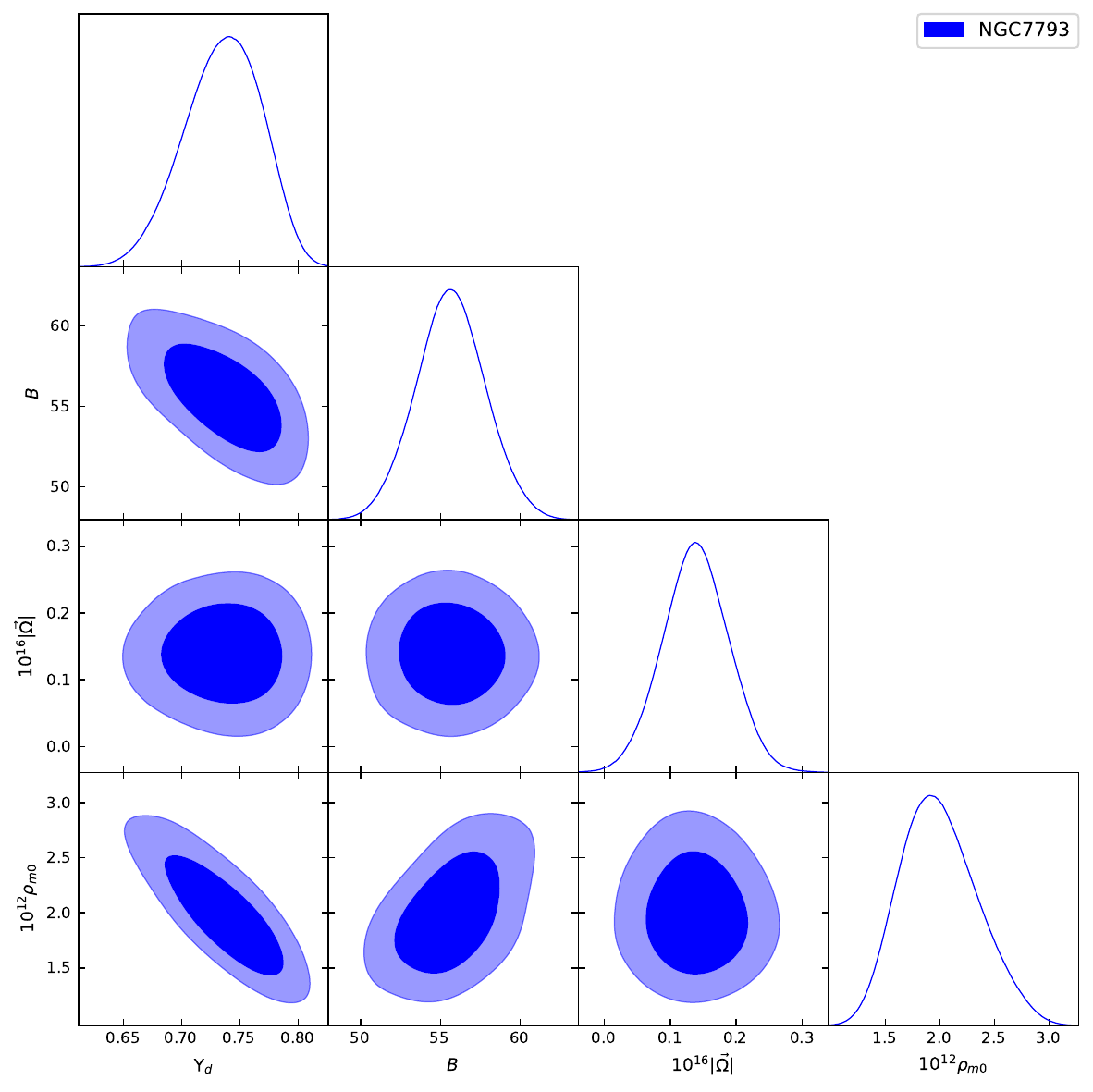}
	\includegraphics[width=0.50\textwidth]{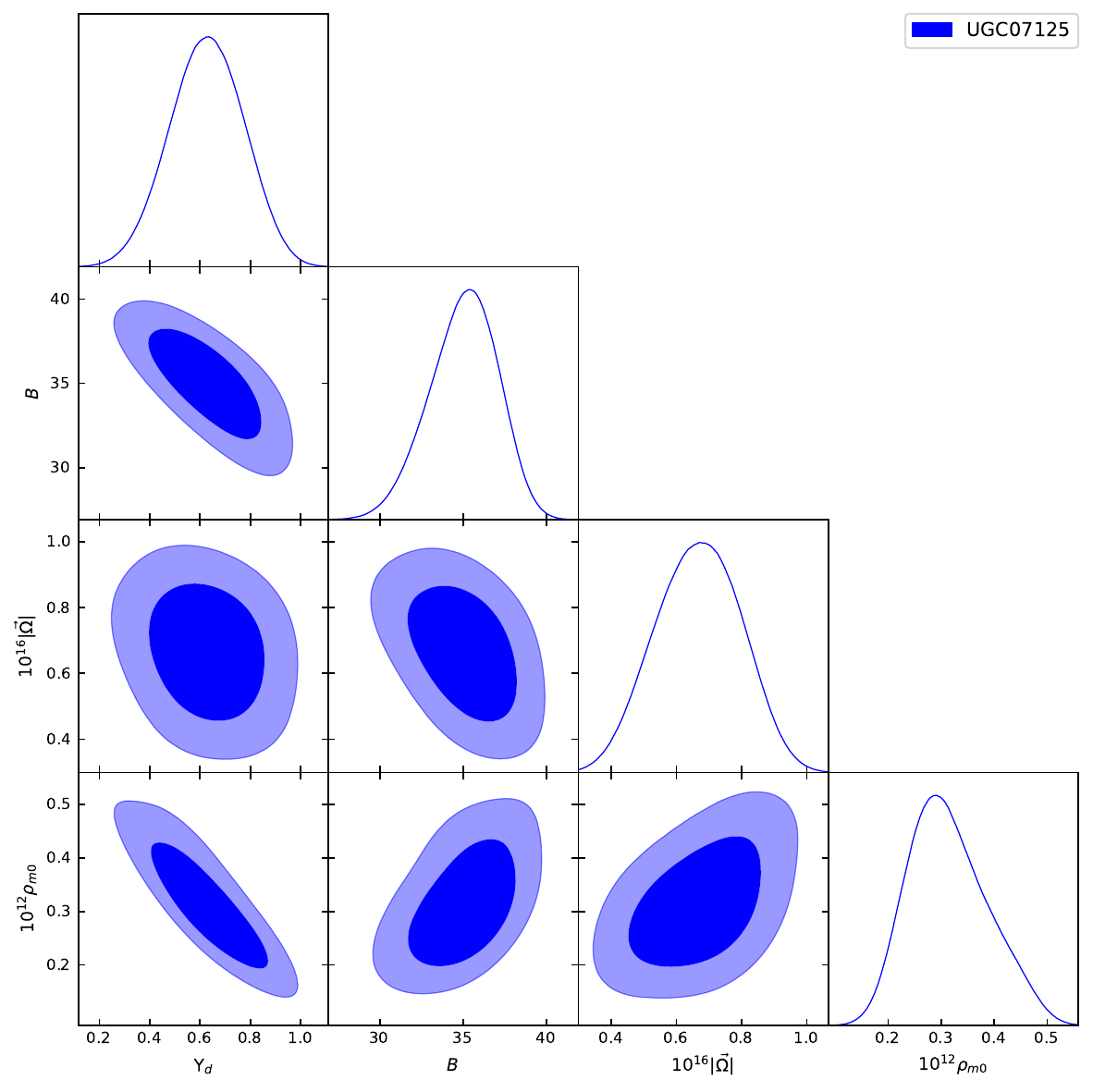}
	\caption{Corner plot of the parameters $(\Upsilon_d, B, |\vec{\Omega}|, \rho_{m0})$ with their $1\sigma$ and $2\sigma$ confidence levels for the NGC7793 (right panel) and UGC07125 (left panel) galaxies, which lack bulge velocity data, in the context of the logarithmic BEC dark matter model. The best-fit values of the parameters are presented in Table~\ref{tab1}.}\label{corner}
\end{figure*}

\paragraph{Corner plots.} Fig.~\ref{corner} presents the corner plots for the best-fit parameters of the logarithmic BEC dark matter model for the two bulgeless galaxies, NGC7793 and UGC07125. These plots illustrate the marginalized posterior distributions and the $1\sigma$ and $2\sigma$ confidence contours for the four fitted parameters. For both galaxies, the presence of tilted and elongated ellipses in the two-dimensional contour plots indicates significant correlations between some of the parameters. 

Notably, a moderate negative correlation is observed between $\Upsilon_d$ and $\rho_{m0}$, as well as between $\Upsilon_d$ and $B$, suggesting degeneracy in how the disk and dark matter contributions balance to reproduce the observed rotation curves. However, the corner plots confirm that the model parameters are well-fitted and statistically meaningful for both galaxies.

	
\begin{table*}[htbp]
	\centering
	\footnotesize
	\renewcommand{\arraystretch}{1.4}
	\renewcommand\cellalign{c}
	\renewcommand\cellgape{\Gape[4pt]}
	\begin{tabular}{|l|c|c|c|c|c|}
		\Xhline{1pt}
		\textbf{Galaxy} &
		\makecell{$\mathbf{\Upsilon_d}$ \\ ($M_\odot/L_\odot$)} &
		\makecell{$\mathbf{B}$ \\ (${\rm km/s})$} &
		\makecell{$|\vec{\mathbf{\Omega}}|$ \\ ($10^{-16}$${\rm s}$$^{-1}$)} &
		\makecell{$\mathbf{\rho_{m0}}$ \\ ($10^{-24}$ ${\rm g/cm}$$^3$)} &
		\makecell{$\boldsymbol{\chi^2_{red}}$} \\
		\Xhline{1pt}
		D631-7 & $0.123^{+0.029}_{-0.028}$ & $45.814^{+1.880}_{-1.667}$ & $0.372^{+0.005}_{-0.005}$ & $0.649^{+0.037}_{-0.040}$ & $0.431^{+0.083}_{-0.077}$ \\
		\hline
		DDO064 & $1.400^{+0.030}_{-0.033}$ & $30.088^{+3.252}_{-2.849}$ & $1.391^{+0.262}_{-0.235}$ & $1.951^{+0.331}_{-0.381}$ & $0.468^{+0.072}_{-0.066}$ \\
		\hline
		DDO154 & $0.773^{+0.048}_{-0.047}$ & $32.555^{+0.186}_{-0.194}$ & $1.190^{+0.052}_{-0.055}$ & $1.127^{+0.021}_{-0.021}$ & $1.006^{+0.097}_{-0.106}$ \\
		\hline
		DDO161 & $0.675^{+0.082}_{-0.078}$ & $44.060^{+0.494}_{-0.474}$ & $0.550^{+0.047}_{-0.049}$ & $0.267^{+0.012}_{-0.012}$ & $0.249^{+0.036}_{-0.037}$ \\
		\hline
		ESO079-G014 & $0.262^{+0.057}_{-0.062}$ & $112.945^{+0.888}_{-0.805}$ & $1.998^{+0.241}_{-0.252}$ & $2.002^{+0.188}_{-0.196}$ & $0.701^{+0.089}_{-0.088}$ \\
		\hline
		ESO116-G012 & $0.811^{+0.036}_{-0.037}$ & $68.735^{+1.209}_{-1.147}$ & $1.795^{+0.262}_{-0.280}$ & $1.901^{+0.070}_{-0.067}$ & $0.778^{+0.071}_{-0.070}$ \\
		\hline
		F568-1 & $0.260^{+0.031}_{-0.031}$ & $83.999^{+0.486}_{-0.473}$ & $3.107^{+0.284}_{-0.288}$ & $4.554^{+0.657}_{-0.649}$ & $0.149^{+0.106}_{-0.104}$ \\
		\hline
		F568-V1 & $0.334^{+0.102}_{-0.101}$ & $80.725^{+2.895}_{-2.848}$ & $2.068^{+0.327}_{-0.337}$ & $4.774^{+0.615}_{-0.600}$ & $0.188^{+0.104}_{-0.098}$ \\
		\hline
		F574-1 & $0.460^{+0.013}_{-0.013}$ & $62.137^{+1.852}_{-1.889}$ & $2.058^{+0.209}_{-0.222}$ & $2.527^{+0.263}_{-0.257}$ & $0.220^{+0.088}_{-0.089}$ \\
		\hline
		F583-1 & $0.142^{+0.052}_{-0.051}$ & $55.661^{+2.406}_{-2.406}$ & $1.325^{+0.222}_{-0.209}$ & $1.364^{+0.101}_{-0.099}$ & $0.181^{+0.057}_{-0.053}$ \\
		\hline
		NGC0024 & $0.743^{+0.047}_{-0.046}$ & $70.180^{+2.451}_{-2.345}$ & $3.384^{+0.125}_{-0.121}$ & $14.207^{+1.366}_{-1.346}$ & $0.600^{+0.042}_{-0.044}$ \\
		\hline
		NGC0055 & $0.380^{+0.080}_{-0.075}$ & $53.921^{+1.150}_{-1.225}$ & $1.002^{+0.064}_{-0.064}$ & $0.641^{+0.049}_{-0.049}$ & $0.219^{+0.064}_{-0.066}$ \\
		\hline
		NGC0100 & $0.428^{+0.049}_{-0.051}$ & $50.258^{+2.877}_{-2.930}$ & $2.556^{+0.283}_{-0.290}$ & $1.690^{+0.196}_{-0.205}$ & $0.127^{+0.051}_{-0.052}$ \\
		\hline
		NGC3109 & $1.188^{+0.186}_{-0.183}$ & $38.769^{+0.988}_{-0.994}$ & $2.794^{+0.069}_{-0.066}$ & $1.060^{+0.077}_{-0.071}$ & $0.278^{+0.039}_{-0.039}$ \\
		\hline
		NGC3741 & $1.422^{+0.065}_{-0.065}$ & $28.630^{+1.939}_{-1.949}$ & $2.256^{+0.261}_{-0.253}$ & $1.185^{+0.113}_{-0.118}$ & $0.643^{+0.059}_{-0.060}$ \\
		\hline
		NGC4214 & $1.143^{+0.067}_{-0.068}$ & $54.991^{+0.323}_{-0.318}$ & $1.302^{+0.050}_{-0.051}$ & $2.970^{+0.006}_{-0.006}$ & $0.200^{+0.053}_{-0.051}$ \\
		\hline
		NGC4559 & $0.354^{+0.010}_{-0.009}$ & $74.524^{+1.270}_{-1.173}$ & $1.349^{+0.151}_{-0.158}$ & $1.285^{+0.149}_{-0.140}$ & $0.786^{+0.035}_{-0.033}$ \\
		\hline
		NGC7793 & $0.736^{+0.035}_{-0.033}$ & $55.650^{+2.127}_{-2.114}$ & $0.139^{+0.051}_{-0.049}$ & $1.992^{+0.373}_{-0.348}$ & $0.697^{+0.023}_{-0.023}$ \\
		\hline
		UGC01281 & $0.229^{+0.080}_{-0.075}$ & $41.526^{+1.471}_{-1.411}$ & $0.077^{+0.031}_{-0.031}$ & $1.605^{+0.144}_{-0.141}$ & $0.188^{+0.043}_{-0.038}$ \\
		\hline
		UGC04278 & $1.073^{+0.107}_{-0.112}$ & $10.518^{+0.509}_{-0.524}$ & $5.029^{+0.126}_{-0.130}$ & $2.992^{+0.323}_{-0.307}$ & $0.507^{+0.039}_{-0.038}$ \\
		\hline
		UGC04325 & $0.600^{+0.048}_{-0.047}$ & $58.050^{+1.592}_{-1.643}$ & $4.025^{+0.377}_{-0.389}$ & $12.740^{+1.005}_{-1.054}$ & $0.456^{+0.236}_{-0.230}$ \\
		\hline
		UGC04499 & $0.453^{+0.222}_{-0.221}$ & $42.699^{+2.515}_{-2.367}$ & $2.015^{+0.096}_{-0.095}$ & $1.871^{+0.427}_{-0.419}$ & $0.660^{+0.195}_{-0.190}$ \\
		\hline
		UGC05005 & $0.622^{+0.053}_{-0.049}$ & $64.980^{+0.312}_{-0.306}$ & $0.678^{+0.113}_{-0.110}$ & $0.283^{+0.054}_{-0.052}$ & $0.103^{+0.084}_{-0.077}$ \\
		\hline
		UGC05721 & $1.017^{+0.086}_{-0.083}$ & $54.803^{+1.454}_{-1.380}$ & $3.259^{+0.215}_{-0.225}$ & $10.507^{+0.370}_{-0.332}$ & $0.319^{+0.054}_{-0.056}$ \\
		\hline
		UGC05750 & $0.133^{+0.005}_{-0.005}$ & $54.964^{+4.221}_{-3.943}$ & $0.623^{+0.052}_{-0.050}$ & $0.460^{+0.060}_{-0.056}$ & $0.166^{+0.122}_{-0.122}$ \\
		\hline
		UGC06399 & $0.552^{+0.120}_{-0.115}$  & $47.553^{+2.394}_{-2.525}$ & $3.069^{+0.098}_{-0.099}$ & $2.514^{+0.418}_{-0.393}$ & $0.312^{+0.151}_{-0.157}$\\
		\hline
		UGC07125 & $0.628^{+0.145}_{-0.141}$ & $35.048^{+2.005}_{-2.213}$ & $0.668^{+0.131}_{-0.134}$ & $0.313^{+0.082}_{-0.076}$ & $0.434^{+0.112}_{-0.121}$ \\
		\hline
		UGC07524 & $1.213^{+0.111}_{-0.122}$ & $46.915^{+1.309}_{-1.321}$ & $0.102^{+0.070}_{-0.073}$ & $0.891^{+0.064}_{-0.061}$ & $0.260^{+0.033}_{-0.033}$ \\
		\hline
		UGC07577 & $0.297^{+0.134}_{-0.130}$ & $12.041^{+1.960}_{-1.952}$ & $0.414^{+0.010}_{-0.010}$ & $0.316^{+0.122}_{-0.121}$ & $0.195^{+0.096}_{-0.095}$ \\
		\hline
		UGC08490 & $1.639^{+0.088}_{-0.092}$ & $48.248^{+1.322}_{-1.339}$ & $1.925^{+0.204}_{-0.204}$ & $2.136^{+0.349}_{-0.324}$ & $0.251^{+0.048}_{-0.045}$ \\
		\hline
		UGC12732 & $0.814^{+0.049}_{-0.049}$ & $48.078^{+0.983}_{-1.060}$ & $2.010^{+0.067}_{-0.064}$ & $1.947^{+0.263}_{-0.250}$ & $0.608^{+0.079}_{-0.084}$ \\
		\hline
		UGCA444 & $0.604^{+0.124}_{-0.129}$ & $13.242^{+1.078}_{-1.062}$ & $5.152^{+0.280}_{-0.293}$ & $5.414^{+1.306}_{-1.260}$ & $0.119^{+0.030}_{-0.027}$ \\
		\hline
	\end{tabular}
	\caption{The best-fit values of the parameters, along with their $1\sigma$ uncertainties, for the galaxies whose rotation curves are shown in Fig.~\ref{velo}.} \label{tab1}
\end{table*}

\begin{figure*}[htbp]
	\centering
	\includegraphics[width=9.5cm]{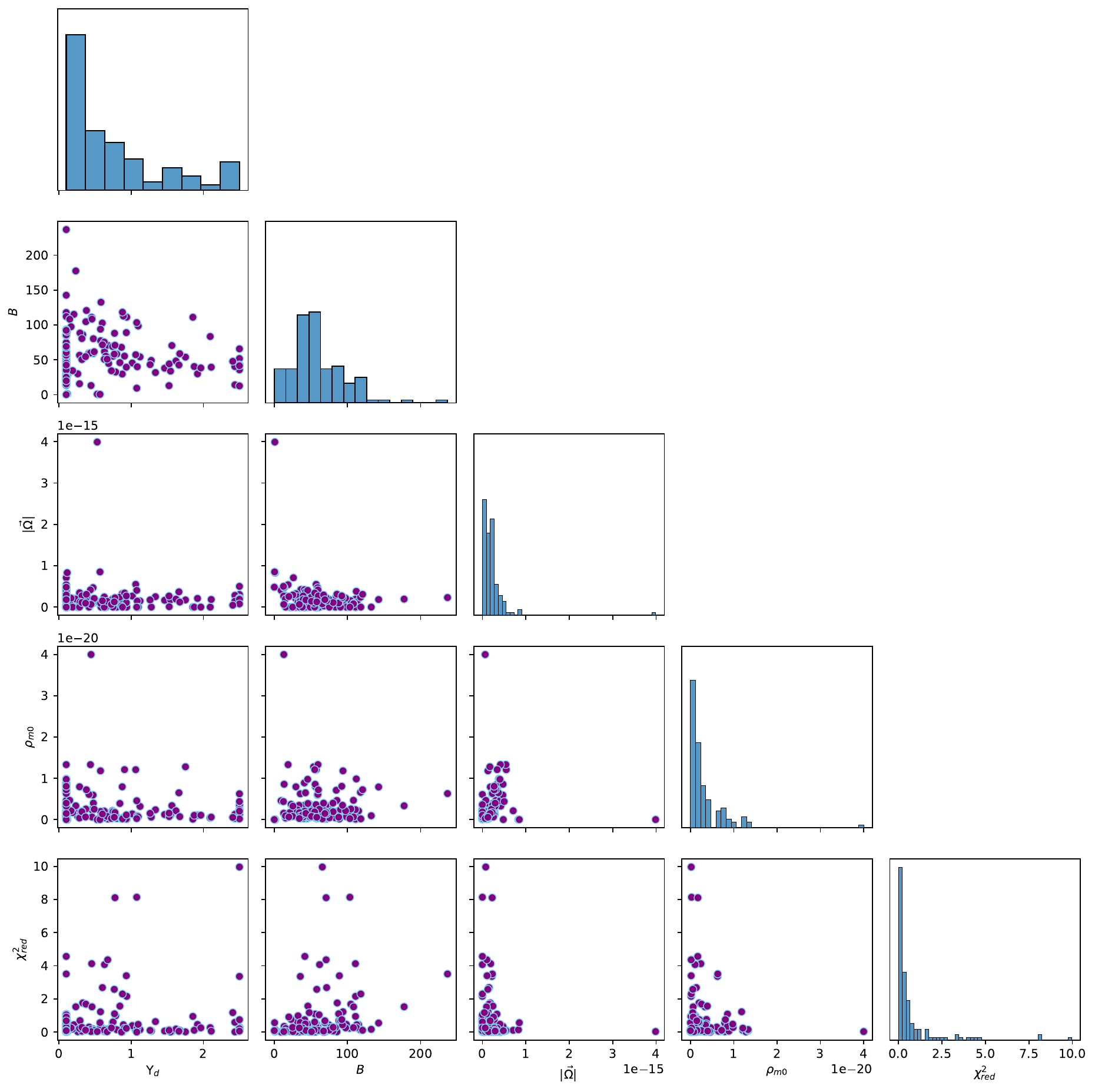}\hspace{.4cm} %
	\includegraphics[width=6.5cm]{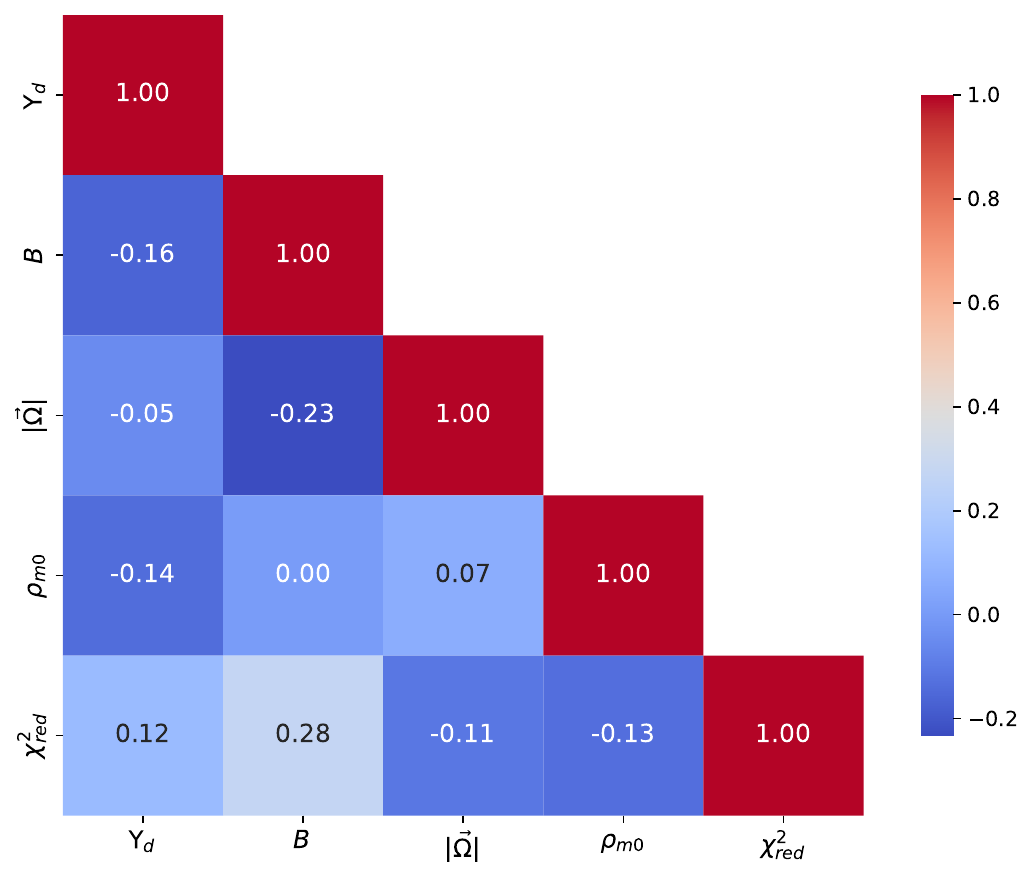}
	\caption{Corner plot (left panel) and correlation heatmap (right panel) of the logarithmic BEC dark matter model parameters, obtained by fitting to the rotation curves of 132 bulgeless galaxies. }
	\label{correl}
\end{figure*}

\paragraph{Fitting results for 132 SPARC galaxies.} To explore the behavior of the logarithmic BEC dark matter model more comprehensively, we fitted the model to 132 bulgeless galaxies with rotation curve data from the SPARC database. The results for all galaxies are summarized in Fig.~\ref{correl}. The left panel displays the corner plot, while the right panel shows the correlation heat map of the best-fit values of the model parameters.

The diagonal panels of the corner plot  (left panel in  Fig.~\ref{correl}) show the distributions (histograms) of the best-fit values for the parameters $\left(\Upsilon_d, B, |\vec{\Omega}|, \rho_{m0}\right)$ and the reduced chi-squared, $\chi^2_{red}$, across the sample. The distribution of $\Upsilon_d$ indicates that 108 out of the 132 bulgeless galaxies (about 82\%) have $\Upsilon_d < 1.5$. The average and median values of this parameter are $0.737$ and $0.508\, (M_\odot/L_\odot)$, respectively.

For the model parameter $B$, the best-fit values span a wide range, from $0.037$ to $236.827\, {\rm km/s}$, with mean and median values of $59.446$ and $54.349\, {\rm km/s}$, respectively. The mean and median values of the best-fit parameter $\left|\vec{\Omega}\right|$ are $2.085\times 10^{-16}\,{\rm s}^{-1}$ and $1.732\times 10^{-16}\,{\rm s}^{-1}$, respectively. 

The distribution of $\rho_{m0}$ shows that 81\% of the sample have values less than $5\times 10^{-20}\,{\rm g/cm}^3$. Finally, the reduced chi-squared indicates that 82\% of the sample have $\chi^2_{red}<1$, suggesting that the model is in good agreement with the observational data. 

The off-diagonal panels of the corner plot indicate the pairwise scatter distributions between the best-fit values of the model parameters and the reduced chi-squared $\chi^2_{\text{red}}$ across the sample. Each point in these panels corresponds to a single galaxy and represents its best-fit parameter values. These panels help identify any potential correlations, trends, or clustering among parameters. The weak or dispersed patterns suggest that there are no strong degeneracies between most parameter pairs. 

More details about the degeneracies can be seen in the right panel of Fig.~\ref{correl} which shows a heatmap of the Pearson correlation coefficients between the model parameters. The color scale ranges from blue (negative correlation) to red (positive correlation). Overall, the correlations are weak, with the strongest being a mild negative correlation between $|\vec\Omega|$ and $B$ ($r \approx -0.23$), and a weak positive correlation between $\chi^2_{red}$ and $B$ ($r \approx 0.28$). These low numerical values of the coefficients confirm that the model parameters are largely independent across the sample. 

It should be noted that the Pearson correlation coefficients  between reduced chi squared $\chi^2_{red}$ and all model parameters are less than $0.3$, indicating that the goodness of fit does not significantly depend on any single parameter.

\subsubsection{Galaxies with a bulge component}

Now we examine the logarithmic BEC dark matter model using observational data for galaxies with bulge components. In this case, there are five parameters $\left(\Upsilon_d, \Upsilon_b, B, |\vec{\Omega}|, \rho_{m0}\right)$,  which are determined by fitting the model to the data. 

\begin{figure*}[htbp]
	\centering
    \includegraphics[width=0.23\textwidth]{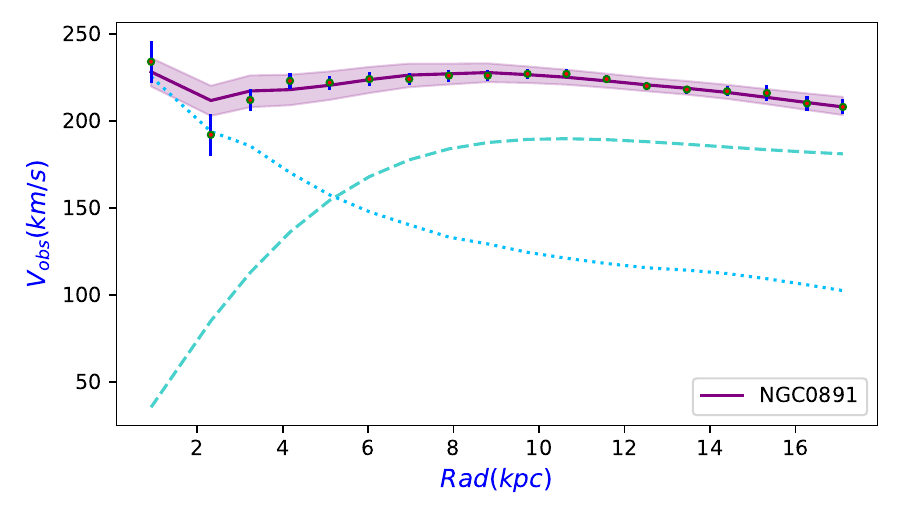} \hspace{0.01\textwidth}
	\includegraphics[width=0.23\textwidth]{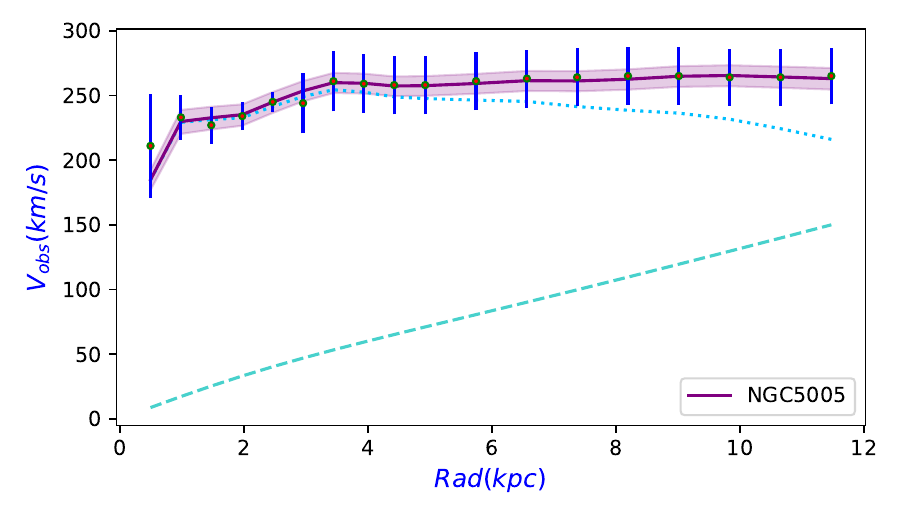} \hspace{0.01\textwidth}
	\includegraphics[width=0.23\textwidth]{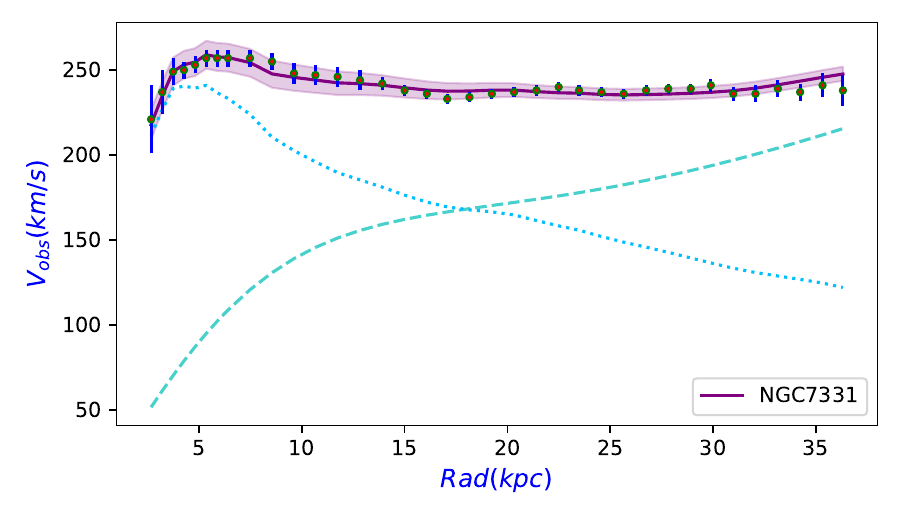} \hspace{0.01\textwidth}
	\includegraphics[width=0.23\textwidth]{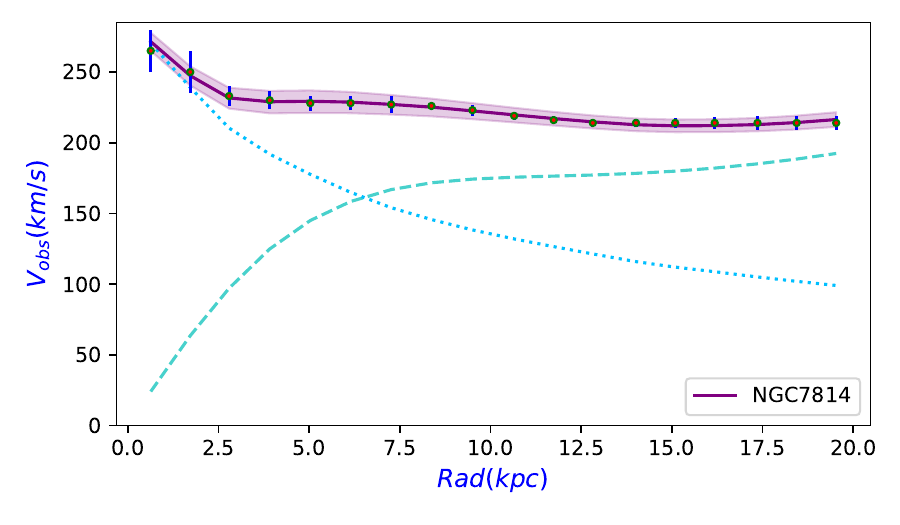} \\[.3cm] 
	\includegraphics[width=0.23\textwidth]{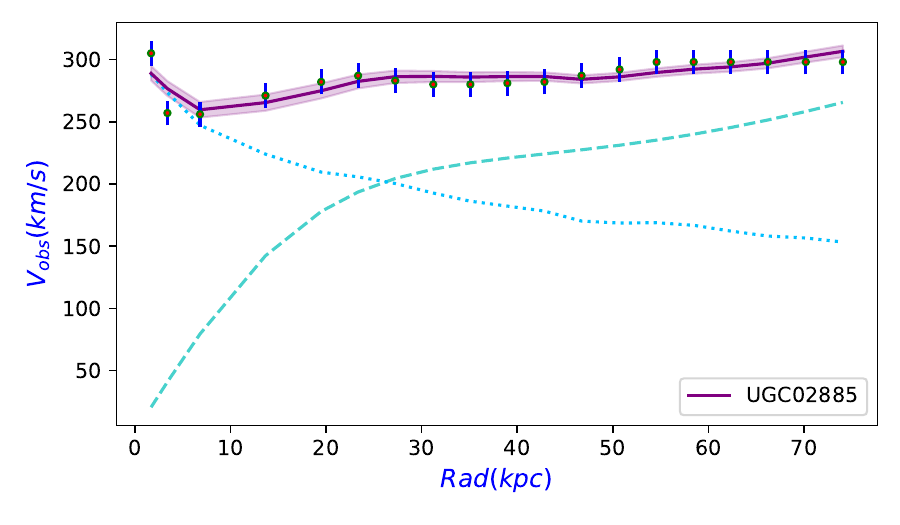} \hspace{0.01\textwidth}
    \includegraphics[width=0.23\textwidth]{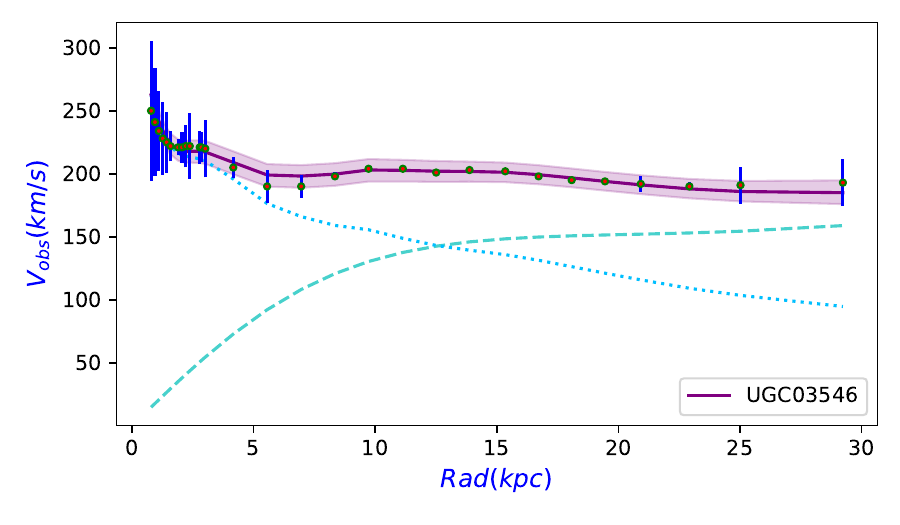} \hspace{0.01\textwidth}
    \includegraphics[width=0.23\textwidth]{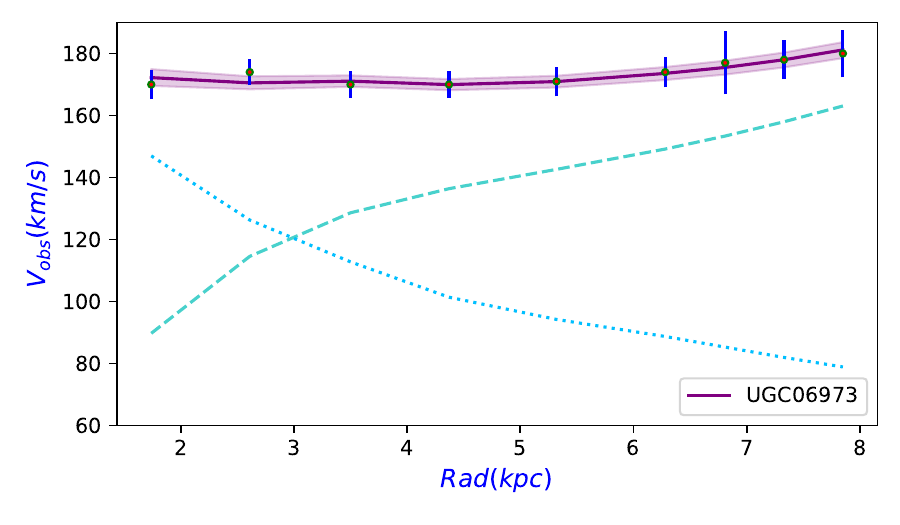} \hspace{0.01\textwidth}	
    \includegraphics[width=0.23\textwidth]{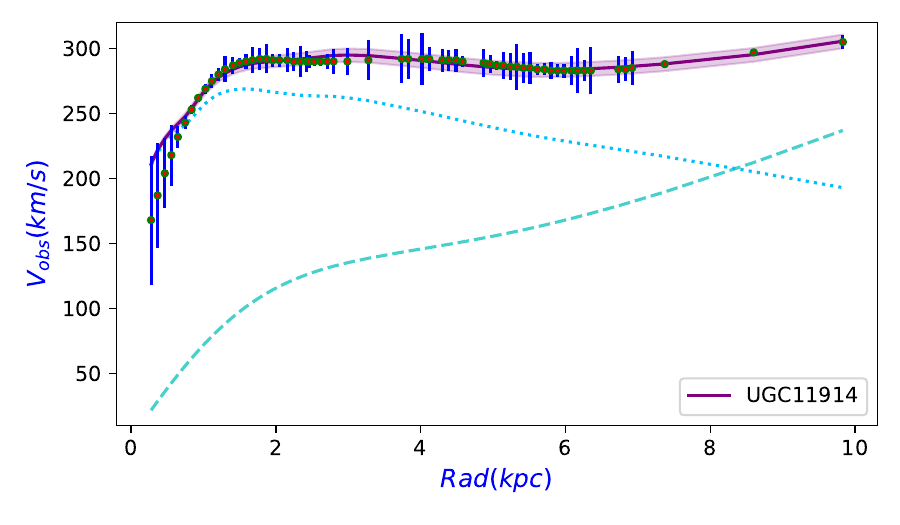} 
	\caption{Rotation curves for 8 SPARC galaxies with bulge velocity components. The observed data points are shown with their respective uncertainties, while the solid lines represent the best-fit model rotation curves based on SPARC data, including the $1\sigma$ confidence interval.  The dashed curves represent the contribution from logarithmic BEC dark matter, whereas the dotted curves indicate the contribution from baryonic matter.}
		\label{velobul}
\end{figure*}

\paragraph{Fits of a small sample of 8 galaxies.} The rotation curves for a sample of 8 galaxies with bulge components are shown in Fig.~\ref{velobul}. The solid line in each panel represents the rotation curve obtained by logarithmic BEC dark matter model, with the $1\sigma$  confidence interval shown as shaded region. The observational data and associated error bars are also displayed. The best-fit values of the parameters for this sample are presented in Table~\ref{tabbul}. To assess the goodness of fit, the reduced chi squared values are also reported in this Table. For all galaxies in the sample the reduced chi squared is less than $1$, indicating good agreement between the model and observational data.

\begin{table*}[htbp]
	\centering
	\renewcommand{\arraystretch}{1.3} 
	\renewcommand\cellalign{c}
	\renewcommand\cellgape{\Gape[4pt]}
	\begin{tabular}{|l|c|c|c|c|c|c|}
	\Xhline{1pt}
		\textbf{Galaxy} & 
		\makecell{$\mathbf{\Upsilon_d}$ \\ $(M_\odot/L_\odot)$} & 
		\makecell{$\mathbf{\Upsilon_b}$ \\ $(M_\odot/L_\odot)$} & 
		\makecell{$\mathbf{B}$ \\$({\rm km/s})$} & 
		\makecell{$|\vec{\mathbf{\Omega}}|$ \\ $(10^{-16}s^{-1})$} & 
		\makecell{$\mathbf{\rho_{m0}}$ \\ ($10^{-24} {\rm g/cm}^3$)} & 
		\makecell{ $\boldsymbol{\chi^2_{red}}$ \\ }\\
	\Xhline{1pt}
		NGC0891   & $0.147^{+0.016}_{-0.015}$ & $0.748^{+0.052}_{-0.053}$ & $141.995^{+0.453}_{-0.431}$ & $2.225^{+0.307}_{-0.296}$ & $5.656^{+0.560}_{-0.545}$ & $0.701^{+0.093}_{-0.096}$ \\
		\hline
		NGC5005   & $0.619^{+0.034}_{-0.036}$ & $0.487^{+0.048}_{-0.049}$ & $35.679^{+0.283}_{-0.281}$  & $5.124^{+0.206}_{-0.201}$ & $1.201^{+0.044}_{-0.048}$ & $0.142^{+0.058}_{-0.060}$ \\
		\hline
		NGC7331   & $0.409^{+0.016}_{-0.015}$ & $0.180^{+0.099}_{-0.101}$ & $113.682^{+2.405}_{-2.292}$ & $2.094^{+0.052}_{-0.054}$ & $1.483^{+0.244}_{-0.243}$ & $0.672^{+0.037}_{-0.038}$ \\
		\hline
		NGC7814   & $0.424^{+0.055}_{-0.056}$ & $0.608^{+0.028}_{-0.028}$ & $125.422^{+2.399}_{-2.529}$ & $3.130^{+0.114}_{-0.120}$ & $5.547^{+0.648}_{-0.663}$ & $0.273^{+0.091}_{-0.091}$ \\
		\hline
		UGC03546  & $0.559^{+0.051}_{-0.052}$ & $0.472^{+0.031}_{-0.031}$ & $107.385^{+2.887}_{-2.855}$ & $1.571^{+0.162}_{-0.157}$ & $1.329^{+0.171}_{-0.174}$ & $0.642^{+0.047}_{-0.053}$ \\
		\hline
		UGC06973  & $0.181^{+0.007}_{-0.006}$ & $0.100^{+0.010}_{-0.010}$ & $92.503^{+0.098}_{-0.098}$  & $6.813^{+0.130}_{-0.144}$ & $13.799^{+0.030}_{-0.029}$& $0.439^{+0.149}_{-0.154}$ \\
		\hline
		UGC11914   & $0.416^{+0.018}_{-0.019}$ &$0.966^{+0.011}_{-0.011}$ & $95.246^{+2.179}_{-1.964}$ & $9.200^{+0.127}_{-0.112}$ & $23.991^{+1.313}_{-1.380}$ & $0.255^{+0.016}_{-0.016}$ \\
		\hline
	\end{tabular}
	\caption{The best-fit values of the parameters, along with their $1\sigma$ uncertainties, for the galaxies whose rotation curves are shown in Fig.~\ref{velobul}.}
		\label{tabbul}
\end{table*}

\begin{figure*}[htb]
\centering
	\includegraphics[width=8cm]{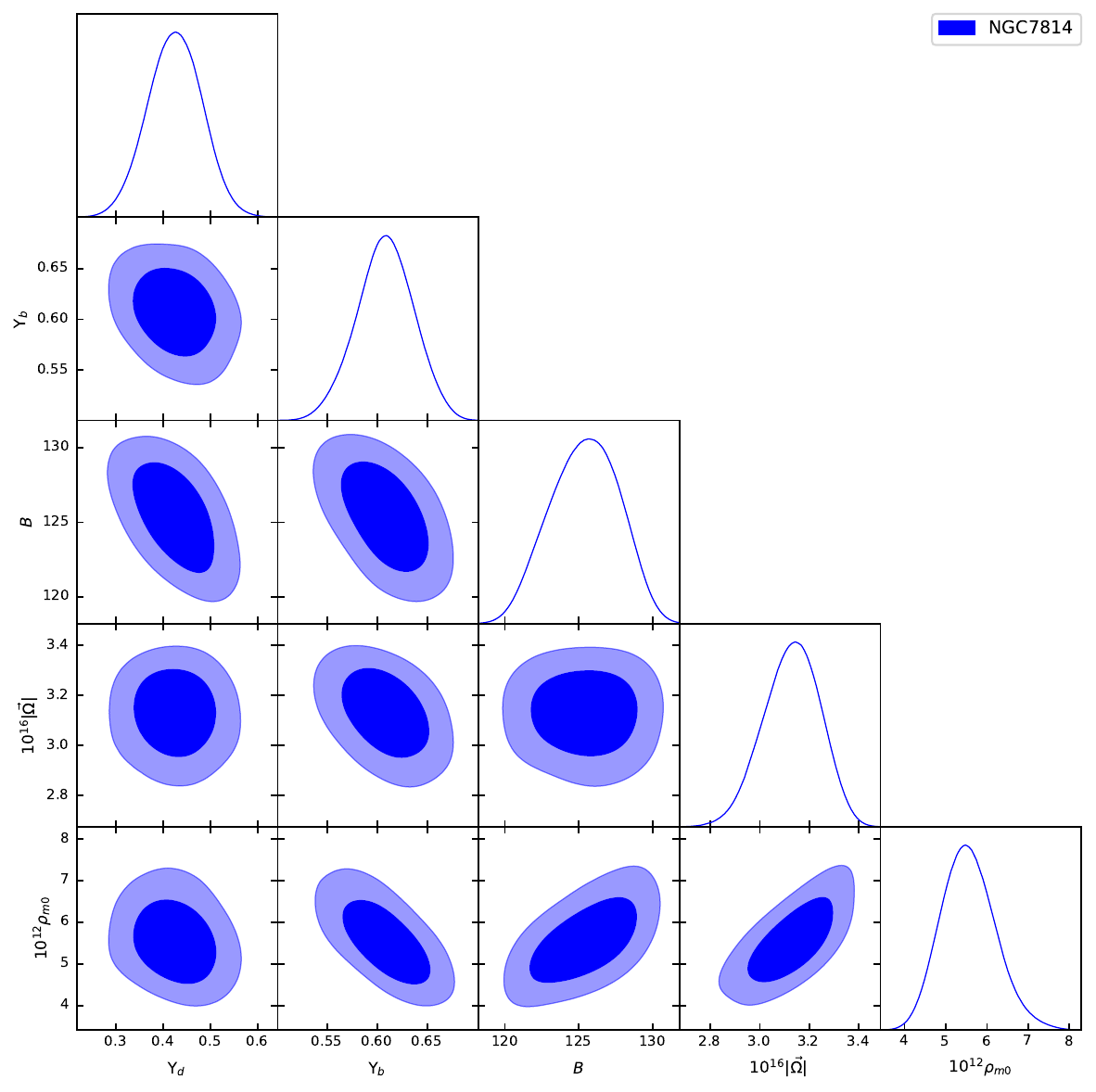}
	\includegraphics[width=8cm]{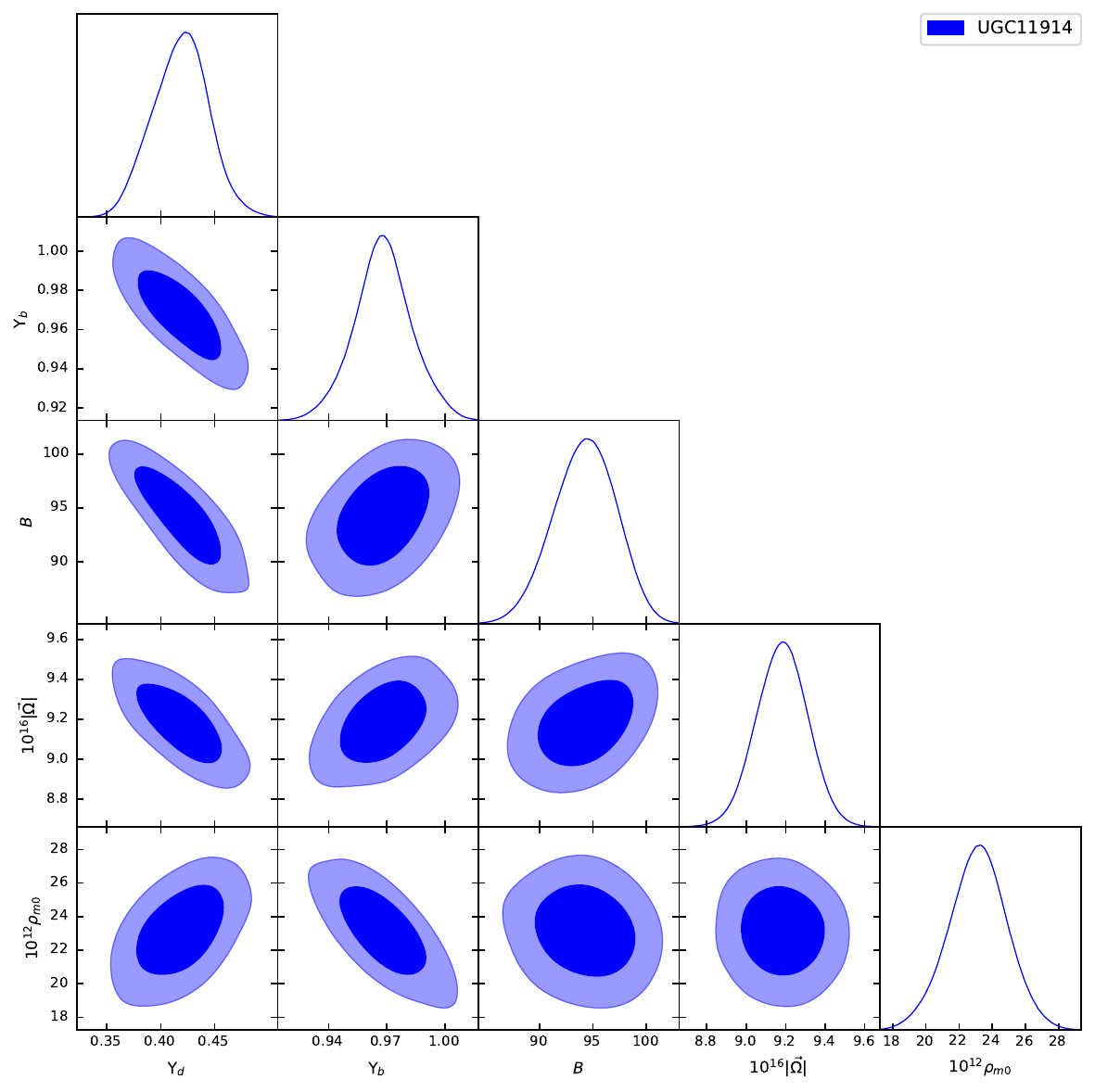}
	\caption{Corner plot of the parameters $(\Upsilon_d, \Upsilon_b, B, |\vec{\Omega}|, \rho_{m0})$ showing the $1\sigma$ and $2\sigma$ confidence regions for the NGC7814 (left panel) and UGC11914 (right panel) galaxies, which include bulge velocity data, under the logarithmic BEC dark matter model. The best fit-values of the parameters are presented in Table~\ref{tabbul}.}\label{cornerbul}
\end{figure*}

\paragraph{Corner plots.} Fig.~\ref{cornerbul} presents the corner plots for two galaxies from the sample shown in Fig.~\ref{velobul}. The corner plots correspond to NGC4814 (upper panel) and UGC11914 (lower panel). The diagonal panels show the marginalized one-dimensional posterior distribution for each parameter, $\left(\Upsilon_d, \Upsilon_b, B, |\vec{\Omega}|, \rho_{m0}\right)$.  The off-diagonal panels display the two-dimensional contour plots at $1\sigma$ and $2\sigma$ confidence levels.  

The tightness and shape of the contours indicate the degree of correlation between parameters. For example, a positive correlation is observed between $B$ and $\rho_{m0}$, while a negative correlation is seen between the $\Upsilon_b$ and $\rho_{m0}$ for NGC7814. Circular counters indicate no correlation between parameter pairs, such as between $\Upsilon_d$ and $|\vec{\Omega}|$. As one can see, in the case of UGC11914 most parameters are mutually correlated.

\begin{figure*}[htbp]
	\centering
	\includegraphics[width=10cm]{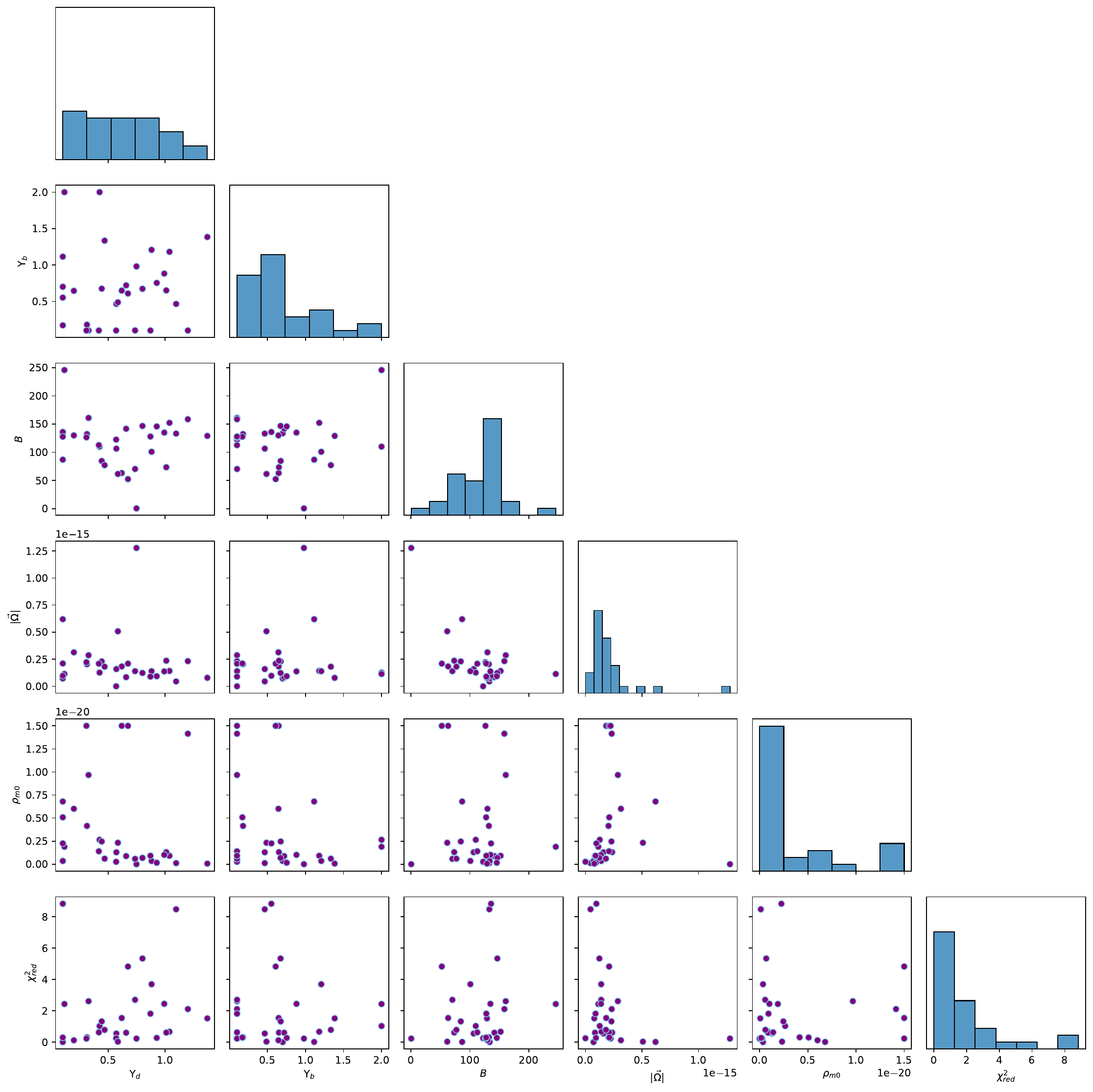}\hspace{.4cm} %
	\includegraphics[width=6.5cm]{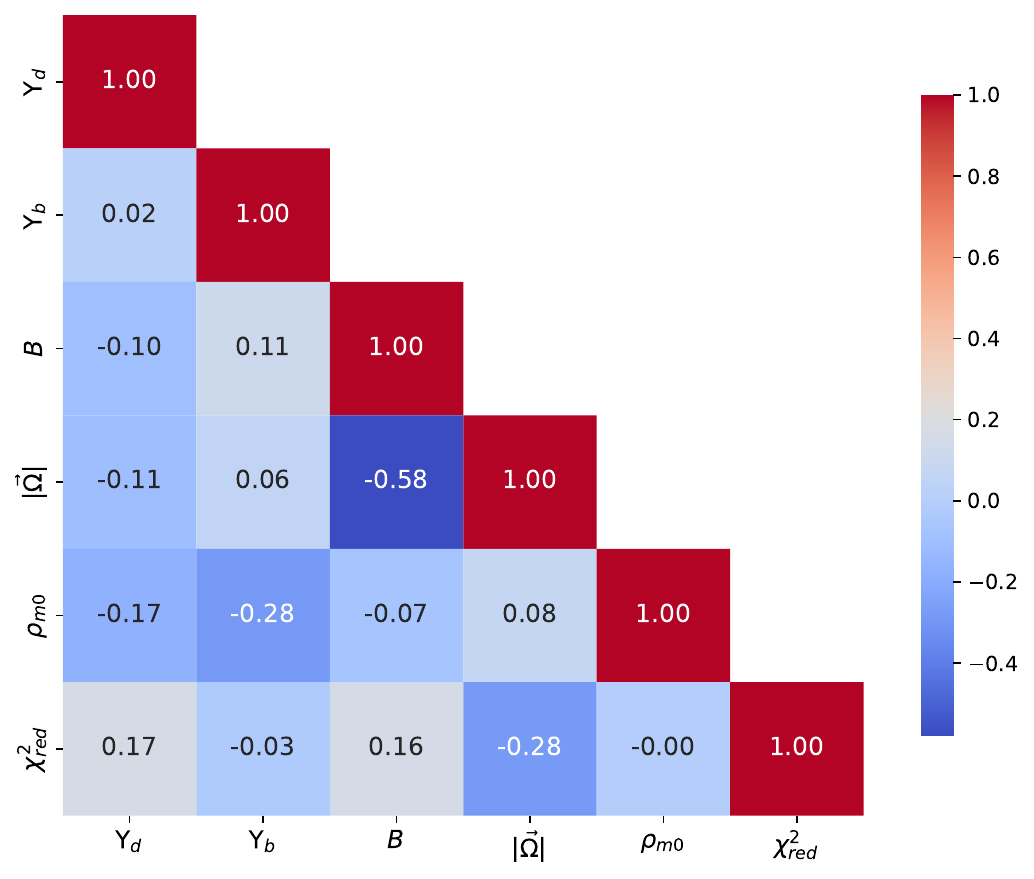}
	\caption{Corner plot and correlation heatmap of the logarithmic BEC dark matter model parameters, obtained by fitting to the rotation curves of 31 galaxies, which include bulge velocity data.}
	\label{correlbul} 
\end{figure*}

\paragraph{Fitting results for a  sample of 32 galaxies with bulge component.} Now, we apply the logarithmic BEC dark matter model to a sample of 32 galaxies with bulge components. The results of fitting the model to observational data for this sample are summarized in Fig.~\ref{correlbul}. 

The left panel presents a corner plot showing the best-fit values of the model parameters and the corresponding reduced chi squared values. The diagonal panels show histogram of the parameters, $(\Upsilon_d, \Upsilon_b, B, |\vec{\Omega}|, \rho_{m0}, \chi^2_{red})$ across the sample. The mean values of these parameters are $(0.605,\, 0.683,\, 116,\, 2.18\times10^{-16},\, 3.65\times 10^{-21},\, 1.81)$ respectively. Each point in the off-diagonal panels corresponds  to the best-fit values of the parameters for each galaxy.  

The panels illustrate the correlations between parameters  across the galaxy sample. The heatmap of Pearson correlation coefficients (left panel in Fig.~\ref{correlbul}) provides a more quantitative assessment of parameter degeneracies. A moderate negative correlation is observed between $B$ and $|\vec{\Omega}|$, with $r = -0.58$, indicating that galaxies with higher $B$ values tend to have lower $|\vec{\Omega}|$, and vice versa. As can be seen, the remaining parameter pairs exhibit either no linear correlation or only weak correlations $(|r| < 0.3)$.

\subsubsection{$\chi ^2$ distribution for the SPARC sample}

\begin{figure}[htbp]
	\centering
	\includegraphics[width=8.3cm]{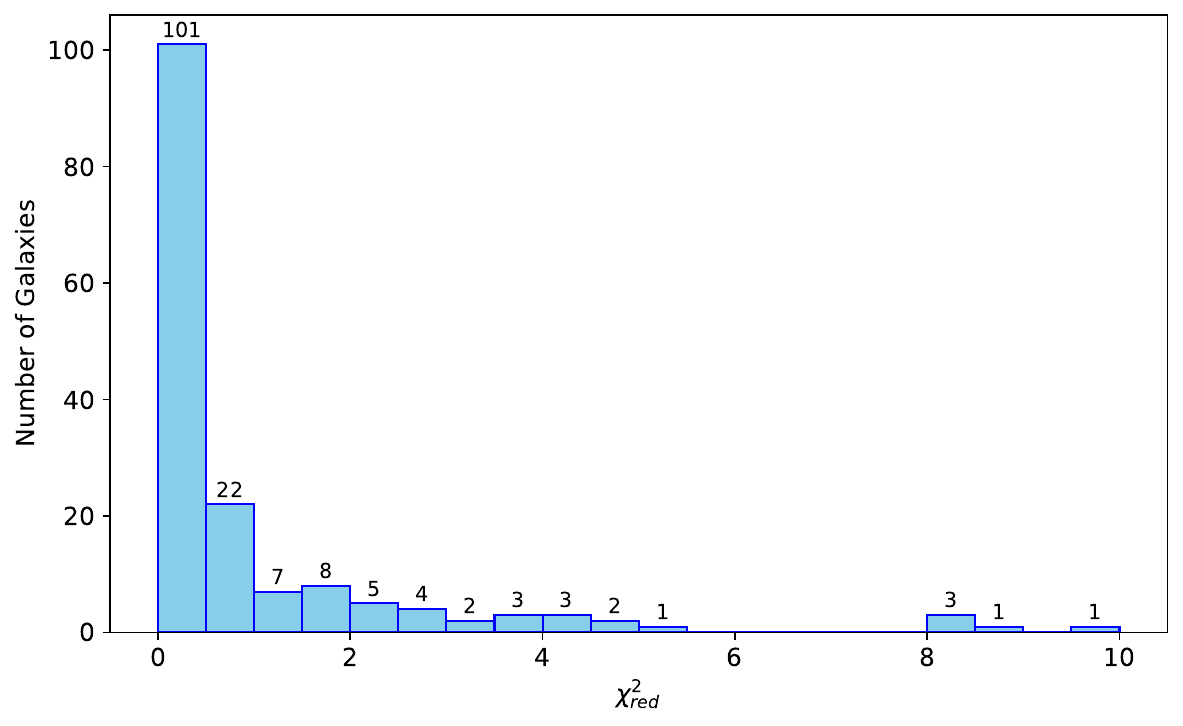}\hspace{.4cm} %
	\caption{Histogram of reduced chi-squared ($\chi^2_{red}$) values for 163 SPARC galaxies fitted with the logarithmic BEC dark matter model. The sample includes 132 bulgeless galaxies and 31 galaxies with a bulge component. }
	\label{allchi}
\end{figure}

Finally, to evaluate the goodness of fit of the logarithmic BEC dark matter model to the observational data, we have plotted the histogram of reduced chi squared  for all galaxies including 132 bulgeless galaxies and 31 with bulge components, as shown in Fig.~\ref{allchi}. This figure shows the number of galaxies in each $\chi^2_{red}$ range. The distribution shows that most galaxies (123) have $\chi^2_{red}$ values less than one, meaning the model fits them very well. Another 15 galaxies have values between  1 and 2 which also suggests good agreement with the data. Only a few have $\chi^2_{red}>5$. These cases may reflect more complex internal structures or larger observational uncertainties. Overall, this plot shows that the model provides good fits for most galaxies in the sample  particularly those with $\chi^2_{\mathrm{red}} < 2$ (about 85\%).

\section{Discussions and final remarks}\label{sect4}

In the present paper we have investigated the possibility that dark matter can exist as a low temperature phase of a logarithmic Bose-Einstein condensate. The logarithmic BEC has the interesting, and intriguing property of satisfying the linear barotropic equation of state of the ideal gas of classical thermodynamics. In order to describe dark matter we have used the coupled system of the nonlinear Gross-Pitaevskii and Poisson equations. By adopting the hydrodynamic representation of the quantum mechanics, the Gross-Pitaevskii-Poisson system can be reduced to the continuity and Euler equations of classical hydrodynamics, in the presence of a quantum potential, with the condensate satisfying the ideal gas equation of state. We have also adopted the Thomas-Fermi approximation, which allows us to neglect the physical effects of the quantum potential. 

In the case of a rotating condensate in the presence of the gravitational field, in the static case, the equilibrium equation of the condensate is described by a generalized Lane-Emden equation given by Eq.~(\ref{basic1}), which does not have an exact solution. Hence, to obtain an insight into its properties approximate methods, numerical or analytical, must be applied. In the present work we have used the Laplace version of the Adomian Decomposition method to obtain an approximate analytical solution of the equilibrium equation of the dark matter condensate. With the use of this solution we have estimated the tangential velocities of massive test particles moving in the dark matter halo. This result allow a comparison of the theoretical predictions of the logarithmic Bose-Einstein Condensate dark matter model with the observations. 

We have performed an extensive comparison of the theoretical predictions of the rotation curves of the present model with a total of 163 galactic curves from the SPARC database \cite{Sparc}. Overall, the fitting results of the logarithmic BEC dark matter model can be characterized as good. Most of the galaxies (75\%) have a $\chi_{red}^2$ smaller than one, which indicates a very good fit of the theoretical model with the data. A number of 138 galaxies (85\%) have a $\chi_{red}^2$ smaller than 2, which represents a good statistical result. 
	\begin{figure}[htbp]
	\centering
	\includegraphics[width=4.2cm]{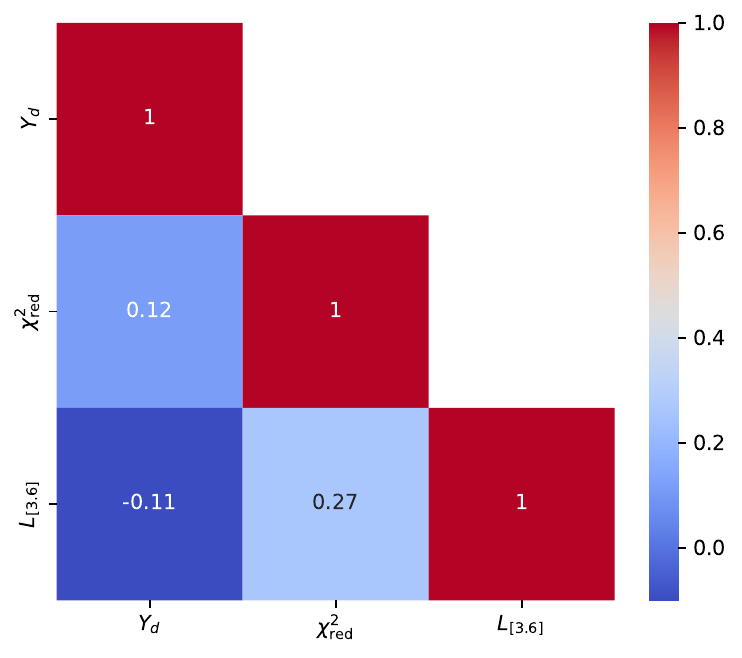}\hspace{.1cm} 
	\includegraphics[width=4.2cm]{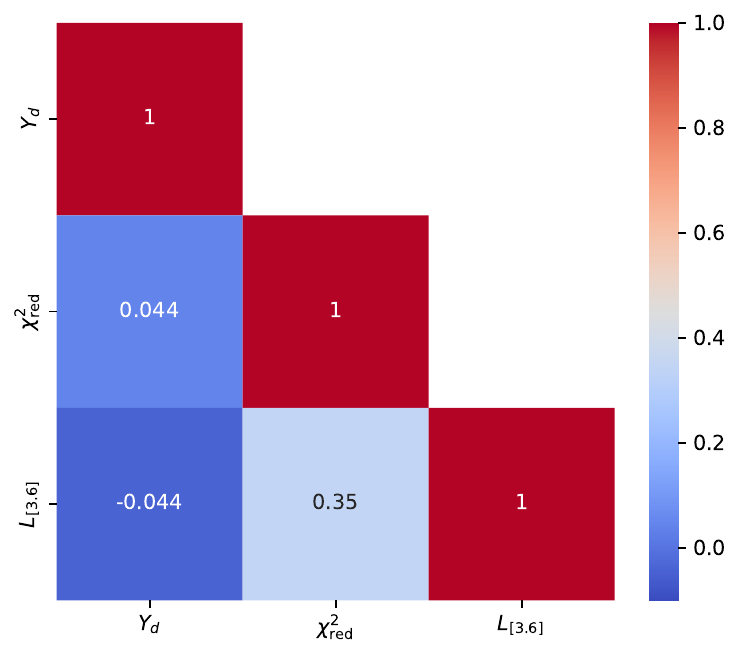} 
	\caption{Correlation heatmap for the sample of 132 bulgeless galaxies, showing the relationships between $(\Upsilon_d, \chi^2_{red}, L_{[3.6]})$. The left panel shows the Pearson linear correlation coefficients, while the right panel shows the Spearman rank correlation coefficients.}
	 \label{LYdcor}
\end{figure}

	We have also analyzed the correlations between  stellar mass-to-light ratio ($\Upsilon_d$),  and $\chi^2_{\rm red}$ obtained from presented dark matter model and the  luminosity ($L_{[3.6]}$) from SPARC database. The sample of $132$ bulgeless galaxies, is the same as that used in Fig.~\ref{correl}.
	The Pearson (left panel) and Spearman (right panel) correlation coefficients are presented in Fig.~\ref{LYdcor}.
	 
 The correlation between $\Upsilon_d$ and $L_{[3.6]}$ is  negligible  (Pearson $r=-0.11$, Spearman $\rho=-0.044$). Similarly, the correlation between $\Upsilon_d$ and $\chi^2_{red}$ is negligible. Therefore, there are no significant correlations between $\Upsilon_d$ and either the goodness of fit or the luminosity. 
	   On the other hand,  $\chi^2_{red}$ has a weak to moderate positive correlation with $L_{[3.6]}$. This indicates that more luminous galaxies tend to have slightly higher reduced chi-squared values, however the overall quality of the fit remains acceptable.  

Consequently, the lack of significant correlation demonstrates that the logarithmic BEC dark matter model is independent of the details of the disk luminosity and the mass-to-light ratio. This feature indicates that the model is applicable to different galaxy types.

Hence, the logarithmic Bose-Einstein Condensate model has at least a phenomenological potential to address the open questions related to the dark matter properties. 

The logarithmic Bose-Einstein Condensate dark matter model we have considered in the present study depends on two intrinsic parameters $b$, the coupling constant in the self-interaction potential in the Gross-Pitaevskii equation, and $m$, the mass of the dark matter particle, and three external (astrophysical) parameters, the central density of the dark matter, its rotation speed, and the mass to light ratios. Our statistical analysis has also allowed to determine the numerical values and the distribution of the astrophysical parameters, as well as of the quantity $B=\sqrt{b/m}$. 

The coupling constant $b$ is the most important parameter characterizing the condensate properties. Unfortunately only the value of the parameter $B$, $B=\sqrt{b/m}$ can be determined from observations. The fittings give for $B$ an average value of the order of  55 km/s. Once a realistic estimate of the mass of the dark matter particle could be obtained, the self-interaction parameter of the condensate dark matter halo could also be determined. The parameter $B$ has a wide variation range, indicating a possible variation from galaxy to galaxy of the dark matter halo properties. However, from a physical point of view such a possibility has been already investigated in the physical literature. In particular, it was pointed out that the coupling parameter $b$ can take different values depending on the local physical conditions, that is, it can change from system to system \cite{Zlo1,Zlo2,Zlo3}. This means that $b$ cannot be considered as a universal constant, but rather a dynamical function $b=b\left(\vec{r},t\right)$, with $b={\rm constant}$ being just an approximation of the general relation, valid only in a finite range of space-time. 

The coupling parameter $b$ can also be related to the background temperature \cite{Zlo1,Zlo2,Zlo3,Zlo4}. In fact $b$ can be directly related to the wave-mechanical temperature $T_\Psi$, which is generally defined as a thermodynamical conjugate of the Everett-Hirschman entropy function,  a quantum information entropy generated by the logarithmic term in the Gross-Pitaevskii equation (\ref{eq:Log interaction})  when one takes the average the wave equation in the Hilbert space of the wave function $\Psi$. The Everett-Hirschman entropy is given by $S_\Psi=-\int_V{ \left|\Psi\right|^2\ln \left(|\Psi|^2\right)/n_c)dV}$ \cite{Zlo3}. Since $T_\Psi$ can be linearly related to the thermodynamic temperature $T$, one can assume that $b \sim T_\Psi \sim  T$, or $b =\alpha \left(T-T_c\right)$, where $\alpha $ is a scale constant, and $T_c$ is a critical temperature, at which the logarithmic term is generated via a phase transition \cite{Zlo3}.  Thus, the coupling  $b$ cannot be considered  as a fixed parameter of the logarithmic BEC model, since its value depends on the environment in which the condensate is located.  Hence the possibility of a space-time varying $b$, taking different values in different astrophysical and galactic environments, as also suggested by the data,  cannot be rejected a priori. 

By using the possible dependence of $b$ on the temperature of the cosmic environment one could obtain a qualitative estimate of the mass of the dark matter particle. Since $b\sim k_BT$, where $k_B$ is Boltzmann's constant, and by taking into account the definition $B=\sqrt{b/m}=\sqrt{k_BT/m}\approx 55\times 10^3$ m/s, we obtain for the mass of the dark matter particle the rough estimate $m\approx 1.368\times 10^{-32}\;{\rm g}=7.67$ eV, where we have taken $T=3$ K as the temperature of the cosmic background. Of course a more precise estimate would require the use of the general relation $b=\alpha k_B\left(T-T_c\right)$, where $\alpha $ is a (background dependent) constant, but presently the range of values of the quantities $\alpha$ and $T_c$ are not known, either on the laboratory, or on cosmic scales.

The distribution of the central densities of the galactic dark matter halo indicates that the values of $\rho_{m0}$ are of the order of $10^{-21}$ g/cm$^3$, also indicating a large spread of this astrophysical parameters. The angular rotation speed of the galactic dark matter halos is also very low, of the order of $10^{-16}$ s$^{-1}$, indicating an almost negligible effect of the rotation on the astrophysical properties of the halos. 

This result can be understood in light of the properties of the condensate wave function, and of the velocity field, which is irrotational, and satisfies the condition $\nabla \times \vec{v}=0$, which follows from the definition of the velocity as $\vec{v}=\nabla \Phi /m$. If the galactic halo is irrotational, then although the dark matter particles possess an angular velocity, they will not induce a global rotation of the condensate. Moreover, phase singularities and quantum
vortices, can also carry angular momentum.

In this work we have pointed out the possibility that the dark matter halos, including the condensate ones, may have a much richer and diverse structure than usually assumed. We have to mention that our results were obtained by assuming a number of simple approximations, which do not necessarily hold for realistic astrophysical systems. Even it obeys the ideal gas type equation of state, the logarithmic Bose-Einstein Condensate, is essentially nonlinear, and its properties are much more closer to those of a quantum liquid than to those of a gas. In our present investigation we have introduced some basic theoretical concepts, which, confronted with the observations, may help in improving our understanding of dark matter phenomenology, and of the physical properties of dark matter.  

\section*{Acknowledgments}

We would like to thank the anonymous reviewer for the careful reading of our manuscript, and for comments and suggestions that helped us to significantly improve our work.    

\appendix

\section{The hydrodynamic representation of the logarithmic BEC's}\label{app1}

In this Appendix, we present the details of the calculations leading to the
hydrodynamic formulation of the Gross-Pitaevskii equation with a logarithmic
self-interaction.

 Starting from the equation (\ref{eq:log SE}), after
substituting the representation of the wave function as given by Eq. (\ref%
{eq:wavefunction}), we first find
\begin{eqnarray}
	i\hbar \dfrac{\partial }{\partial t}\left( \sqrt{n(\overrightarrow{r},t)}e^{i%
		\frac{\Phi(\overrightarrow{r},t)}{\hbar }}\right) &=&-\dfrac{\hbar ^{2}}{2m}%
	\nabla ^{2}\left( \sqrt{n(\overrightarrow{r},t)}e^{i\frac{\Phi(\overrightarrow{r%
			},t)}{\hbar }}\right)  \notag \\
	&&+V\left( \sqrt{n(\overrightarrow{r},t)}e^{i\frac{\Phi(\overrightarrow{r},t)}{%
			\hbar }}\right) .\nonumber\\
\end{eqnarray}

Then, after some simple calculations, we obtain the result
\begin{eqnarray}
	&&i\hbar \left( e^{\frac{i\Phi}{\hbar }}\frac{\partial \sqrt{n}}{\partial t}+%
	\sqrt{n}e^{\frac{i\Phi}{\hbar }}\frac{i}{\hbar }\frac{\partial \Phi}{\partial t}%
	\right) =-\frac{\hbar ^{2}}{2m}\nabla ^{2}\left( \sqrt{n}e^{\frac{i\Phi}{\hbar }%
	}\right) +  \notag \\
	&&V\sqrt{n}e^{\frac{i\Phi}{\hbar }}=-\frac{\hbar ^{2}}{2m}\nabla \cdot \left(
	e^{\frac{i\Phi}{\hbar }}\nabla \sqrt{n}+\sqrt{n}e^{\frac{i\Phi}{\hbar }}\frac{i}{%
		\hbar }\nabla \Phi\right) +  \notag \\
	&&V\sqrt{n}e^{\frac{i\Phi}{\hbar }}=-\frac{\hbar ^{2}}{2m}e^{\frac{i\Phi}{\hbar }}%
	\Bigg(\frac{2i}{\hbar }\nabla \Phi\cdot \nabla \sqrt{n}+\nabla ^{2}\sqrt{n}-
	\notag \\
	&&\frac{1}{\hbar ^{2}}\sqrt{n}|\nabla \Phi|^{2}+\frac{i}{\hbar }\sqrt{n}\nabla
	^{2}\Phi\Bigg)+V\sqrt{n}e^{\frac{i\Phi}{\hbar }}.  \label{eq:reduced logSE}
\end{eqnarray}%

After cancelling the phase factor $e^{\frac{i\Phi}{\hbar }}$ on both sides of Eq.~(\ref%
{eq:reduced logSE}), and after separating the imaginary and real part of it, we arrive to two 
independent equations of motion.

The equation corresponding to the imaginary part of the equation (\ref%
{eq:reduced logSE}) is given by
\begin{equation}
	\hbar \frac{1}{2\sqrt{n}}\frac{\partial n}{\partial t}=-\frac{\hbar ^{2}}{2m}%
	\left( \frac{2}{\hbar }\nabla \Phi\cdot \nabla \sqrt{n}+\frac{1}{\hbar }\sqrt{n}%
	\nabla ^{2}\Phi\right) .
\end{equation}

With the definition $\overrightarrow{v}=\nabla \Phi/m$, the above equation
becomes,
\begin{equation}
	\frac{1}{2\sqrt{n}}\frac{\partial n}{\partial t}=-\overrightarrow{v}\cdot
	\nabla \sqrt{n}-\frac{\sqrt{n}}{2}\nabla \cdot \overrightarrow{v},
\end{equation}%
or
\begin{equation}
	\frac{\partial n}{\partial t}=-(\overrightarrow{v}\cdot \nabla n+n\nabla
	\cdot \overrightarrow{v})=-\nabla \cdot (n\overrightarrow{v}).
\end{equation}%
Hence, we obtain
\begin{equation*}
	\frac{\partial \rho _{m}}{\partial t}+\nabla \cdot (\rho _{m}\overrightarrow{%
		v})=0,
\end{equation*}%
which is the continuity equation in fluid dynamics.

The equation obtained from the real part of the equation (\ref%
{eq:reduced logSE}) gives
\begin{eqnarray}
	\sqrt{n}\frac{\partial \Phi}{\partial t} &=&\frac{\hbar ^{2}}{2m}\left( \nabla
	^{2}\sqrt{n}-\frac{1}{\hbar ^{2}}\sqrt{n}|\nabla \Phi|^{2}\right) -V\sqrt{n}, \nonumber\\
	\frac{\partial \Phi}{\partial t} &=&-\left( -\frac{\hbar ^{2}}{2m}\frac{\nabla
		^{2}\sqrt{n}}{\sqrt{n}}+\frac{1}{2m}|\nabla \Phi|^{2}+V\right)  \notag \\
	&=&-\left( -\frac{\hbar ^{2}}{2m}\frac{\nabla ^{2}\sqrt{\rho _{m}}}{\sqrt{%
			\rho _{m}}}+\frac{mv^{2}}{2}+V\right) ,
\end{eqnarray}%
leading to
\begin{equation}  \label{A7}
	\frac{\partial S}{\partial t}=-(Q+K+V),
\end{equation}%
where $Q=-\left( \hbar ^{2}/2m\right) \nabla ^{2}\sqrt{\rho _{m}}/\sqrt{\rho
	_{m}}$ is the quantum potential, $K=mv^{2}/2$ is the kinetic energy,
and $V=mV_{ext}+b\ln \left( |\Psi |^{2}/|\Psi _{0}|^{2}\right) $ is the
total potential energy. 

With the help of the expression of the fluid velocity, Eq. (\ref{A7})
becomes
\begin{equation}
	\frac{\partial \vec{v}}{\partial t}=-\frac{1}{m}\nabla \left(Q+K+V\right),
\end{equation}
or, equivalently,
\begin{equation}
	\frac{\partial\overrightarrow{v}}{\partial t}+(\overrightarrow{v}\cdot\nabla)%
	\overrightarrow{v}=-\frac{1}{m}\nabla \left(Q+V\right),
\end{equation}
which represents the Euler equation of quantum mechanics.

\end{document}